\documentclass[%
 reprint,
 amsmath,amssymb,
 aps,
]{revtex4-2}

\usepackage{graphicx}
\usepackage{dcolumn}
\usepackage{bm}
\usepackage{xcolor}

\usepackage[subrefformat=parens]{subcaption}
\usepackage{multirow}
\usepackage[disable,color=lightgray]{todonotes}
\let\oldtodo\todo
\renewcommand{\todo}[1]{\oldtodo[inline]{#1}}
\begin{document}

\preprint{APS/123-QED}

\title{Manifestation of spurious currents and interface regularization in wind turbulence over fast-propagating waves}

\author{Hanul Hwang}
\author{Catherine Gorl\'{e}}%
\altaffiliation[Also at ]{Civil and Environmental Engineering, Stanford University, California, USA 94305}
\affiliation{%
Center for Turbulence Research, Stanford University, California, USA 94305
}%


\date{\today}

\begin{abstract}
Accurate simulation of wind turbulence over fast-propagating waves requires interface-capturing methods that suppress numerical artifacts while accurately resolving momentum transfer across the interface. In high wave-age regimes, numerical errors at the air–water interface can reach magnitudes comparable to the physical flow, directly affecting predicted turbulence statistics. This study examines widely used interface-capturing techniques to evaluate \textcolor{black}{how curvature estimation and flux discretization influence wind–wave simulations through the resulting spurious currents and interface regularization.}
A systematic assessment is performed using static and translating droplet benchmarks, together with solitary and monochromatic wave cases, to identify and quantify the dominant numerical error mechanisms. In addition, comparison with experimental measurements reveals how these primary error sources manifest in coupled wind–wave simulations. These findings clarify the numerical origin of the observed discrepancies and underscore the importance of accurate curvature and flux treatment in high wave-age regimes, \textcolor{black}{without which numerical artifacts risk being misattributed to genuine wind–wave physics}.
\end{abstract}

\maketitle


\section{Introduction\label{sec:introduction}}

\textcolor{black}{The study of wind–wave interaction has become increasingly important because of its central role in modern weather forecasting models and its influence on the resilience of coastal communities~\citep{huang2009,chini2010impact,dodet2019contribution}. This interaction arises from the continuous exchange of mass, momentum, and energy across the air–water interface, spanning scales from long-fetch winds and gravity waves down to the small-scale processes that govern local momentum transfer~\citep{breivik2015surface}. Accurate parameterization of these small-scale fluxes provides essential boundary conditions for numerical weather prediction~\citep{powers2017weather} and ocean forecasting systems~\citep{hara2002wind,edson2013exchange}, motivating continued efforts to understand the underlying momentum-transfer mechanisms at scales accessible to numerical simulation. \textcolor{black}{Realizing this potential, however, depends on interface-capturing methods that resolve momentum transfer accurately rather than introducing numerical artifacts of their own.}}

Computational approaches have played an important role in advancing this field by addressing limitations inherent to field measurements and laboratory experiments. Numerical simulations offer detailed and controllable environments for studying complex wind–wave interactions and have yielded insights not accessible through direct observation. These studies have validated existing theories and have also revealed new perspectives on the mechanisms that govern the transfer of momentum and energy across the air–sea interface. Direct numerical simulations (DNS) have, for example, examined wave-induced turbulent structures near the interface~\citep{sullivan2000simulation,yang2010direct}. The analysis of turbulent kinetic energy and Reynolds stress by \citet{yang2010direct} demonstrated strong agreement with experimental observations. Large-eddy simulations have further expanded these capabilities. \citet{sullivan2014large} introduced a broadband spectrum of time-varying finite-amplitude surface waves at the lower boundary to explore a wide range of wind conditions. The resulting wave-induced and turbulent wind stresses, and their contributions to total drag, have been examined in detail~\citep{hara2015wave}. In addition, \citet{cao2021numerical} identified the vertical component of wave orbital velocity as the principal source of airflow and pressure perturbations, which play a pivotal role in shaping the spatial structure of the coupled system.

Despite these advances, most DNS and high-fidelity LES studies remain limited by their treatment of the air–water interface. Many investigations have focused exclusively on the wind flow while omitting explicit resolution of the interface~\citep{yang2010direct,cao2021numerical,lin2008direct}. Single-phase frameworks often approximate the interface through simplified boundary conditions~\citep{sullivan2000simulation}, commonly imposing a lower boundary motion to mimic mild-sloped waves. In such formulations, the Navier–Stokes equations are solved in a wave-following coordinate system that restricts allowable wave slopes due to stability constraints associated with the Jacobian operator and that cannot represent surface drift. In other cases, wave effects are represented through localized form drag~\citep{suzuki2013impact} or through empirical and potential-flow-based models derived from observations~\citep{sullivan2018turbulent}. This simplification frequently excludes critical two-phase processes such as wave breaking, bubble entrainment, sea spray generation, and spume formation~\citep{andreas1992sea,holthuijsen2012wind,makin2005note}. Neglecting these processes is particularly problematic in high-energy conditions, where the dynamics of breaking waves differ fundamentally from those of non-breaking waves~\citep{banner1976separation,belcher1998turbulent,banner1998tangential}. High-fidelity simulations that explicitly address such small-scale dynamics remain limited, with only a few examples such as \citet{yang2018direct}. Although two-phase simulations have been applied to more complex problems, including breaking waves and wave–structure interactions, many of these efforts are dominated by liquid-phase dynamics~\citep{song2004numerical,iafrati2013modulational} or rely on Reynolds-averaged Navier–Stokes models~\citep{Ren2014,Liu2017,Park2018}, which restrict their ability to represent detailed wind–wave coupling. \textcolor{black}{Even among the two-phase, interface-resolving simulations that do exist, the sensitivity of the resulting flow field to the choice of interface-capturing method is rarely quantified. This includes the accuracy of curvature estimation and the discretization of the interface-advection flux. Numerical artifacts can arise from these choices, such as spurious currents near the interface and the artificial momentum flux introduced by interface regularization. \textcolor{black}{The former have been recognized since the earliest continuum-surface-force computations of interfacial flow~\citep{brackbill,lafaurie1994modelling}, where an imbalance between the discrete surface-tension force and the pressure gradient was shown to drive a persistent vortical velocity field near the interface, with a characteristic magnitude set by the ratio of surface tension to dynamic viscosity. Their sensitivity to the surface-tension discretization, the curvature estimator, and the interface representation has since been characterized systematically~\citep{denner2014fully,abadie2015combined} and reviewed by \citet{popinet2018numerical}, though largely in canonical test configurations.} Left unquantified, these artifacts risk being mistaken for genuine wind–wave turbulence physics rather than recognized as numerical error.}

Realistic wind–wave simulations require accurate modeling of interface advection and surface-tension discretization. Such accuracy depends on reliable resolution of the air–water interface, which can be achieved through either explicit interface tracking or implicit interface-capturing. \textcolor{black}{Each of the principal interface-capturing families carries a characteristic weakness, and each has seen sustained development directed at that weakness.}

\textcolor{black}{The level-set method~\citep{osher1988fronts} represents the interface as the zero level of a smooth signed-distance function. It gives accurate curvature and normal estimates, but it does not conserve mass exactly without additional coupling to a volume-of-fluid (VOF)-like field~\citep{sussman2000coupled}.} \textcolor{black}{This limitation has motivated substantial subsequent development. For example, reference map methods represent interface deformation through a map to the reference configuration rather than by repeatedly reinitializing a signed-distance field, which substantially reduces the associated mass loss~\citep{rycroft2020reference,theillard2021volume}.}

\textcolor{black}{The phase-field method~\citep{liu2003phase} instead diffuses the interface over a finite thickness governed by a Cahn--Hilliard-type equation, giving smooth curvature evaluation. However, this thickness must remain resolved by the computational grid, so in practice it is often set far larger than the true physical interface thickness. On coarse grids this can introduce artifacts whose magnitude is difficult to separate from genuine interfacial dynamics, a limitation conceptually similar to the numerical sensitivities examined in this study for VOF-type methods.} \textcolor{black}{Conservative formulations address this directly by introducing a regularization flux that balances sharpening against diffusion, holding the interface thickness fixed at a prescribed value, and extend the approach from the spectral discretization of the original formulation to unstructured grids~\citep{hwang2024robust}.}

\textcolor{black}{The VOF method~\citep{hirt1981volume} tracks a conserved phase-indicator field instead of a smooth auxiliary function, offering exact mass conservation and comparatively low cost, at the expense of the curvature accuracy addressed throughout this study.} \textcolor{black}{Mitigation here has developed along the same two axes this study examines. Addressing the capillary force directly, formulations that fit a higher-order surface to the phase indicator or that confine the surface-tension force to a narrow band at the interface, such as the parabolic reconstruction of surface tension~\citep{renardy2002prost} and the sharp surface force model~\citep{francois2006balanced}, target the curvature error at its source. Addressing the advection instead, bounded compressive schemes that hold the phase indicator sharp, among them the multidimensional universal limiter with explicit solution (MULES) employed in this study and the high-resolution interface capturing scheme, have been reported to lower the residual interfacial velocity~\citep{hoang2013benchmark,georgoulas2015numerical,aboukhedr2018simulation}, and are surveyed by \citet{mulbah2022review}. Assessment of these strategies has nonetheless concentrated on bubble-driven configurations, ranging from quasi-static growth and detachment at submerged orifices to confined Taylor-bubble transport, and the OpenFOAM isoAdvector solvers used here have likewise been benchmarked against single-bubble cases~\citep{gamet2020validation}. Wind–wave turbulence poses the question differently, because the interface is deformed continuously by an external shear flow over long integration times, and what matters is not the absolute spurious velocity but its magnitude relative to the air-side wave-coherent and turbulent velocities the simulation is meant to resolve.}

\textcolor{black}{
In practice, the choice of interface-capturing method for wind–wave simulations is more constrained than this general comparison suggests. Realistic domains often involve irregular bathymetry, sediment or particle transport, and fluid--structure interaction. These favor unstructured-mesh or immersed-boundary-compatible formulations over methods that rely on a smooth, globally defined auxiliary field. At the same time, resolving small-scale interface deformation while accommodating a large, open wind-side domain requires local mesh refinement, through either adaptive mesh refinement or spatially varying grid spacing, without excessive global cost. The method must also remain computationally efficient over long simulation durations and robust across the large density ratio between air and water. 
The interface-capturing methods introduced in this study are VOF-type methods, which are widely adopted as compatible with these requirements. The specific curvature and flux treatments investigated here represent the practical standard for this class of problem.
}

\textcolor{black}{
Within the VOF framework itself, two broad reconstruction strategies exist. Geometric VOF methods reconstruct the interface explicitly within each cell using piecewise-linear interface calculation (PLIC) schemes, in which a planar or polygonal interface segment is oriented according to an estimated interface normal and positioned to match the local volume fraction. Algebraic VOF methods instead advect the phase indicator directly without explicit geometric reconstruction. Various hybrid methods have also been proposed to combine the strengths of these approaches with those of other interface representations. For example, hybrid level-set/VOF (CLSVOF) methods retain the level-set field to improve curvature accuracy while using the VOF field to enforce exact mass conservation~\citep{sussman2000coupled}. 
}

In this study, we focus on these widely adopted interface-capturing methods as implemented in multiphase-flow solvers such as OpenFOAM. We investigate the manifestation of spurious currents arising from \textcolor{black}{curvature-estimation} errors and from interface-advection schemes used in flux evaluation, and assess their impact on wind–wave flow characteristics. By identifying the sensitivities associated with these numerical treatments, this study provides an essential step toward improving wind–wave interaction models and enhancing predictive capability for environmental and engineering applications. \textcolor{black}{These numerical sensitivities are not purely algorithmic. As shown later, the same errors distort the wave-coherent and turbulent stress fields when validated against laboratory wind–wave measurements. They change which momentum-transfer and turbulence-generation mechanisms the simulation appears to capture. Left unexamined, such algorithm-dependent artifacts can be mistaken for genuine wind–wave physics. Separating the two is therefore a critical physical question, not only a numerical one.}

This paper is organized as follows. Section~\ref{sec:method} describes the governing equations and numerical models, including the interface-capturing techniques and post-processing procedures employed in this study. Section~\ref{sec:results:canonical} presents canonical test cases, beginning with static drop simulations to evaluate \textcolor{black}{curvature-estimation} errors and assess grid-resolution effects, followed by moving-drop cases. Section~\ref{sec:results:waves} investigates interface advection using solitary-wave simulations and practical monochromatic wave cases, with comparisons to experimental data. Conclusions are presented in Section~\ref{sec:conclusions}.

\section{Governing equations and interface-capturing methods\label{sec:method}}
In this section, we describe the governing equations for fluid motion, followed by a comprehensive overview of interface-capturing methods used in this study. We then introduce the post-processing procedure for wind and wave simulation data using triple decomposition.

\subsection{Governing equations\label{sec:method:governing equations}}
The incompressible Navier-Stokes equations governing two immiscible fluids (phases$1$ and $2$), \textcolor{black}{using the Einstein summation convention for repeated indices,} are given by
\begin{equation}
    \frac{\partial u_i}{\partial x_i} = 0,
    \label{eq:governing equation, continuity equation}
\end{equation}
\begin{eqnarray}
    &\displaystyle
    \frac{\partial \rho u_i}{\partial t} + \frac{\partial \rho u_i u_j}{\partial x_j} = - \frac{\partial p^*}{\partial x_i} + \frac{\partial}{\partial x_j} \left( \mu_{\text{eff}} \frac{\partial u_i}{\partial x_j} \right) \nonumber \\
    &\displaystyle
    - g_j X_j \frac{\partial \rho}{\partial x_i} 
    + \sigma \kappa \frac{\partial \phi}{\partial x_i},
    \label{eq:governing equation, momentum equation}
\end{eqnarray}
where $\rho$ and $u_i$ denote the density and velocity components, respectively. The pressure $p^*$ is defined as $p^* = p - \rho g_j X_j$, where $p$ represents the total pressure and $g_j$ the gravitational acceleration. The term $X_j$ corresponds to the position vector relative to a specified reference point. 
The effective viscosity, $\mu_{\text{eff}}$, is expressed as $\mu_{\text{eff}} = \mu + \rho \nu_{t}$, where $\mu$ is the dynamic viscosity and $\nu_{t}$ denotes the eddy viscosity.
The phase indicator function, $\phi$, is used to distinguish between the two immiscible fluid phases. A value of $\phi = 1$ denotes that the control volume is fully occupied by phase$1$, whereas $\phi = 0$ corresponds to a region entirely filled with phase$2$.
Based on this formulation, the local fluid properties are computed as 
\begin{equation}
    \rho = \rho_1 \phi + \rho_2 (1 - \phi),
\end{equation}
\begin{equation}
    \mu = \mu_1 \phi + \mu_2 (1 - \phi),
\end{equation}
for density and viscosity, respectively.

The governing equations, Eqs.~\eqref{eq:governing equation, continuity equation} and \eqref{eq:governing equation, momentum equation}, are discretized using a finite-volume method on an unstructured mesh. \textcolor{black}{The domains used in this study consist predominantly of hexahedral cells, with local mesh refinement introducing non-hexahedral transition cells between different refinement levels.} Spatial discretization is carried out using a second-order \textcolor{black}{central} difference scheme, while temporal integration is performed using an implicit \textcolor{black}{first-order Euler method. This choice follows the default temporal scheme of the interFoam family of solvers in OpenFOAM, one of the most widely used open source frameworks for wind–wave and free-surface simulations, underlying toolboxes such as waves2Foam~\citep{jacobsen2012waves2foam} and IHFoam~\citep{higuera2013}, and adopted in large-scale wind–wave turbulence studies such as \citet{cimarelli2023}. A blended Crank–Nicolson/Euler scheme is available for second-order temporal accuracy. However, pure Crank–Nicolson is unstable when coupled with the built-in interface-capturing schemes considered here, and the blending coefficient must be tuned for each case to avoid solution blow-up. The first order implicit Euler scheme is therefore retained as the robust default.
This default configuration is the one whose numerical sensitivities are quantified throughout this study, making the results directly relevant to separating physical wind–wave dynamics from numerical artifacts in simulations that use the same framework.
}
The pressure-velocity coupling is achieved employing the pressure implicit with split operator (PISO) algorithm.
The surface tension effect is included in the last term of Eq.~\eqref{eq:governing equation, momentum equation}, where $\sigma$ denotes the surface tension coefficient, $\kappa$ represents the interface curvature.
We employ the continuum surface force (CSF) formulation proposed by \citet{brackbill}, together with the balanced-force algorithm of \citet{francois2006balanced}.

\subsection{Interface modeling\label{sec:method:interface modeling}}
The interface between two immiscible fluids is modeled using the VOF method, an interface-capturing method categorized under one-fluid models.
We consider two variants of the VOF method: a geometric formulation based on isoAdvector~\cite{roenby2016computational,scheufler2019accurate} and an algebraic formulation~\cite{higuera2013} employing the MULES limiter~\cite{marquez2013extended}. Within the geometric formulation, two subclasses are examined, distinguished by their interface-reconstruction procedures. Although detailed algorithmic descriptions are available in the original references, the interface-capturing methods are outlined here to clarify differences in interface treatment. The reconstruction methods associated with the geometric formulation are further described in Appendix~\ref{sec:method:reconstruction of the interface}.

\subsubsection{Geometric VOF with isoAdvector\label{sec:method:geometric VOF}}
For the geometric VOF formulation isoAdvector, developed by \citet{roenby2016computational}, the phase indicator $\phi$ is transported according to
\begin{equation}
    \frac{\partial \phi}{\partial t} + \frac{\partial u_{i}\phi}{\partial x_i} = 0.
    \label{eq:phase indicator transport equation, isoAdvector}
\end{equation}
Unlike the algebraic VOF transport equation below (e.g., see Eq.\eqref{eq:phase indicator transport equation, algebraic VoF}), Eq.\eqref{eq:phase indicator transport equation, isoAdvector} does not include an artificial compression term.

\begin{figure}
    \centering
    \includegraphics[width=0.6\columnwidth]{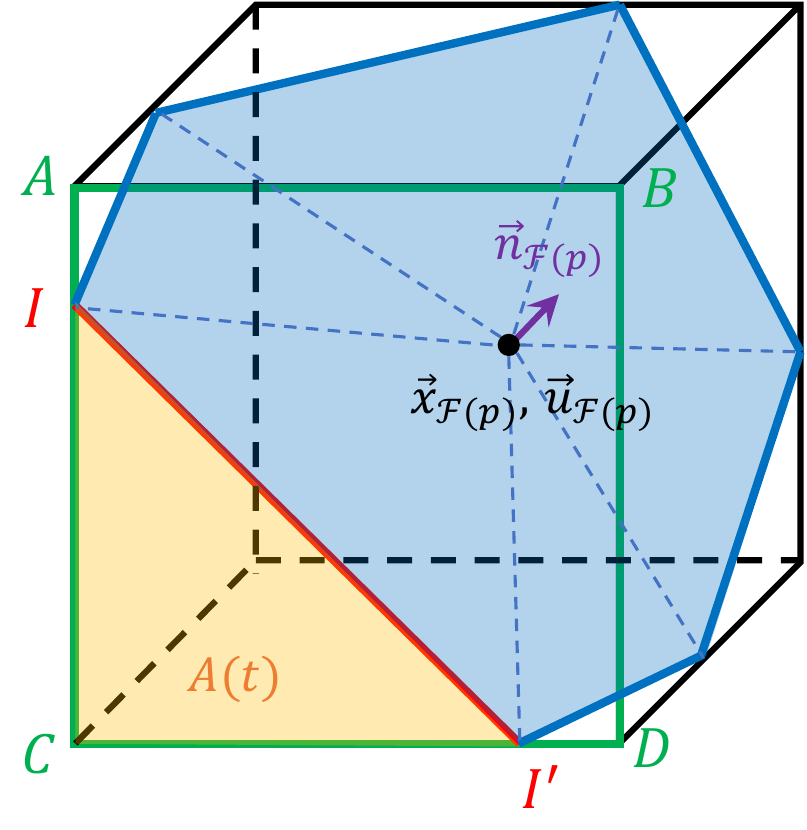}
    \caption{A schematic of $\phi_p\text{-isoface}$.}
    \label{fig:isoface schematic}
\end{figure}

In this approach, the face flux is computed based on geometric considerations of the interface, and the evolution of $\phi$ is advanced in two steps.
First, the interface is reconstructed within each computational cell. In a collocated framework, the phase indicator $\phi$ is stored at the cell center. The reconstruction process seeks to determine a plane that divides the cell into two subregions, each occupied entirely by one phase.
The objective is to ensure that the phase indicator, defined as the volume fraction of phase$1$ within a subcell, is consistent with the cell-centered volume fraction $\phi_p$, such that $\phi_p = V_{\text{phase}1} / V_{cv}$, \textcolor{black}{where $V_{cv}$ is the volume of the computational cell (control volume)}.
As illustrated in Fig.~\ref{fig:isoface schematic}, a cubic cell $p$ (outlined in black) is intersected by iso-lines (blue lines), which constitute the periphery of the $\phi_p$-isoface. The geometric properties of the $\phi_p$-isoface, including its face center $\mathbf{x}_{\mathcal{F}(p)}$, face normal $\mathbf{n}_{\mathcal{F}(p)}$, and velocity vector $\mathbf{u}_{\mathcal{F}(p)}$, are then determined. Further details regarding the reconstruction schemes used in this study are provided in Section~\ref{sec:method:reconstruction of the interface}.

In the second step, the face flux is computed based on the motion of the $\phi_p$-isoface. To update $\phi$ from time step $t_n$ to $t_{n+1} = t_n + \Delta t$, Eq.~\eqref{eq:phase indicator transport equation, isoAdvector} is integrated over the control volume $V_{\text{cv}}$ and in time as follows.
\begin{equation}
    \phi^{n+1} = \phi^n - \frac{1}{V_{\text{cv}}} \sum_{f} \Delta V^n_{f}.
    \label{eq:isoAdvector:discretized phase indicator}
\end{equation}
Here, $\Delta V^n_{f}$ represents the volumetric face flux, which is approximated as
\begin{equation}
    \Delta V^n_{f} \approx \frac{\Phi_{f}^n}{|S_{f,i}|} \int_{t_n}^{t_{n+1}} A(\tau) \textnormal{d}\tau.
    \label{eq:isoAdvector:total volume of fluid flux}
\end{equation}
The face flux, $\Phi_{f}$, is given by $\Phi_{f} = u_{f,i}S_{f,i}$, where $u_{f,i}$ and $S_{f,i}$ represent the face-interpolated velocity and face normal vectors, respectively.
In Eq.~\eqref{eq:isoAdvector:total volume of fluid flux}, $A(\tau)$ represents the instantaneous area submerged by phase$1$.

As depicted in Fig.~\ref{fig:isoface schematic}, if we calculate the volumetric face flux $\Delta V_{f}$ in the green solid-line ABCD plane, the intersection between the ABCD plane and the $\phi_p$-isoface is marked as the red line $\overline{II'}$. The submerged area (highlighted in yellow) varies over time as the $\phi_p$-isoface propagates. The face-interface intersection $\overline{II'}$ sweeps a quadrilateral (green square), leading to an expression for the submerged area $A(\tau) = P_k \tau^2 + Q_k \tau + A_j(\widetilde{t}_{k})$. \textcolor{black}{We emphasize that the cubic cell shown in Fig.~\ref{fig:isoface schematic} is used purely as an illustrative example. The isoAdvector algorithm itself is formulated for arbitrary polyhedral cells and is not restricted to cubic or hexahedral geometries.}
Thus, the time integration of the submerged area $A(\tau)$ in Eq.~\eqref{eq:isoAdvector:total volume of fluid flux} is expressed as
\begin{equation}
    \int_{t_{n}}^{t_{n+1}} A_j(\tau) \textnormal{d}\tau = \sum_{k=1}^{N_{v}-1} \int_{\widetilde{t}_{k}}^{\widetilde{t}_{k+1}} A_j(\tau) \textnormal{d}\tau,
    \label{eq:isoAdvector:isoface area sum}
\end{equation}
where $N_{v}$ denotes the total number of vertices in the corresponding cell. The time instances $\widetilde{t}_k$ represent the moments at which the $\phi_p$-isoface passes through each $k^{\text{th}}$ vertex of the cell and are approximated as
\begin{equation}
    \widetilde{t}_k \approx t_n + (x_{k,i} - x_{\mathcal{F}(p),i}) \frac{n_{\mathcal{F}(p),i}}{u_{\mathcal{F}(p),i} n_{\mathcal{F}(p),i}}.
    \label{eq:t tilde}
\end{equation}
Here, $x_{k,i}$ denotes the spatial coordinate of the $k^\text{th}$ vertex, with $k = 1,\dots,N_{v}$. The quantities $n_{\mathcal{F}(p),i}$ and $u_{\mathcal{F}(p),i}$ refer to the components of the interface normal and the velocity of the $\phi_p$-isoface at its center. The definitions of these quantities are provided in Eq.~\eqref{eq:normal of polygonal face} for the interface normal and Eq.~\eqref{eq:center of polygonal face} for the centroid in the following section. The velocity $u_{\mathcal{F}(p),i}$ is obtained by linear interpolation of the velocities at the vertices of the tetrahedron that contains the point $\vec{x}_{\mathcal{F}(p)}$.
The coefficients $P_k$ and $Q_k$ are polynomial terms computed from the geometric configuration of the $\phi_p\text{-isoface}$ near each vertex. Additional details regarding their evaluation can be found in \citet{roenby2016computational}. Using this information, each integral term in Eq.~\eqref{eq:isoAdvector:isoface area sum} can be evaluated analytically as
\begin{eqnarray}
    &\displaystyle \int_{\widetilde{t}_{k}}^{\widetilde{t}_{k+1}} A_j(\tau) \textnormal{d}\tau = \frac{1}{3}\left( \widetilde{t}^3_{k+1} - \widetilde{t}^3_{k} \right) P_k \nonumber \\
    &\displaystyle + \frac{1}{2}\left( \widetilde{t}^2_{k+1} - \widetilde{t}^2_{k} \right) Q_k
    + \left( \widetilde{t}_{k+1} - \widetilde{t}_{k} \right) A_j(\widetilde{t}_{k}).
    \label{eq:isoAdvector:isoface area change}
\end{eqnarray}

The total flux associated with interface transport is necessary for the convection term in the momentum equation, Eq.\eqref{eq:governing equation, momentum equation}. In isoAdvector approach, it is computed as
\begin{equation}
    F_{\text{ISO}} = \frac{\Delta V_f}{\Delta t} (\rho_1 - \rho_2) + \Phi_f \rho_2.
    \label{eq:total flux, geometric VOF}
\end{equation}
The superscript $n$ is dropped for simplicity. Since Eq.\eqref{eq:isoAdvector:discretized phase indicator} integrates the volumetric flux in time, the first term in Eq.\eqref{eq:total flux, geometric VOF} is divided by $\Delta t$.

\todo{Think about this paragraph again, whether to include or not. Also, acronyms}
\todo{clarify isoAdvector and isoAlpha}
As a brief note on the isoAdvector method, it offers improved computational efficiency on unstructured grids compared to conventional geometric VOF reconstruction approaches. Traditional methods, such as the Mixed Youngs-Centered scheme or the Efficient Least-Square Volume-of-Fluid Interface Reconstruction Algorithm, involve more complex interface reconstruction procedures~\citep{roenby2016computational}. In addition, isoAdvector bypasses the need for operator splitting or unsplit advection schemes, both of which can introduce significant parallel load imbalances in unstructured grid frameworks~\citep{Jofre2015}. Despite these advantages, isoAdvector has been reported to generate larger spurious currents near the interface, an issue that is further analyzed below.

\subsubsection{Algebraic VOF with MULES limiter\label{sec:method:algebraic VOF}}
For the algebraic VOF approach, the transport equation for the phase indicator $\phi$ is given by
\begin{equation}
    \frac{\partial \phi}{\partial t} + \frac{\partial u_{i}\phi}{\partial x_i} + \frac{\partial \mathcal{U}_{i}\phi (1-\phi)}{\partial x_i} = 0.
    \label{eq:phase indicator transport equation, algebraic VoF}
\end{equation}
Here, $\mathcal{U}_{i}$ represents the relative velocity between the two phases. The last term, commonly referred to as the compression term, is introduced to maintain a sharp interface representation. This term becomes active only in interface cells, where the phase indicator takes intermediate values, $\phi \in (0,1)$, and operates in the direction normal to the interface.

To discretize the compression term, the normal component of $\mathcal{U}_{i}$ is defined at the cell faces as
\begin{equation}
    \mathcal{U}_{fn} = n_{f} C_{\alpha}\frac{|\Phi_{f}|}{|S_{f,i}|}.
    \label{eq: relative velocity definition}
\end{equation}
The subscript $f$ denotes values evaluated at cell faces. The coefficient $C_{\alpha}$ controls the intensity of interface compression and is set to unity in the present study, unless stated otherwise. The term $n_{f}$, representing the dot product of the interface normal and the face normal vector $S_{f,i}$, is defined as
\begin{equation}
    n_f = \left( \frac{\partial \phi}{\partial x_i}  \bigg/ \left|\frac{\partial \phi}{\partial x_i} \right| \right) S_{f,i}.
    \label{eq: definition of nf}
\end{equation}

The discretized form of Eq.~\eqref{eq:phase indicator transport equation, algebraic VoF} within a finite-volume framework is written as
\begin{equation}
    \frac{\phi^{n+1}-\phi^{n}}{\Delta t} V_{cv} + \sum_{f} \left\{ \phi_f^{n} F_{L}^{n} + \phi_f^{n} (1-\phi_f^{n}) F_{NL}^{n} \right\} = 0.
    \label{eq:discretized algebraic VOF equation}
\end{equation}
Here, $V_{cv}$ represents the control volume of a computational cell. The terms $F_L$ and $F_{NL}$ correspond to the flux contributions from the linear and nonlinear terms, respectively, and are defined as $F_L = \Phi_f$ and $F_{NL} = \mathcal{U}_{fn}$.
The face-interpolated phase indicator, $\phi_f$, is computed using a high-resolution scheme to maintain boundedness and accuracy of $\phi$. The superscripts $n$ refers to the $n^{th}$ time step.
Equation~\eqref{eq:discretized algebraic VOF equation} is solved iteratively using the MULES solver~\citep{marquez2013extended} to preserve boundedness of the phase indicator, as accurate maintenance of a sharp interface is essential in two-phase flow simulations.
This approach is based on Zalesak’s limiter~\citep{zalesak1979fully}, where the weighting factors of the high-order (or antidiffusive) flux are determined iteratively.

The total flux associated with the transport of $\phi$, corresponding to the second term in Eq.~\eqref{eq:discretized algebraic VOF equation}, is calculated by solving
\begin{equation}
    F_{\phi}^{n} = \phi_f^{n} F_{L}^{n} + \phi_f^{n} (1-\phi_f^{n}) F_{NL}^{n}.
    \label{eq:subcycle flux, algebraic VOF}
\end{equation}
Further details on the MULES solver can be found in \citet{marquez2013extended}.
%
The values of $\phi$ and $F_{\phi}$ are updated at the end of each subcycle. Accordingly, the total flux for MULES, denoted by $F_{\text{MULES}}$, which accounts for both transport and compression of $\phi$, is computed as
\begin{equation}
    F_{\text{MULES}} = \sum_{i=1}^{n_{sc}} \frac{\Delta t_{sc}}{\Delta t} F_{sc,i},
    \label{eq:total flux, algebraic VOF}
\end{equation}
where $F_{sc,i}$ is given by
\begin{equation}
    F_{sc,i} = F_{\phi} (\rho_1 - \rho_2) + F
    _{L} \rho_2. 
    \label{eq:intermediate total flux, algebraic VOF}
\end{equation}
The superscript $n$ is dropped to be consistent with the expression Eq.~\eqref{eq:total flux, geometric VOF}.
Finally, the computed total flux $F_{\text{MULES}}$ is used to update the momentum equation.

\subsection{Phase-averaging and triple decomposition\label{sec:method:triple decomposition}}

To analyze the wind flow field over surface waves, a phase-averaging approach based on the Hilbert transform is employed to quantify the statistical properties of wind turbulence~\cite{gardner1989statistical}. Assuming that the wave-coherent component and the turbulent fluctuations of the wind are sufficiently uncorrelated, the velocity field is decomposed into three distinct components: a phase-independent mean, a wave-coherent component, and a turbulent fluctuation component.

\begin{figure}
    \centering
     \begin{subfigure}[b]{0.9\columnwidth}
         \centering
         \caption{}
         \includegraphics[width=\textwidth]{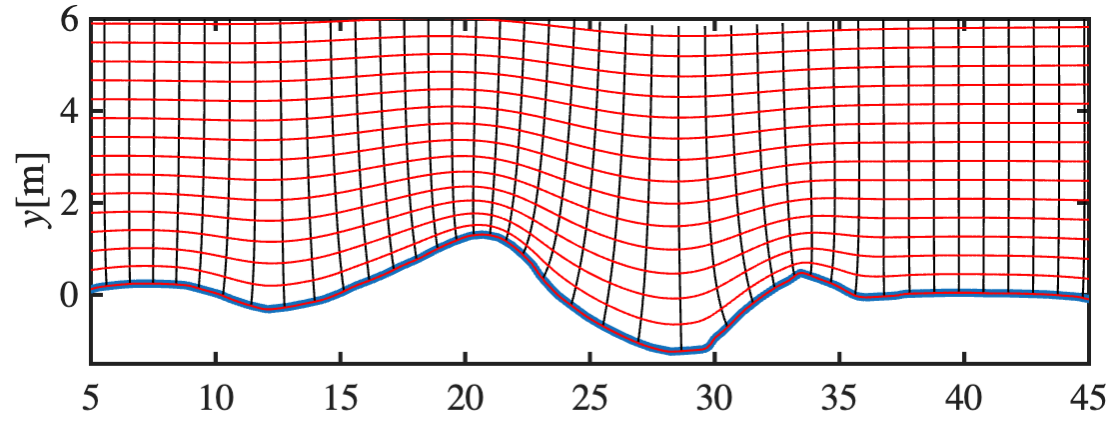}
     \end{subfigure}
     \\
     \begin{subfigure}[b]{0.9\columnwidth}
         \centering
         \caption{}
         \includegraphics[width=\textwidth]{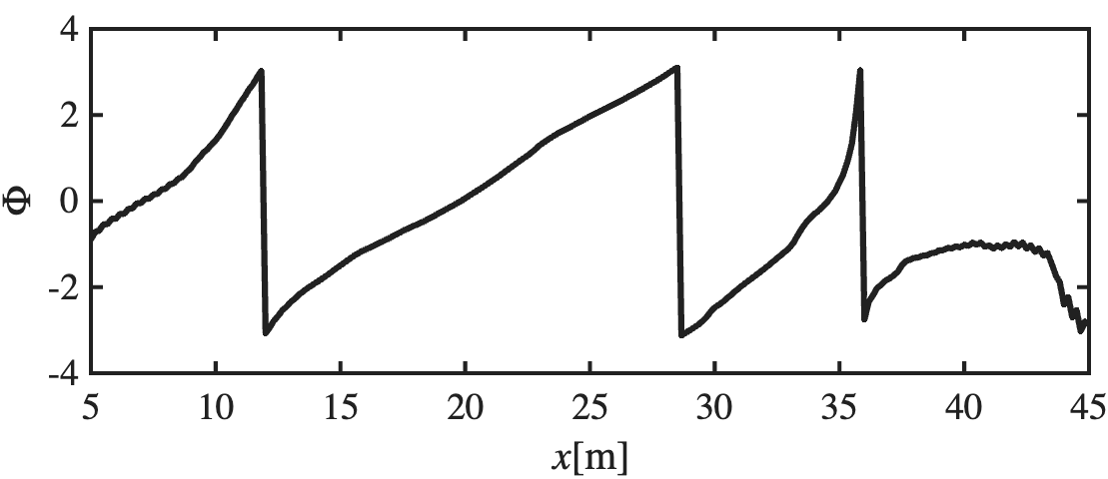}
     \end{subfigure}
    \caption{\textcolor{black}{(a) Illustration of the wave-following coordinate system $(\xi,\zeta)$, used only in the post-processing phase-averaging procedure of Section~\ref{sec:method:triple decomposition} and not the computational mesh. Red solid lines denote curves of constant $\zeta$, which follow the wave surface, and black solid lines denote curves of constant $\xi$, which are locally orthogonal to the surface and become vertical, parallel lines away from the interface. The wave surface itself is indicated by the thicker blue solid line.} (b) Corresponding wave phase $\Phi$ (Eq.~\eqref{eq:phase identification}) of the signal shown in (a).}
    \label{fig:wave following coordinate}
\end{figure}





\todo{Move some portion to appendix for record purpose?}

Consider a wave elevation field $\eta(\mathbf{x}, t)$. The corresponding analytic signal is defined as $\eta(\mathbf{x}, t) + \textnormal{i}\mathcal{H}_\eta(\mathbf{x}, t)$, where $\mathcal{H}_\eta(\mathbf{x}, t)$ denotes the inverse transform to physical space of the Hilbert transform of the wave signal evaluated in spectral space. The wave phase $\Phi(\mathbf{x}, t)$ is then defined as
\begin{equation}
\Phi(\mathbf{x}, t) = \arg\left[\eta(\mathbf{x}, t) + \mathrm{i}\mathcal{H}\eta(\mathbf{x}, t)\right].
\label{eq:phase identification}
\end{equation}

\begin{figure*}
    \centering
    \includegraphics[width=0.8\textwidth]{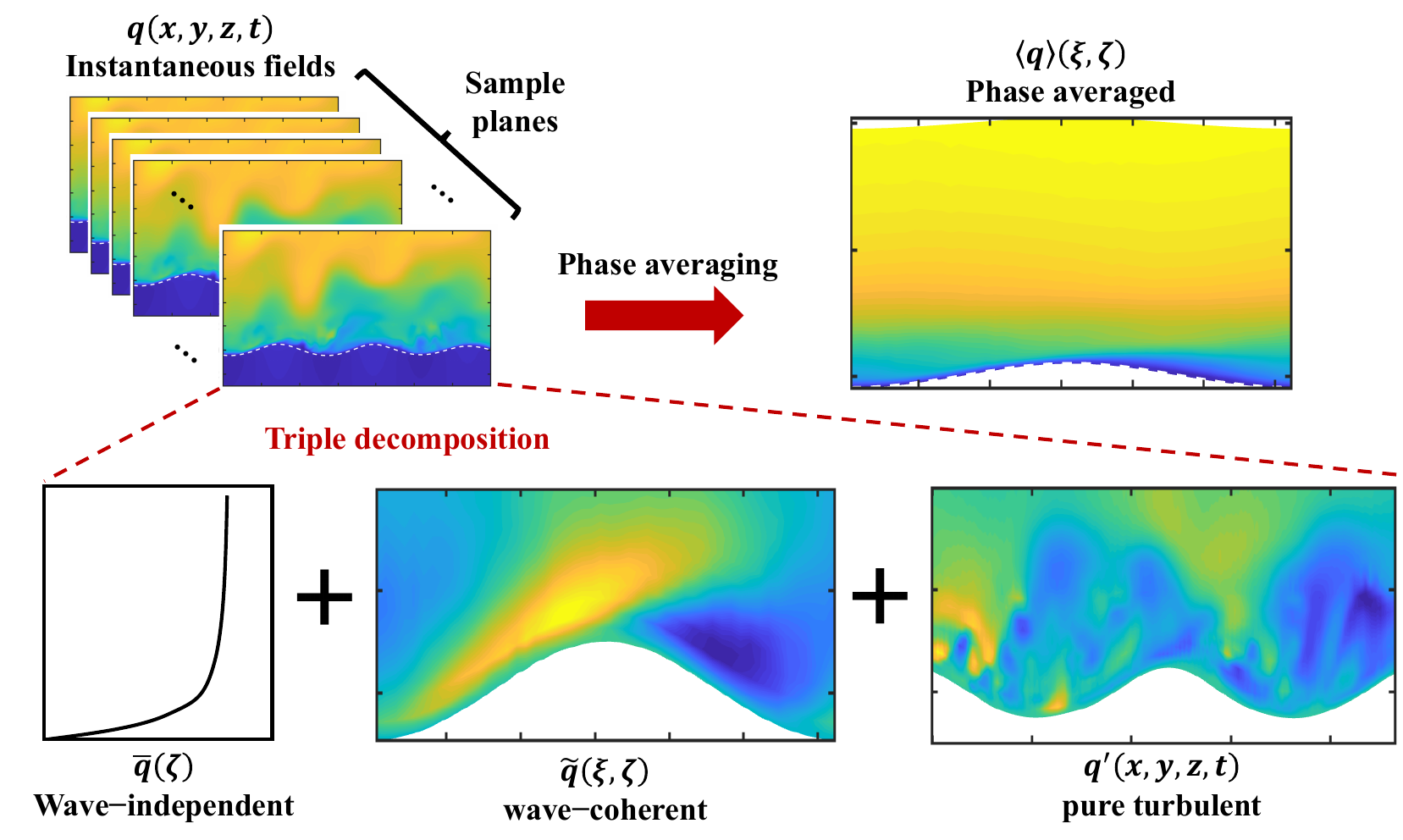}
    \caption{Schematic of the triple decomposition. The phase-averaged field $\langle \mathbf{u} \rangle (\xi,\zeta)$ is obtained from $N_{\text{samples}}$ snapshots, and the velocity field is decomposed into the wave-independent component $\mathbf{\overline{u}}(\zeta)$ (Eq.~\eqref{eq:wave independent component}), the wave-coherent component $\mathbf{\widetilde{u}}(\xi,\zeta)$ (Eq.~\eqref{eq:wave coherent component}), and the turbulent component $\mathbf{u'}(x,y,z,t)$ (Eq.~\eqref{eq:definition of pure turbulence})}
    \label{fig:triple decomposition schematic}
\end{figure*}

Once the unique phase of the wave signal is identified, the turbulent component $\mathbf{u'}$ of the flow field $\mathbf{u}$ is defined as \begin{equation}
\mathbf{u'}(x,y,z,t) = \mathbf{u}(x,y,z,t) - \langle \mathbf{u} \rangle (\xi,\zeta),
\label{eq:definition of pure turbulence}
\end{equation}
where $\langle \mathbf{u} \rangle(\xi,\zeta)$ denotes the phase-averaged velocity field in a wave-following coordinate system, and $(x, y, z)$ and $t$ represent the Cartesian coordinates and time, respectively. Fig.~\ref{fig:wave following coordinate} illustrates the wave-following coordinate system along with the corresponding wave phase.

The phase-averaged component is obtained as
\begin{equation}
\langle \mathbf{u} \rangle (\xi,\zeta) = \frac{1}{N_{t} N_{z}} \sum_{i=1}^{N_{t}} \sum_{j=1}^{N_{z}} \mathbf{u}(\xi - c t_i, y_j, \zeta, t_i),
\label{eq:phase averaging}
\end{equation} where $N_t$ and $N_z$ are the number of temporal snapshots and the number of sampling planes in the spanwise direction, respectively. $c$ is the wave celerity. It is assumed that the flow is statistically homogeneous in the spanwise direction. The total number of samples used in the averaging is $N_{\text{samples}} = N_t \times N_z$.
The wave-following coordinate system and the procedure for computing the phase-averaged velocity field are described in detail in Section~\ref{sec:wave-following coordinate systeme}.


The phase-averaged velocity field $\langle \mathbf{u} \rangle(\xi, \zeta)$ is further decomposed into two components: a phase-independent mean $\overline{\mathbf{u}}(\zeta)$ and a wave-coherent fluctuation $\widetilde{\mathbf{u}}(\xi, \zeta)$. These are defined respectively as
\begin{equation} \overline{\mathbf{u}}(\zeta) = \frac{1}{N_\xi} \sum_{k = 1}^{N_\xi} \langle \mathbf{u} \rangle (\xi_k, \zeta), \label{eq:wave independent component} \end{equation}
and 
\begin{equation} \widetilde{\mathbf{u}}(\xi, \zeta) = \langle \mathbf{u} \rangle (\xi, \zeta) - \overline{\mathbf{u}}(\zeta), \label{eq:wave coherent component} \end{equation}
where $N_\xi$ is the total number of discretized points in the $\xi$ direction.

As a result, the instantaneous velocity field in the air domain can be expressed using the triple decomposition as
\begin{equation}
\mathbf{u}(x, y, z, t) = \overline{\mathbf{u}}(\zeta) + \widetilde{\mathbf{u}}(\xi, \zeta) + \mathbf{u}'(x, y, z, t),
\label{eq:trilple decomposition}
\end{equation}
as summarized in Fig.~\ref{fig:triple decomposition schematic}.
The first term represents the phase-independent mean flow, the second term accounts for the wave-coherent modulation, and the third term corresponds to the turbulent fluctuations.

It is important to note that the triple decomposition is applicable only to statistically stationary flows. Furthermore, accurate statistical convergence of the phase-averaged and wave-coherent components requires a sufficiently large number of samples within each phase bin. In this study, we employed $N_{\text{bins}} = 100$ and ensured a total sample size of $N_{\text{samples}} > 1000$.


\section{Canonical test cases and results\label{sec:results:canonical}}
In the discussion of results, the following abbreviations are adopted for the three interface-capturing methods used throughout the subsequent sections.
The geometric VOF-type method with $\phi$-based interface reconstruction is referred to as the \texttt{isoPhi} method. The reconstructed distance function (RDF)-based geometric VOF method is denoted as \texttt{plicRDF}. The algebraic VOF-type method using the MULES limiter is referred to as \texttt{gradPhi}. 
\todo{Throughout the paper, including all the figures, keep the order of the schemes}

\textcolor{black}{
For the canonical test cases in this section, where density, viscosity, surface tension, and the characteristic velocity and length scales are all prescribed a priori, the relevant nondimensional groups, such as the Laplace, Weber, and Reynolds numbers, are well defined, and we report them where relevant.
}

\subsection{Static drop\label{sec:result:spurious current in static bubble cases}}
We conduct classical static circular interface simulations to assess the accuracy of interface representation across different interface-capturing methods. In the presence of an interface, inaccuracies in its representation can lead to numerical errors in the velocity field near the interface, commonly referred to as spurious currents. These spurious currents originate from an imbalance between the surface tension force and the pressure gradient across the interface~\citep{brackbill}. When the surface tension and pressure gradient terms are consistently discretized~\citep{francois2006balanced}, they should ideally cancel each other out, provided the interface curvature is computed exactly. This canonical test case is thus employed to verify the correctness of the code implementation and to quantify curvature-related errors through the manifestation of spurious currents across different interface-capturing methods. 
In what follows, we refer to the static circular interface simply as a drop for brevity. 
\subsubsection{Simulation setup}
The drop test case configuration follows that of \citet{abadie2015combined}. A two-dimensional domain of size $[-1,1] \times [-1,1]~\textnormal{m}$ is considered, containing a stationary circular drop of diameter $D_{\text{drop}} = 0.8~\textnormal{m}$ centered at $(x, y) = (0, 0)~\textnormal{m}$. The surface tension coefficient is set to unity, \textcolor{black}{$\sigma = 1~\textnormal{kg}/\textnormal{s}^2$}. Both viscosity and density ratios are set to one, with $\rho_{\text{drop}} = 1~\textnormal{kg}/\textnormal{m}^3$ and $\mu_{\text{drop}} = 8.165 \times 10^{-3}~\textnormal{Pa}\cdot\textnormal{s}$, corresponding to a Laplace number of $La = \rho_{\text{drop}} D_{\text{drop}} \sigma / \mu_{\text{drop}}^2 = 12000$.
The domain is discretized using $N = 64$ grid points in each direction, and a time step of $\Delta t = 0.00065~\textnormal{s}$ is used. Symmetry boundary conditions are imposed on the lateral boundaries, and periodic boundary conditions are applied on the top and bottom boundaries. Simulations are performed up to $t = \textcolor{black}{102.86\,t_\sigma}$, where $t_\sigma = \sqrt{\rho_{\text{drop}} D_{\text{drop}}^3/\sigma}$ is the characteristic capillary time, corresponding to $t = 72~\textnormal{s}$.

\begin{figure}
     \centering
     \begin{subfigure}[b]{0.7\columnwidth}
         \centering
         \caption{}
         \includegraphics[width=\textwidth]{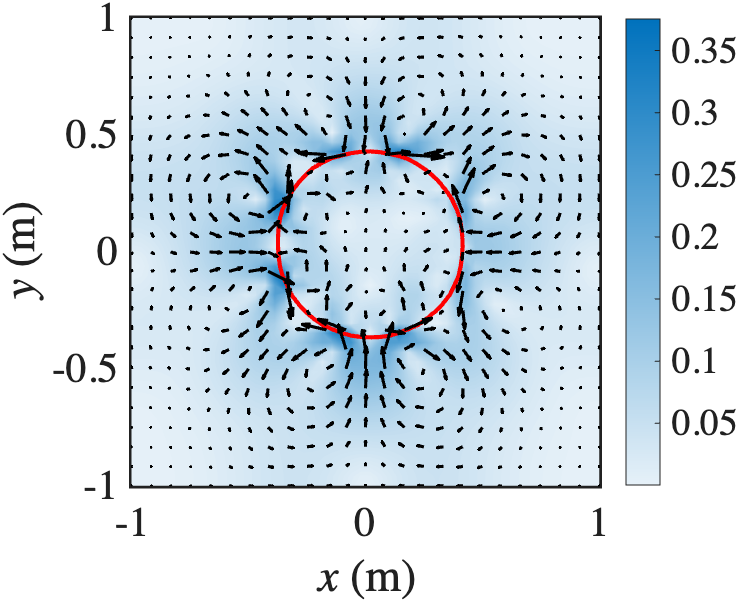}
     \end{subfigure}
     \begin{subfigure}[b]{0.7\columnwidth}
         \centering
         \caption{}
         \includegraphics[width=\textwidth]{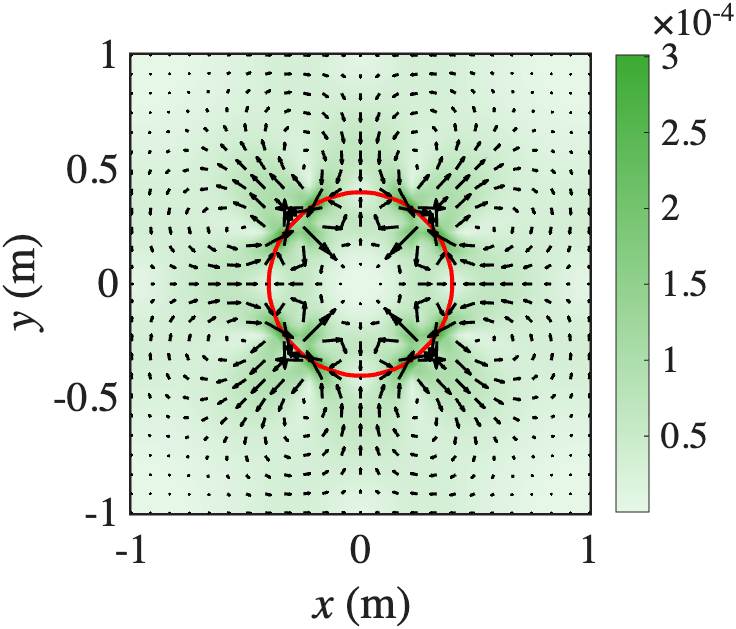}
     \end{subfigure}
     \begin{subfigure}[b]{0.7\columnwidth}
         \centering
         \caption{}
         \includegraphics[width=\textwidth]{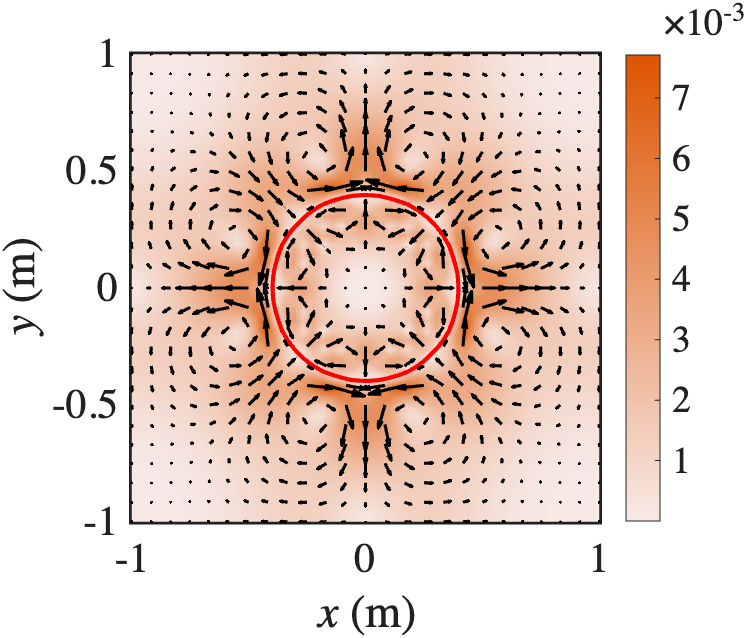}
     \end{subfigure}
        \caption{
        Snapshots of static drop simulations using (a) \texttt{isoPhi}, (b) \texttt{plicRDF}, and (c) \texttt{gradPhi} at $t = 10.21t_\sigma$. Colormaps indicate the velocity magnitude \textcolor{black}{in m/s}, and black arrows represent the velocity vector field. The drop interface, defined by the $\phi = 0.5$ contour, is shown with red solid lines.
        }
    \label{fig:static drop: velocity field}
\end{figure}

\subsubsection{Spurious current}
Fig.~\ref{fig:static drop: velocity field} shows the velocity fields at $t = 10.21t_\sigma$, visualized using colormaps for magnitude and vectors for direction. Note that the colorbar scale differs between subfigures to reflect the range of spurious current magnitudes. 
The maximum spurious velocity in the \texttt{isoPhi} method is on the order of $O(10^{-1})~\textcolor{black}{\textnormal{m/s}}$ (Fig.~\ref{fig:static drop: velocity field}(a)). In comparison, the \texttt{plicRDF} and \texttt{gradPhi} methods yield significantly lower maximum spurious velocities, on the order of $O(10^{-4})~\textcolor{black}{\textnormal{m/s}}$ and $O(10^{-3})~\textcolor{black}{\textnormal{m/s}}$, respectively.
The \texttt{plicRDF} method achieves three orders of magnitude lower spurious velocity than the \texttt{isoPhi} method due to its more accurate curvature estimation by the RDF.
The lower spurious current in the \texttt{gradPhi} method is attributed to the smearing of the interface, controlled by the compression velocity $\mathcal{U}_i$ in Eq.~\eqref{eq:phase indicator transport equation, algebraic VoF}, which facilitates more stable \textcolor{black}{curvature estimation}.

\begin{figure}
     \centering
     \begin{subfigure}[b]{0.9\columnwidth}
         \centering
         \caption{}
         \includegraphics[width=\textwidth]{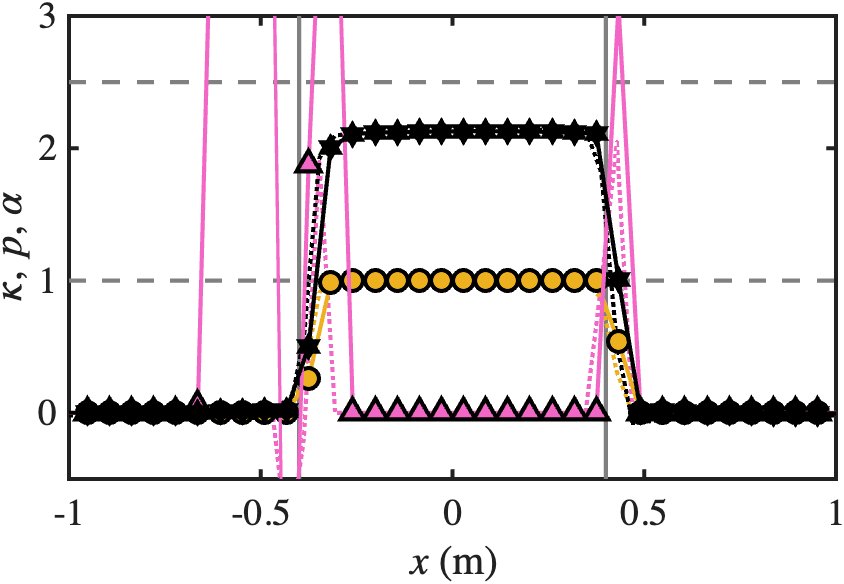}
     \end{subfigure}
     \begin{subfigure}[b]{0.9\columnwidth}
         \centering
         \caption{}
         \includegraphics[width=\textwidth]{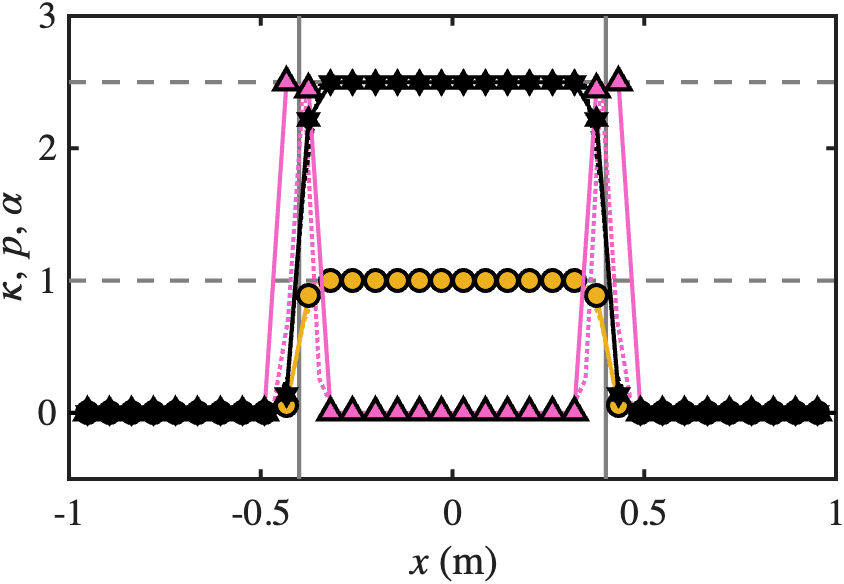}
     \end{subfigure}
     \begin{subfigure}[b]{0.9\columnwidth}
         \centering
         \caption{}
         \includegraphics[width=\textwidth]{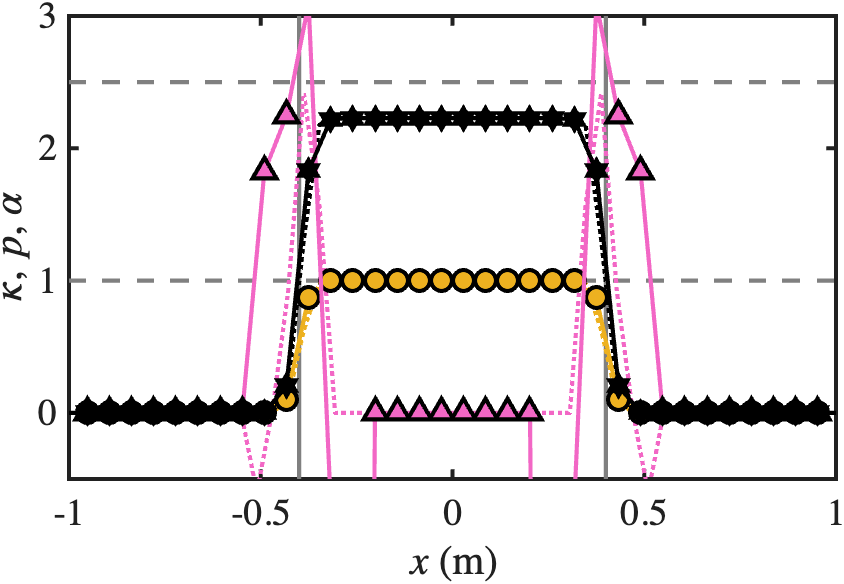}
     \end{subfigure}
        \caption{
        Curvature (pink triangle lines) and pressure jump (black hexagram lines) extracted at $t = 10.21t_\sigma$ along the horizontal line $y = 0~\textnormal{m}$ (dotted lines) and the diagonal line $y = x$ (solid lines) for (a) \texttt{isoPhi}, (b) \texttt{plicRDF}, and (c) \texttt{gradPhi}. The phase indicator $\phi$ is shown in yellow circles. Vertical gray solid lines denote the expected interface location ($\phi = 0.5$), and horizontal gray-dashed lines indicate reference values for curvature and pressure jump.
        }
    \label{fig:static drop: pressure}
\end{figure}

For a more quantitative assessment, Fig.~\ref{fig:static drop: pressure} presents curvature and pressure profiles along the horizontal centerline ($y = 0$) and the diagonal line ($y = x$). The interface location is marked with vertical gray lines, and the initial phase indicator $\phi$ is shown in yellow circles. The exact curvature is $\kappa = 2 / D_{\text{drop}} = 2.5~\textnormal{m}^{-1}$, and the corresponding pressure jump is given by the Young–Laplace law as $\Delta p = \sigma \kappa = 2.5~\textnormal{Pa}$. Reference values for $\phi$, $\kappa$, and $\Delta p$ are indicated by horizontal gray-dashed lines.

The pressure jump for the \texttt{isoPhi} method is approximately $\Delta p \approx 2.1~\textnormal{Pa}$, with curvature $\kappa \simeq 2~\textnormal{m}^{-1}$ along $y = 0$, but significant deviation is observed along $y = x$. Moreover, the symmetry of the solution breaks down, as the large spurious currents distort the drop shape. \textcolor{black}{This behavior originates from the \textcolor{black}{curvature estimation} itself, which is computed as the divergence of the interface normal $n_i = (\partial \phi/\partial x_i)/|\partial \phi/\partial x_i|$. Because the underlying phase-indicator field is not smooth, this normalization can amplify local gradients into large curvature overshoots, a well-known limitation of gradient-based curvature estimation~ \citep{abadie2015combined}. The resulting spurious currents are large enough to advect the drop itself, and at later times this drift moves the \texttt{isoPhi} drop entirely away from the domain center.}
The most accurate results are obtained using the \texttt{plicRDF} method, which yields an exact pressure jump of $\Delta p = 2.5~\textnormal{Pa}$ and excellent agreement in curvature along both $y = 0$ and $y = x$, as shown in Fig.~\ref{fig:static drop: velocity field}(b).
The \texttt{gradPhi} method preserves symmetry across both directions, with a slightly improved pressure jump of $\Delta p \approx 2.2~\textnormal{Pa}$. The curvature is accurately predicted along $y = 0$, though it deviates by approximately $20\%$ along $y = x$.

\begin{figure}
    \centering
     \begin{subfigure}[b]{0.9\columnwidth}
         \centering
         \caption{}
         \includegraphics[width=\textwidth]{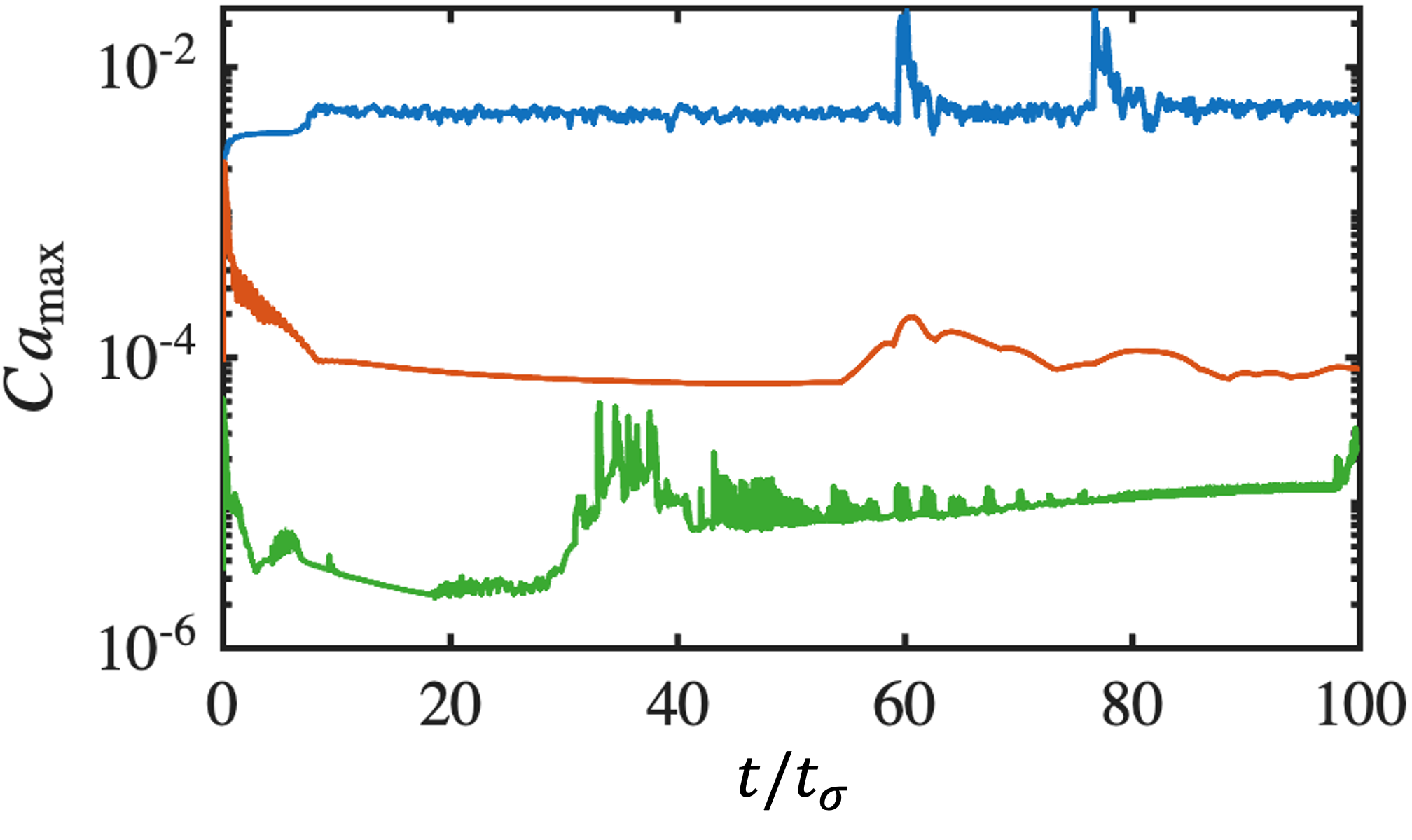}
     \end{subfigure}
     \hfill
     \begin{subfigure}[b]{0.9\columnwidth}
         \centering
         \caption{}
         \includegraphics[width=\textwidth]{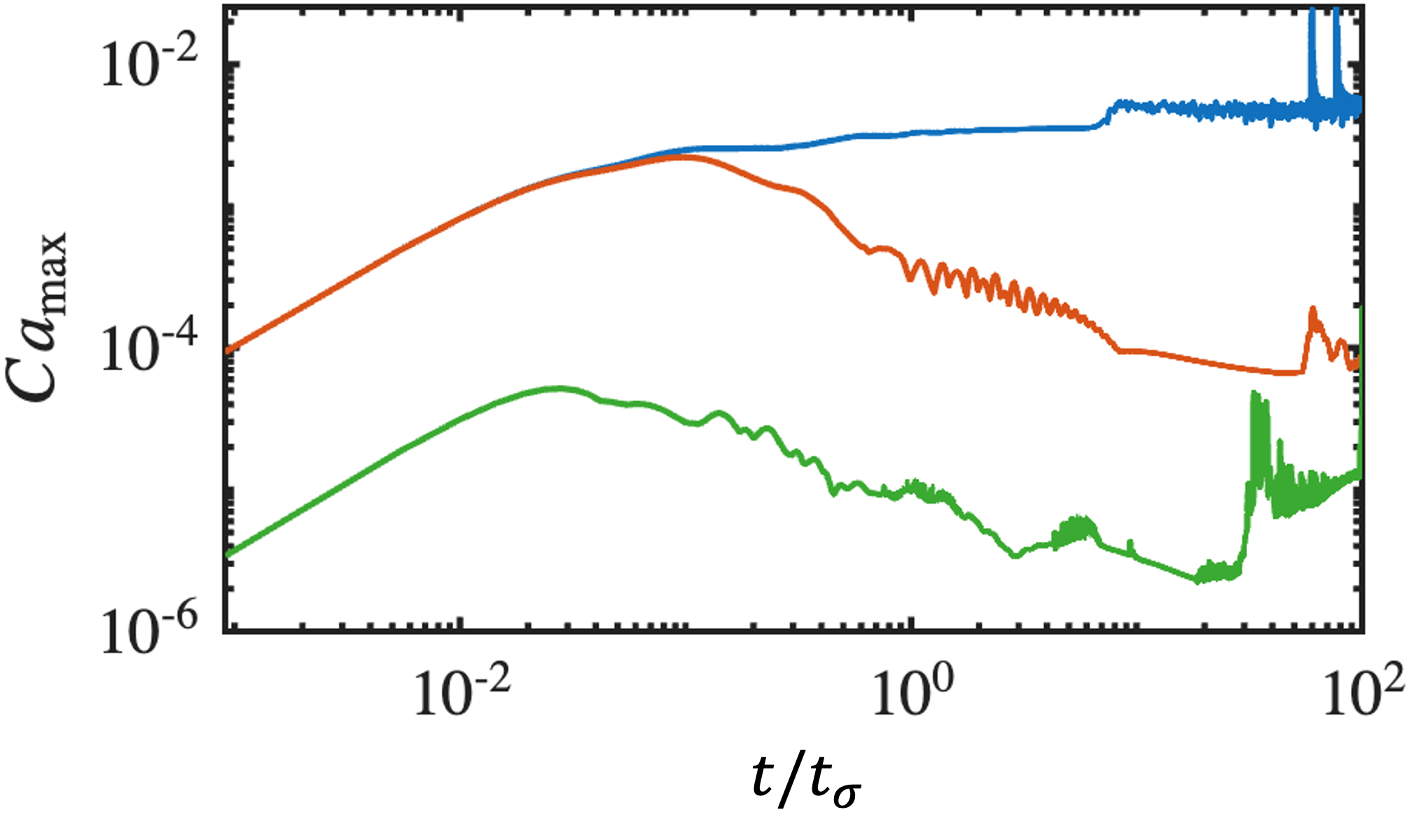}
     \end{subfigure}
    \caption{
    Time evolution of the maximum capillary number $Ca_{\text{max}}$ as a function of \textcolor{black}{normalized time $t/t_\sigma$}, for the static drop simulations shown in (a) linear and (b) logarithmic scale. Results are shown for \texttt{isoPhi} (blue), \texttt{plicRDF} (green), and \texttt{gradPhi} (red) methods.
    }
    \label{fig:static drop: cmax}
\end{figure}

The long-term effects of spurious currents can be significant in wind–wave configurations, as they continuously influence the wind field within a confined region.
To further evaluate the spurious current behavior in a long time duration, the nondimensional capillary number is defined as $Ca_{\text{max}} = \mu_{\text{drop}} U_{\text{max}} / \sigma$, where $U_{\text{max}}$ is the maximum velocity magnitude in the domain. Fig.~\ref{fig:static drop: cmax} shows the evolution of $Ca_{\text{max}}$ for each method. 
The observed magnitudes are consistent with those reported in \citet{gamet2020validation} for a Laplace number of $La = 12000$. It is also noted that the initial time steps of both \texttt{isoPhi} and \texttt{gradPhi} exhibit identical capillary numbers.
Although not shown here, we note that the magnitude of the spurious current scales proportionally with the Laplace number. Specifically, the spurious current increases with higher drop density and surface tension, and decreases with higher drop viscosity. This trend is consistent with results from additional static drop simulations performed using non-unity viscosity and density ratios.

\subsubsection{Curvature error scaling} \label{sec:results:curvature error scaling}
As an extension of the classical static drop analysis, we define errors associated with curvature, interface normal, and pressure jump calculations, and investigate how these errors scale with grid resolution. 

This analysis is particularly important for wind–wave simulations, where the common approach to resolving small-scale wind turbulence near the interface is to refine the grid locally. In this context, it is desirable for the spurious current magnitude to decrease, or at least not increase, as the grid is refined. 
However, increased resolution can amplify curvature errors and associated spurious currents, especially in models lacking curvature converging techniques ~\citep{cummins2005estimating}. 
This issue becomes even more critical in unstructured grid frameworks, where accurate curvature estimation is not only more difficult but also computationally expensive.

Given the simple circular geometry of the drop, exact analytical values of curvature, interface normal, and pressure jump are known. The curvature error is defined using both infinity-norm and 1-norm-like metrics as follows:

\begin{equation}
    \mathcal{E}_{\kappa}^{\infty} = \max_i  \frac{\left| \kappa_i - \kappa_{\textnormal{exact}} \right|}{\kappa_{\textnormal{exact}}},~ \mathcal{E}_{\kappa}^{1} = \frac{1}{N_i} \sum_i  \frac{ \left| \kappa_i - \kappa_{\textnormal{exact}} \right| }{ \kappa_{\textnormal{exact}} },
\end{equation}
where $i$ denotes the index of interfacial cells and $N_i$ is the total number of such cells. Similarly, the pressure error is defined as:
\begin{equation}
\mathcal{E}_{p}^{\infty} = \max_i \frac{ \left| p_i - p_{\textnormal{exact}} \right| }{  p_{\textnormal{exact}} } ,~\mathcal{E}_{p}^{1} = \frac{1}{N_i} \sum_i \frac{ \left| p_i - p_{\textnormal{exact}} \right| }{ p_{\textnormal{exact}} }.
\end{equation}
The interface normal error is quantified by:
\begin{eqnarray}
    &\displaystyle\mathcal{E}_n^{\infty} = \textnormal{max}_i \left( 1 - n_i\cdot n_{\textnormal{exact}, i}\right), \nonumber \\
    &\displaystyle\mathcal{E}_n^{1} = \frac{1}{N_i} \sum_i \left( 1 - n_i\cdot n_{\textnormal{exact}, i}\right).
\end{eqnarray}

\begin{figure*}
     \centering
     \begin{subfigure}[b]{0.8\columnwidth}
         \centering
         \caption{}
         \includegraphics[width=\textwidth]{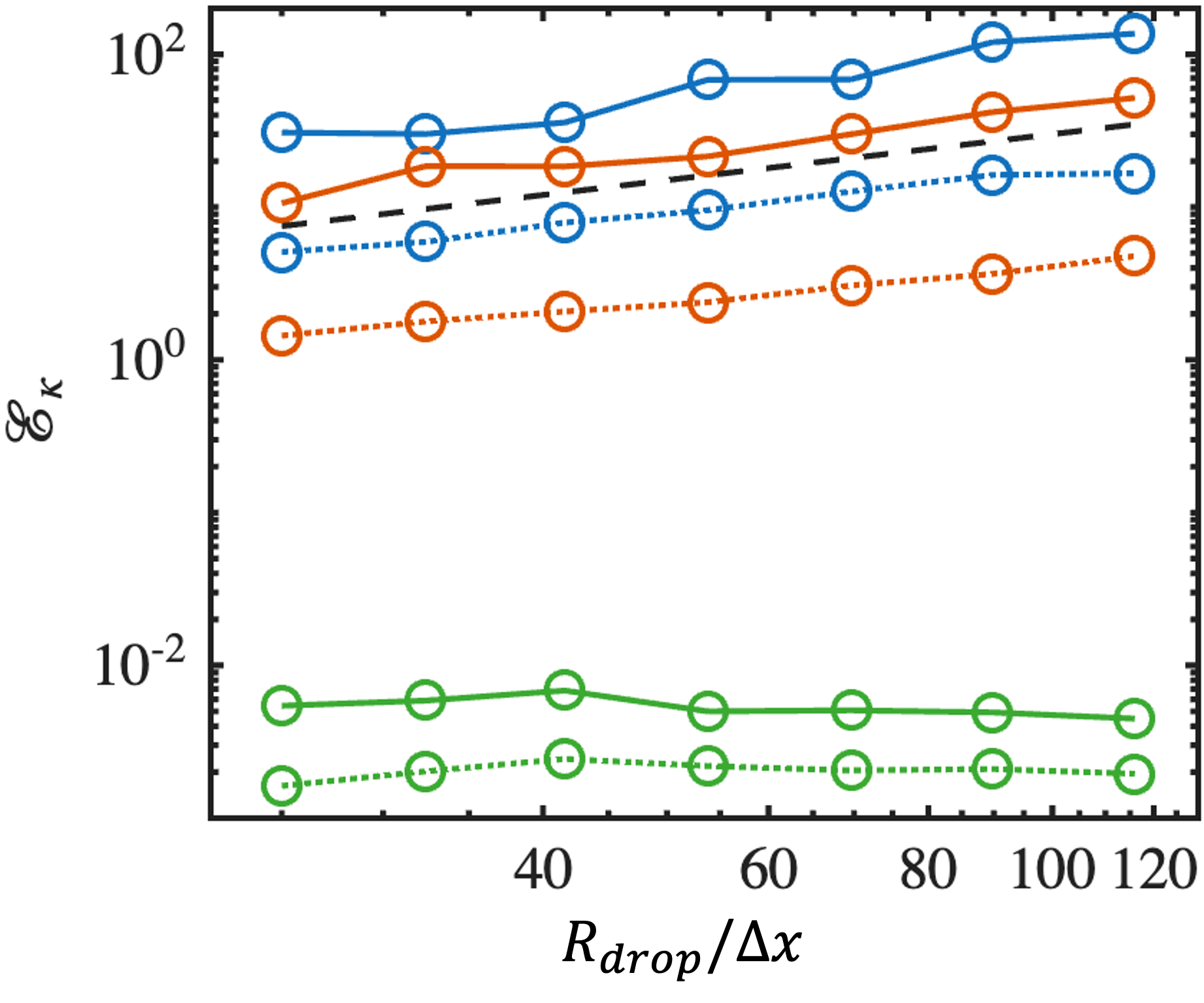}
     \end{subfigure}
     \begin{subfigure}[b]{0.8\columnwidth}
         \centering
         \caption{}
         \includegraphics[width=\textwidth]{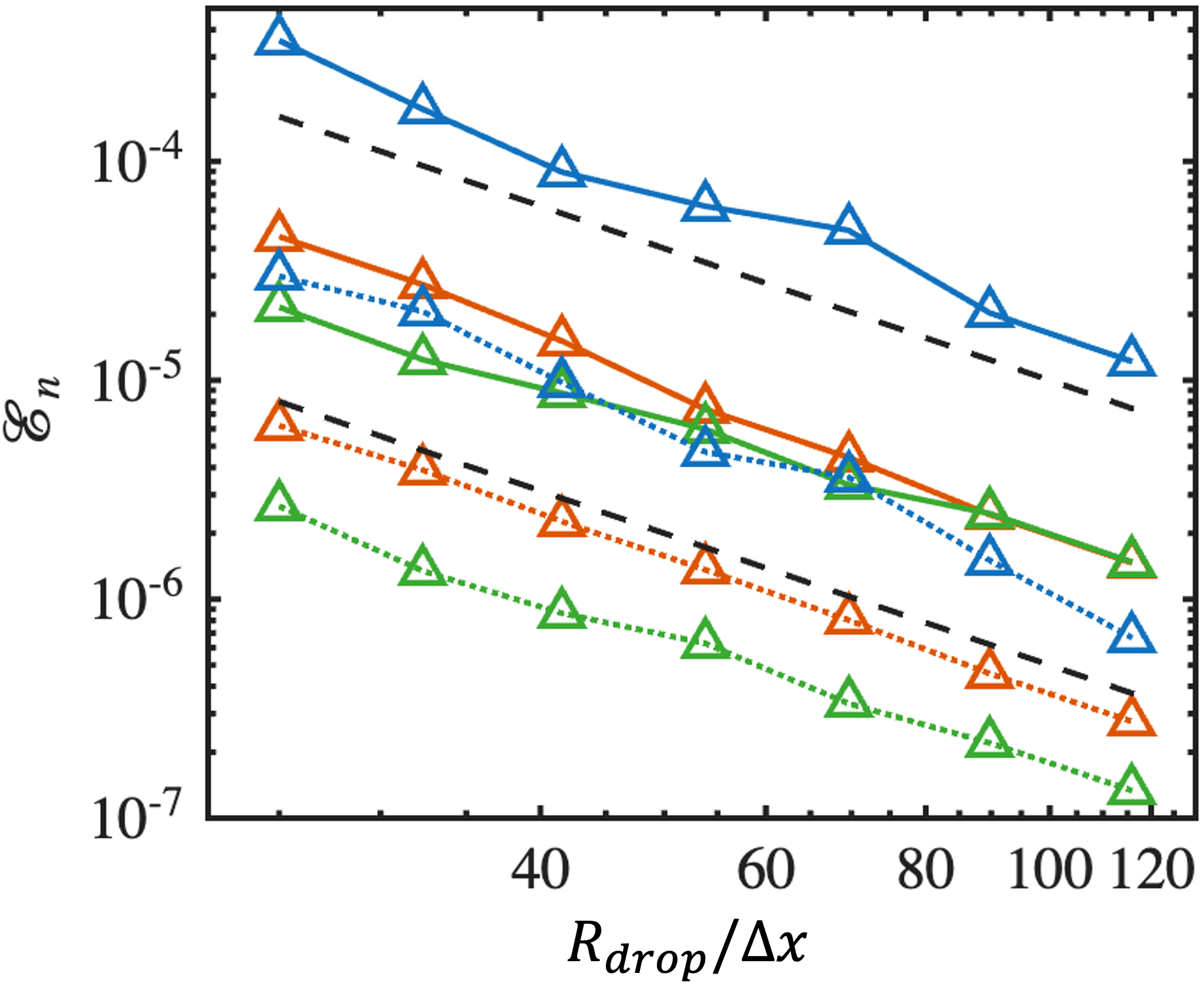}
     \end{subfigure}
     \\
     \begin{subfigure}[b]{0.8\columnwidth}
         \centering
         \caption{}
         \includegraphics[width=\textwidth]{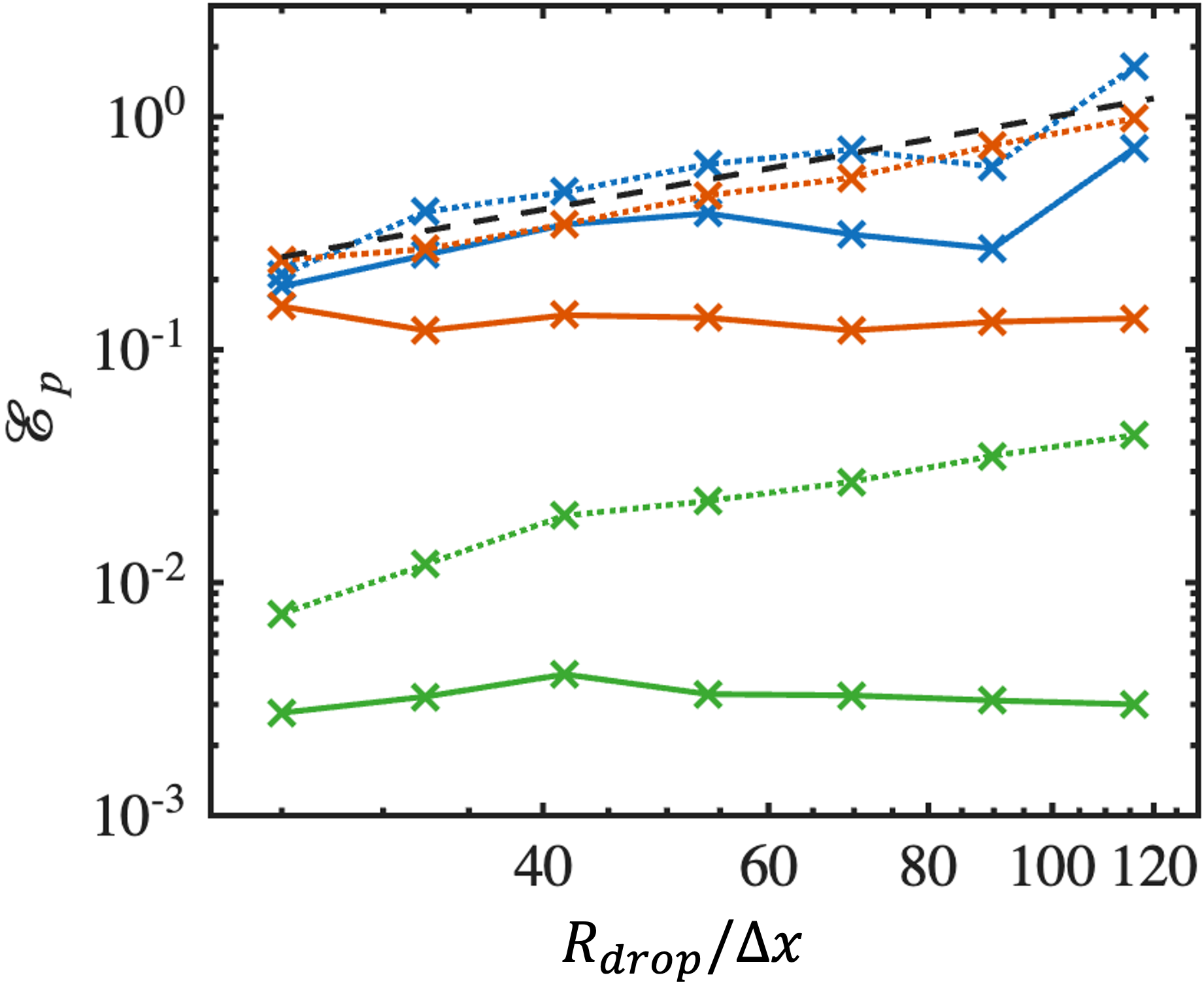}
     \end{subfigure}
     \begin{subfigure}[b]{0.8\columnwidth}
         \centering
         \caption{}
         \includegraphics[width=\textwidth]{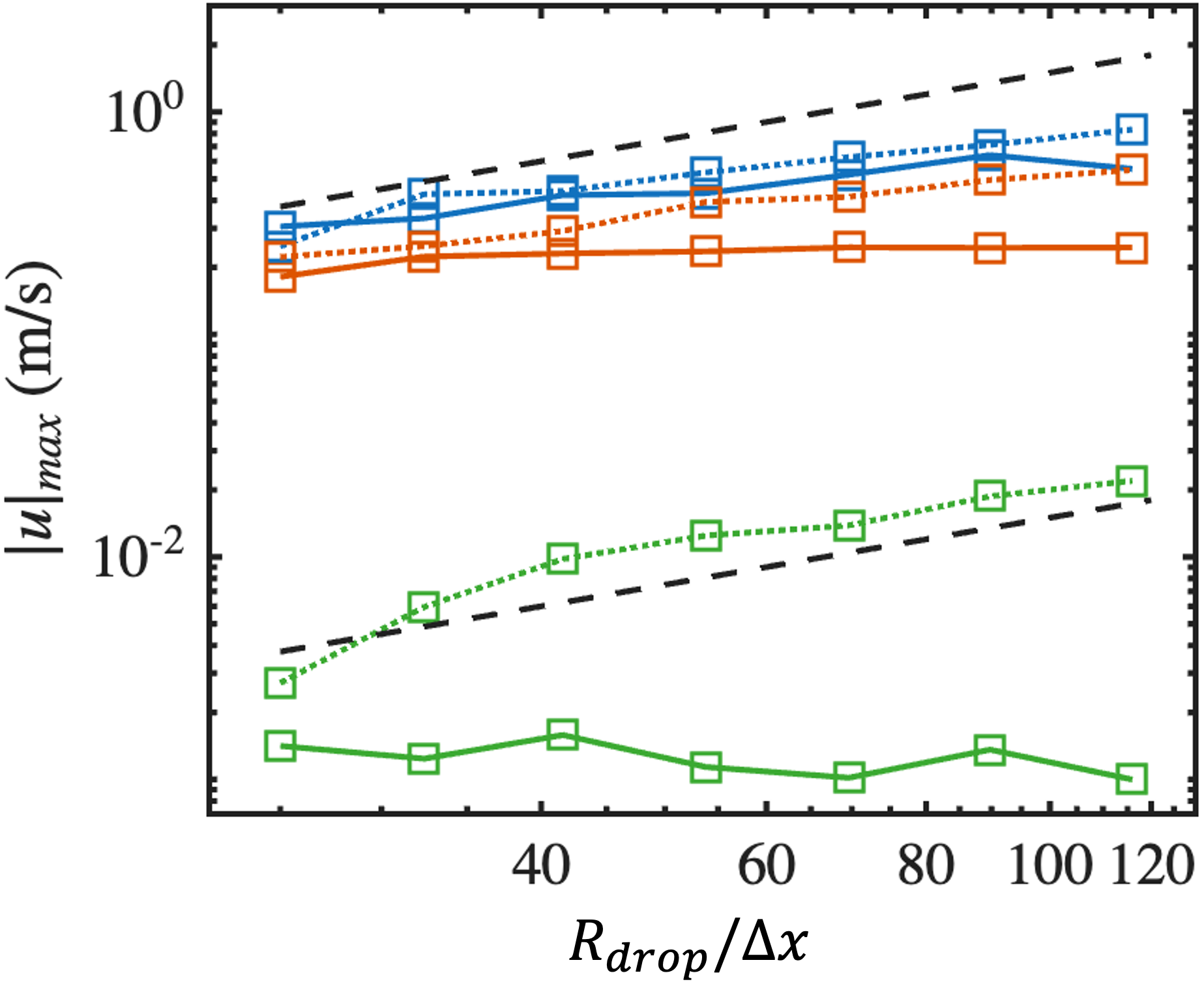}
     \end{subfigure}
        \caption{
        Error metrics for (a) curvature, (b) interface normals, (c) pressure jump, and (d) maximum magnitude of spurious current as functions of grid resolution for \texttt{isoPhi} (blue), \texttt{plicRDF} (green), and \texttt{gradPhi} (red). Solid lines indicate infinity-norm-like errors, while dotted lines represent 1-norm-like errors. Reference scaling trends are shown with black dashed lines: $\Delta x^{-1}$ in panels (a), (c), and (d), and $\Delta x^{2}$ in panel (b). \textcolor{black}{The horizontal axis in all panels is $R_{\text{drop}}/\Delta x = 1/(\kappa_{\text{exact}}\Delta x)$.}
        }
    \label{fig:error scaling}
\end{figure*}

Fig.~\ref{fig:error scaling} presents these errors, along with the maximum spurious current magnitude $|u|_{\text{max}}$ and its spatial average $\overline{|u|}_{\text{max}}$, across various grid resolutions. All other simulation parameters are kept constant while the number of grid points $N$ is varied logarithmically across \textcolor{black}{seven} cases: $N = 125$, $162$, $208$, $269$, $348$, $449$, and $580$. \textcolor{black}{Beyond $N = 580$, maintaining numerical stability required reducing the time step below the capillary time-step constraint used at coarser resolutions. Without this adjustment, the amplified spurious currents at finer resolution destroy the interface and mix the two phases, so no higher-resolution case is included in this sweep.} Data is collected at $100$-th time steps, allowing interface-capturing methods to converge sufficiently before the onset of large spurious motion.

Fig.~\ref{fig:error scaling}(a) shows that the curvature errors for the \texttt{isoPhi} and \texttt{gradPhi} methods increase linearly with grid refinement, following a $\Delta x^{-1}$ scaling (black dashed line). In contrast, the \texttt{plicRDF} method exhibits much smaller errors, two orders of magnitude lower, and remains largely insensitive to grid resolution.
Interface normal errors, shown in Fig.~\ref{fig:error scaling}(b), converge with a rate close to $\Delta x^2$ across all methods. However, the \texttt{isoPhi} method displays one order of magnitude higher error than the \texttt{gradPhi} and \texttt{plicRDF} methods. 
For pressure jump errors, the \texttt{gradPhi} method slightly outperforms \texttt{isoPhi}, while the \texttt{plicRDF} method again shows the most accurate results, with two orders of magnitude lower error. These trends are mirrored in the spurious current magnitudes, which increase with grid resolution for all methods, though not as steeply as the curvature errors. The most noticeable increase is observed in the \texttt{isoPhi} method. 
Across all resolutions, the \texttt{isoPhi} method exhibits the highest errors, while the \texttt{plicRDF} method consistently yields the most accurate results.

\subsection{Moving drop\label{sec:result:moving drop}}
The previous test cases considered static interfaces with unity viscosity and density ratios. In this section, we investigate large density and viscosity ratio settings along with non-zero surface tension.
The interface is advected over a short time interval to evaluate the importance of the induced velocity manifested differently across interface-capturing methods, with emphasis placed on relative differences rather than on the absolute magnitude of spurious currents.

\subsubsection{Simulation setup}
The numerical setup for the moving drop cases is identical to that of the static drop cases in Section~\ref{sec:result:spurious current in static bubble cases}, except for four modifications. First, the density of the fluid inside the circular drop is set to $\rho_{\textnormal{circle}} = 998~\textnormal{kg}/\textnormal{m}^3$, and that of the surrounding fluid is $\rho_{\textnormal{sur}} = 1.2~\textnormal{kg}/\textnormal{m}^3$, representing water and air, respectively, \textcolor{black}{with corresponding dynamic viscosities of $\mu_{\textnormal{circle}} = 1.0\times10^{-3}~\textnormal{Pa}\cdot\textnormal{s}$ and $\mu_{\textnormal{sur}} = 1.8\times10^{-5}~\textnormal{Pa}\cdot\textnormal{s}$, giving a density ratio $\rho_{\textnormal{circle}}/\rho_{\textnormal{sur}} = 831.7$ and a viscosity ratio $\mu_{\textnormal{circle}}/\mu_{\textnormal{sur}} = 55.6$}. Second, the surface tension is set to $\sigma = 0.072~\textnormal{N}/\textnormal{m}$. Third, the initial velocity of the drop is assigned a finite uniform value of \textcolor{black}{$0.1~\textnormal{m}/\textnormal{s}$} in the $-\hat{y}$ direction. Lastly, the time step is fixed at $\Delta t = \textcolor{black}{0.0005}~\textnormal{s}$. \textcolor{black}{This time step is chosen to remain well below the convective, interface-compression, and capillary time-step restrictions across all cases considered, giving a maximum Courant number of approximately $0.2$ for the fastest drop speed ($|u|_{\textnormal{center}} = 1.0~\textnormal{m}/\textnormal{s}$) on the finest grid ($N = 1024$).}
Simulations are run for $\textcolor{black}{100}$ time steps, corresponding to a physical time of $t = 0.05~\textnormal{s}$. This brief duration ensures the drop maintains an approximately circular shape while minimizing the influence of the interface advection term. The drop shifts slightly downward, and its new center is identified. We verified that the drop's geometry remains nearly circular at this time.

\todo{Rather than, error, use something different. For example, difference..? Explain why we use isoPhi as a reference.}
The \texttt{isoPhi} results are set as a reference and define the velocity difference metric relative to it as 
\begin{equation}
\mathcal{D}_{\textnormal{md}} = \frac{\left\{ \frac{1}{N_{\textnormal{cv}}} \sum_i \left( u_i - u_{\textnormal{\texttt{isoPhi}},i} \right)^2 \right\}^{1/2}}{u_{\textnormal{center}}},
\label{eq:error metric for moving drop}
\end{equation}
where $N_{\textnormal{cv}}$ is the number of control volumes and $u_{\textnormal{center}}$ is the bulk velocity of the drop.
We emphasize that Eq.~\eqref{eq:error metric for moving drop} does not represent an error, but rather a metric quantifying the difference between \texttt{plicRDF} and \texttt{gradPhi} relative to \texttt{isoPhi}. Because \texttt{plicRDF} provides enhanced \textcolor{black}{curvature-estimation} accuracy, whereas \texttt{gradPhi} employs a different numerical flux formulation, comparisons with \texttt{isoPhi} isolate how these factors manifest in the present setting. \textcolor{black}{We acknowledge that \texttt{isoPhi} is itself not free of numerical error, and Sec.~\ref{sec:result:spurious current in static bubble cases} shows it to be the least accurate of the three methods in curvature and pressure-jump estimation. However, unlike the static drop, this transient moving-interface configuration is a manufactured test case with no known ground-truth solution against which velocity errors could be independently quantified. \texttt{isoPhi} is therefore not selected as a reference because it approximates the true solution, but because both \texttt{plicRDF} and \texttt{gradPhi} are direct algorithmic modifications of \texttt{isoPhi}. Specifically, \texttt{plicRDF} enhances its \textcolor{black}{curvature estimation}, while \texttt{gradPhi} modifies its flux discretization through interface compression. \texttt{isoPhi} thus shares a common algorithmic root with both methods, making it a natural baseline for isolating the effect of each modification individually.}

\begin{figure*}
     \centering
     \begin{subfigure}[b]{0.8\columnwidth}
         \centering
         \caption{}
         \includegraphics[width=\textwidth]{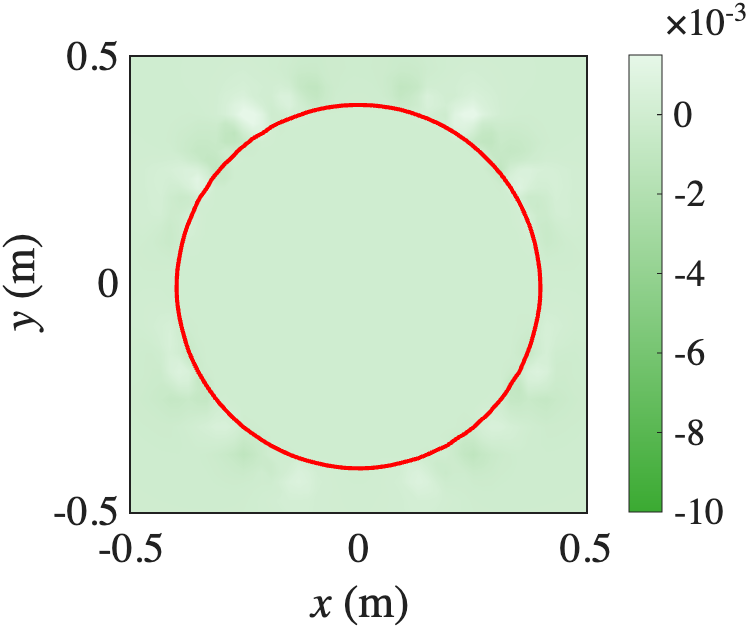}
     \end{subfigure}
     \begin{subfigure}[b]{0.8\columnwidth}
         \centering
         \caption{}
         \includegraphics[width=\textwidth]{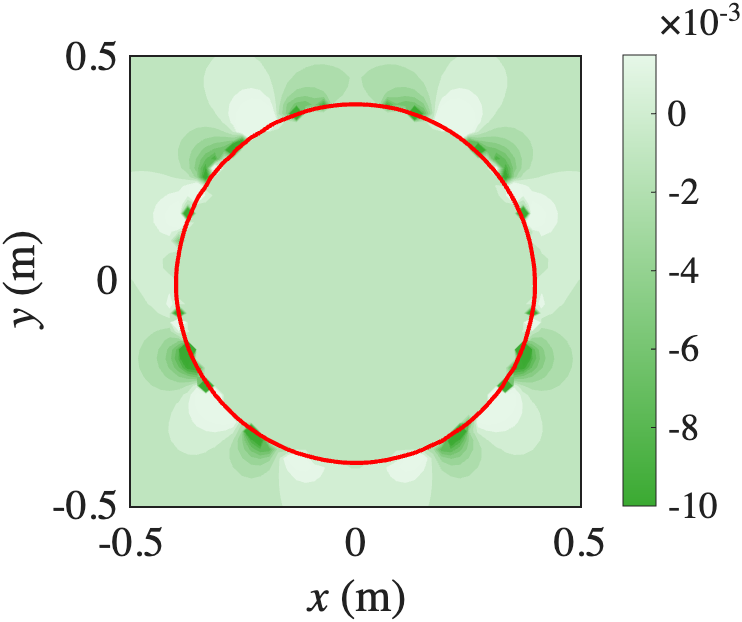}
     \end{subfigure}
     \\
     \begin{subfigure}[b]{0.8\columnwidth}
         \centering
         \caption{}
         \includegraphics[width=\textwidth]{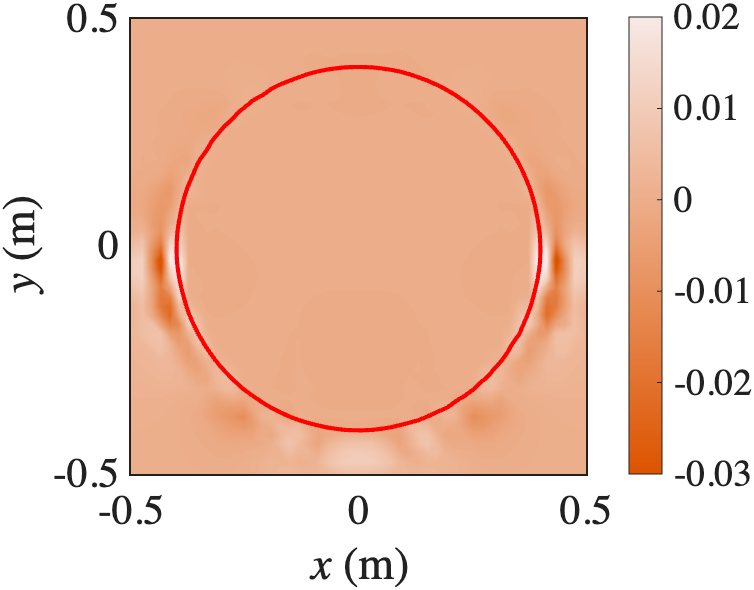}
     \end{subfigure}
     \begin{subfigure}[b]{0.8\columnwidth}
         \centering
         \caption{}
         \includegraphics[width=\textwidth]{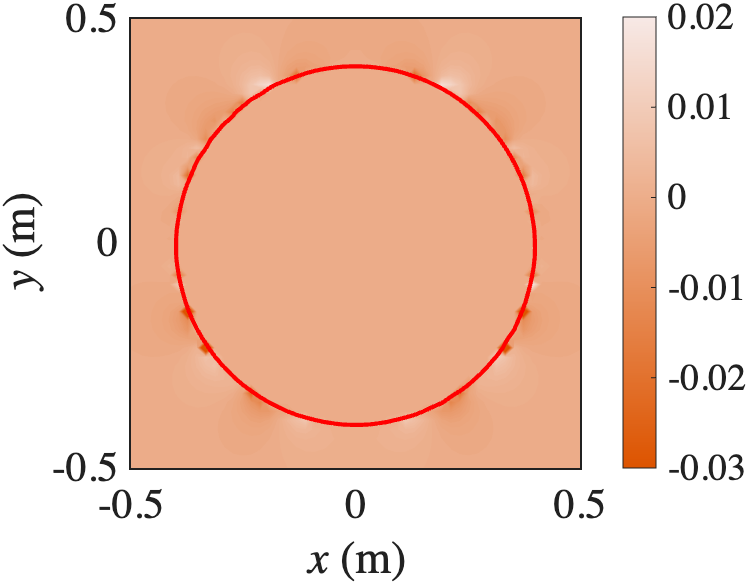}
     \end{subfigure}
        \caption{
        Velocity difference fields for \texttt{plicRDF} (first row) and \texttt{gradPhi} (second row) after $100$ time steps, relative to the \texttt{isoPhi} approach. The first and second columns correspond to grid resolutions of $N_x = 64$ and $N_x = 1024$, respectively. The interface, defined by the $\phi = 0.5$ contour, is indicated by the red solid line.
        }
        \label{fig:moving drop error fields}
\end{figure*}

\begin{figure}
     \centering
     \begin{subfigure}[b]{0.9\columnwidth}
         \centering
         \caption{}
         \includegraphics[width=\textwidth]{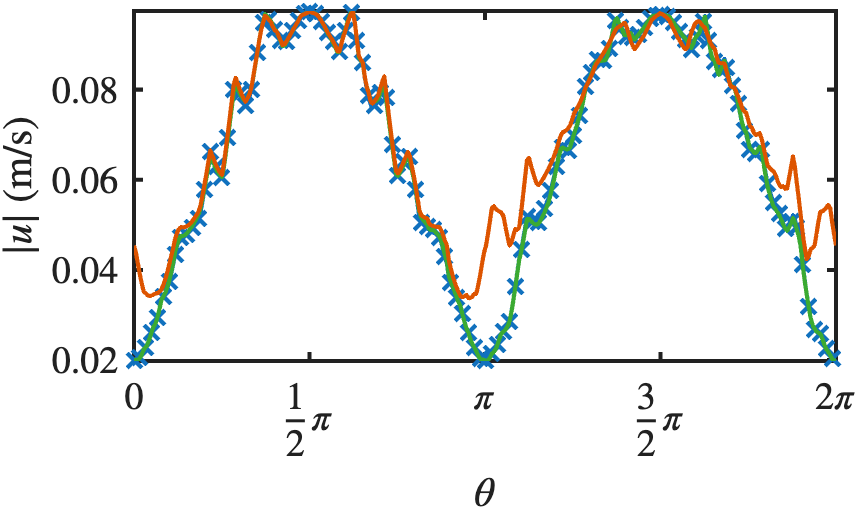}
     \end{subfigure}
     \hfill
     \begin{subfigure}[b]{0.9\columnwidth}
         \centering
         \caption{}
         \includegraphics[width=\textwidth]{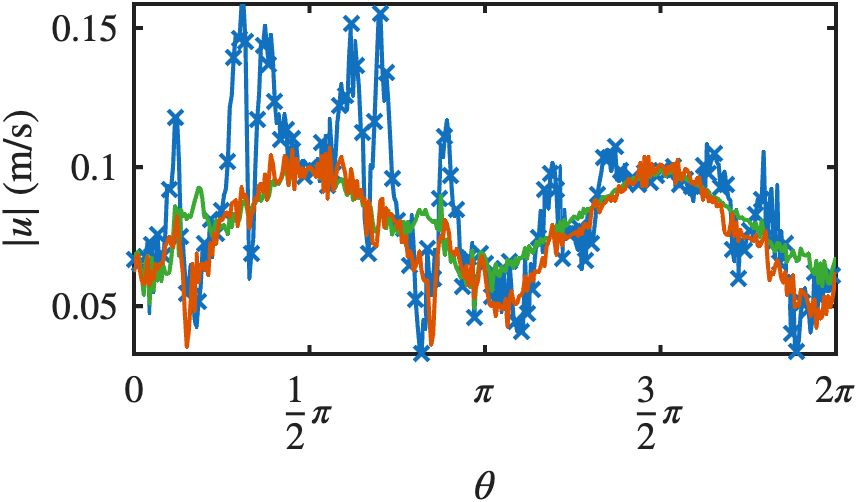}
     \end{subfigure}
        \caption{
        Velocity magnitude measured along a circular path located at $r = 1.01R_{\textnormal{drop}}$ from the drop center after $100$ time steps. The angular position is measured counterclockwise from the vector $\mathbf{n} = (1,0)$. The \texttt{isoPhi} result is shown as a blue solid line with cross symbols, while \texttt{plicRDF} and \texttt{gradPhi} results are shown in green and red solid lines, respectively. 
        }
        \label{fig:moving drop error offset}
\end{figure}

\subsubsection{Velocity field comparison}
\todo{Come up with difference section name}
Fig.~\ref{fig:moving drop error fields} presents the velocity field differences for the \texttt{plicRDF} and \texttt{gradPhi} methods relative to \texttt{isoPhi}, with the first and second rows corresponding to each method, respectively. The first and second columns correspond to coarse ($N_x = 64$) and fine ($N_x = 1024$) grid resolutions. Additionally, Fig.~\ref{fig:moving drop error offset} shows the velocity magnitude along a ring at $r = 1.01R_{\textnormal{drop}}$, where $R_{\textnormal{drop}}$ is the initial drop radius, as a function of the polar angle $\theta$, measured counterclockwise from $\vec{n} = (1,0)$.

For the coarse grid, \texttt{plicRDF} (Fig.~\ref{fig:moving drop error fields}(a)) shows negligible velocity difference, with a maximum difference $\mathcal{D}_{\textnormal{md}} \sim O(10^{-3})$. In contrast, the \texttt{gradPhi} method (Fig.~\ref{fig:moving drop error fields}(c)) exhibits pronounced velocity differences near $\theta = 0$ and $\pi$, as well as in the region $\theta\in[\pi,2\pi]$, with magnitudes an order of magnitude larger. This difference persists regardless of grid resolution and is addressed in the following section. These trends are confirmed in Fig.~\ref{fig:moving drop error offset}(a), where \texttt{gradPhi} displays larger velocity magnitudes at $\theta = 0$ and $\pi$, while all other methods show consistent velocity distributions.

With grid refinement, the difference between \texttt{isoPhi} and \texttt{plicRDF} increases significantly, as shown in Fig.~\ref{fig:moving drop error fields}(b). This increase in $\mathcal{D}_{\textnormal{md}}$ is attributed to amplified spurious currents in the \texttt{isoPhi} method, consistent with curvature error trends discussed in Section~\ref{sec:results:curvature error scaling}. This is further supported by Fig.~\ref{fig:moving drop error offset}(b), where \texttt{isoPhi} shows higher velocity fluctuations across all angles. In contrast, the velocity difference for the \texttt{gradPhi} method at $\theta = 0$ and $\pi$ decreases with grid refinement, gradually converging toward the velocity profile of the \texttt{plicRDF} method.

\begin{figure}
     \centering
     \begin{subfigure}[b]{0.8\columnwidth}
         \centering
         \caption{}
         \includegraphics[width=\textwidth]{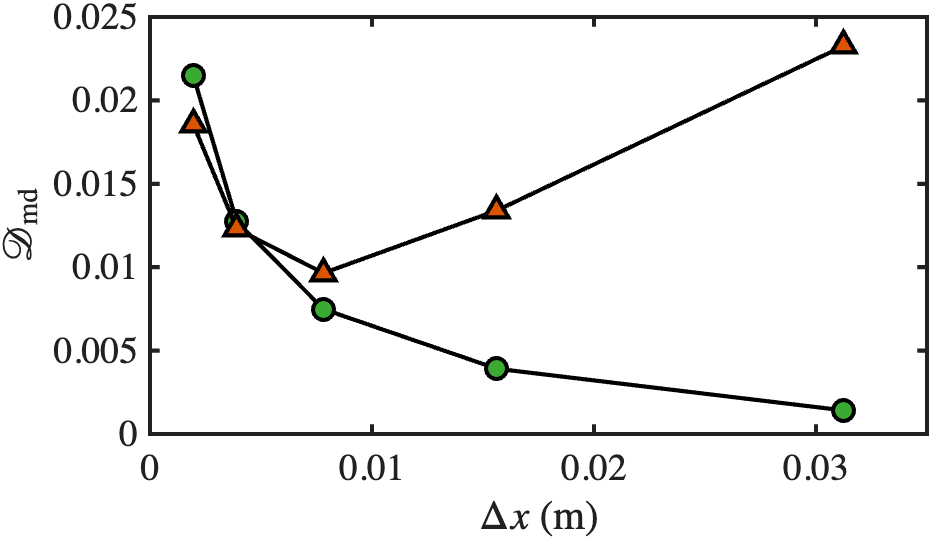}
     \end{subfigure}
     \hfill
     \begin{subfigure}[b]{0.8\columnwidth}
         \centering
         \caption{}
         \includegraphics[width=\textwidth]{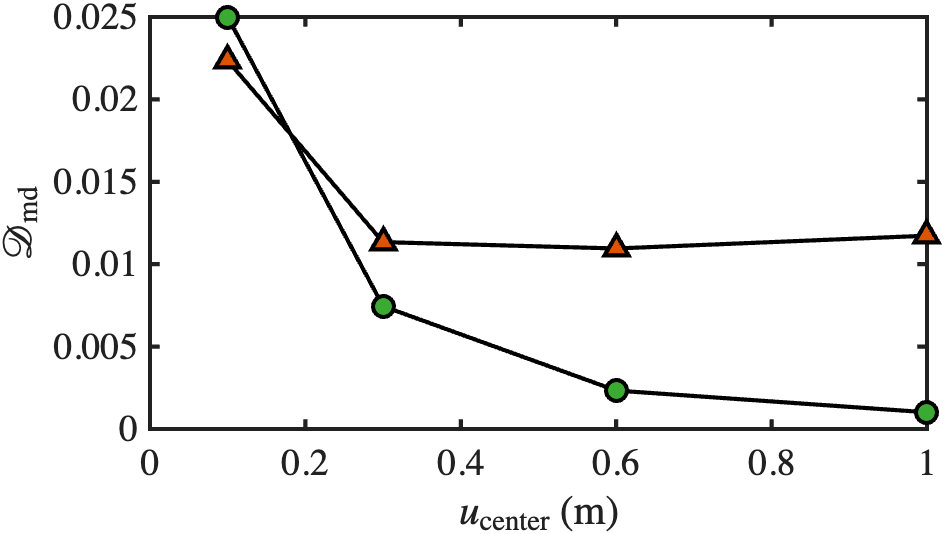}
     \end{subfigure}
        \caption{
        Velocity difference metric $\mathcal{D}_{\textnormal{md}}$ for \texttt{plicRDF} (green circles) and \texttt{gradPhi} (red triangles) as a function of (a) grid resolution and (b) drop velocity $|u_{\textnormal{center}}|$.
        }
        \label{fig:moving drop error}
\end{figure}

Given the sensitivity of $\mathcal{D}_{\textnormal{md}}$ to resolution, we further evaluate its behavior for grid resolutions $N_x = 64$, $128$, $256$, $512$, and $1024$ in Fig.~\ref{fig:moving drop error}(a). For \texttt{plicRDF}, the velocity difference increases with grid refinement, whereas coarse grids yield nearly identical velocity fields ($\mathcal{D}_{\textnormal{md}} \simeq 10^{-3}$). This behavior, combined with earlier findings, highlights the strong link between curvature error and velocity fluctuations near the interface.
For the \texttt{gradPhi} method, the trend is non-monotonic. $\mathcal{D}_{\textnormal{md}}$ decreases initially, then increases again with further refinement. At coarse resolution, the dominant velocity error stems from compression-induced artifacts near $\theta = 0$ and $\pi$. At finer resolution, the primary contribution to $\mathcal{D}_{\textnormal{md}}$ shifts toward increased spurious currents in the \texttt{isoPhi} reference solution.

To further examine the importance of drop speed, we vary the initial speed of the drop while fixing the grid resolution at $N_x = 512$ and $N_y = 256$. The drop speed is changed to $|u|_{\textnormal{center}} = 0.1$, $0.3$, $0.6$, and $1.0$ m/s, \textcolor{black}{corresponding to Weber numbers $We = \rho_{\textnormal{circle}} |u|_{\textnormal{center}}^2 D_{\textnormal{drop}} / \sigma = 111$, $998$, $3992$, and $11089$, and Reynolds numbers $Re = \rho_{\textnormal{circle}} |u|_{\textnormal{center}} D_{\textnormal{drop}} / \mu_{\textnormal{circle}} = 7.98\times10^{4}$, $2.40\times10^{5}$, $4.79\times10^{5}$, and $7.98\times10^{5}$, respectively}. The resulting velocity difference $\mathcal{D}_{\textnormal{md}}$ are shown in Fig.~\ref{fig:moving drop error}(b). For the \texttt{plicRDF} method, as the drop velocity increases, the relative importance of velocity fluctuation errors decreases significantly. This indicates that numerical errors from inaccurate curvature estimation become significant only when the spurious current is on the same order of magnitude as the bulk velocity. In other words, when the interface moves sufficiently fast compared to the magnitude of the spurious current, its impact becomes negligible.
For the \texttt{gradPhi} method, however, $\mathcal{D}_{\textnormal{md}}$ remains nearly constant regardless of drop velocity. Since the grid is fixed, this invariance implies that $\mathcal{D}_{\textnormal{md}}$ is not caused by spurious currents, but rather by excessive velocities introduced by the compression term. This issue will be addressed in more detail in the following section.

\section{Wave test cases and results \label{sec:results:waves}}
We next examine the influence of interface advection on the flow field through the numerical discretization of the flux term. The \texttt{isoPhi} and \texttt{plicRDF} schemes employ the flux formulation given in Eq.~\eqref{eq:total flux, geometric VOF}, whereas the \texttt{gradPhi} method with MULES adopts the formulation in Eq.~\eqref{eq:total flux, algebraic VOF}.

\textcolor{black}{We report simulation parameters in dimensional form rather than through a fixed set of nondimensional groups. For the wind-wave cases in this section, a complete characterization of the flow state would require several independent nondimensional parameters, for example a wave Reynolds number, a wave Weber number, a Froude number, and the wave age, and quantities such as the friction velocity that enter these groups are not prescribed a priori. They are instead diagnosed a posteriori from the simulated velocity profile after the wind and wave fields reach approximate equilibrium.
This diagnosed value is sensitive to the simulation domain, the averaging procedure, and the degree of wind-wave equilibration, and small differences in how it is computed can produce large apparent changes in nondimensionalized velocity or stress magnitudes. 
The relationship between these nondimensional numbers and the underlying physics is not yet fully established. No single choice of reference length or velocity scale, for example air-side versus water-side quantities, or wave height versus wavelength, has been established as standard for defining these groups in this configuration~\citep{wu2022revisiting}.
Reporting a fixed set of nondimensional groups therefore risks obscuring rather than clarifying comparisons between cases. We provide nondimensional numbers where they can be unambiguously defined, but retain dimensional quantities as the primary reporting convention in this section for this reason.
}

\subsection{Sensitivity on interface advection: solitary wave\label{sec:result:two-dimensional solitary wave}}
A solitary wave is a single, isolated wave propagating under conditions relevant to wind–wave interaction. It exhibits relatively high propagation speed over a large spatial scale, with negligible surface-tension effects. To isolate advection effects, we simulate a two-dimensional solitary wave in a periodic domain over a short duration. \textcolor{black}{Under these conditions, a full three-dimensional simulation would be expected to exhibit negligible three-dimensional effects, which justifies restricting the present simulations to a two-dimensional domain, where such effects are eliminated by construction.}

\subsubsection{Simulation setup}
We consider a rectangular domain of dimensions $L_x = 72.6~\textnormal{m}$ (horizontal) and $L_y = 2.2~\textnormal{m}$ (vertical), discretized with uniform grid resolution of $N_x = 2112$ and $N_y = 110$ in the horizontal and vertical directions, respectively. 
\textcolor{black}{All simulations throughout Section~\ref{sec:results:waves} are performed with finite viscosity, representative of realistic wind–wave conditions. The only exception is the validation of the wave elevation profile against the inviscid potential-flow theory of \citet{laitone1960second} in Fig.~\ref{fig:2D inviscid solitary wave}(b) (see below), for which an inviscid simulation is used for direct comparison with the theoretical solution. Because the density ratio between water and air is $O(10^3)$ and the simulation duration is short relative to the wave inertia, the resulting wave shape is nearly identical between the inviscid and viscous cases, so this exception does not affect the viscous results presented elsewhere.}

Boundary conditions are prescribed as follows. \textcolor{black}{A} Neumann boundary condition is applied at the top \textcolor{black}{for the inviscid validation case}, while a slip boundary condition is applied at the top of the viscous domain used for all other cases, \textcolor{black}{including} the kinetic-energy analysis discussed below. Periodic conditions are imposed on the lateral boundaries, and a no-slip condition is enforced at the bottom wall.

\begin{figure*}
     \centering
     \begin{subfigure}[b]{0.6\textwidth}
         \centering
         \caption{}
         \includegraphics[width=\textwidth]{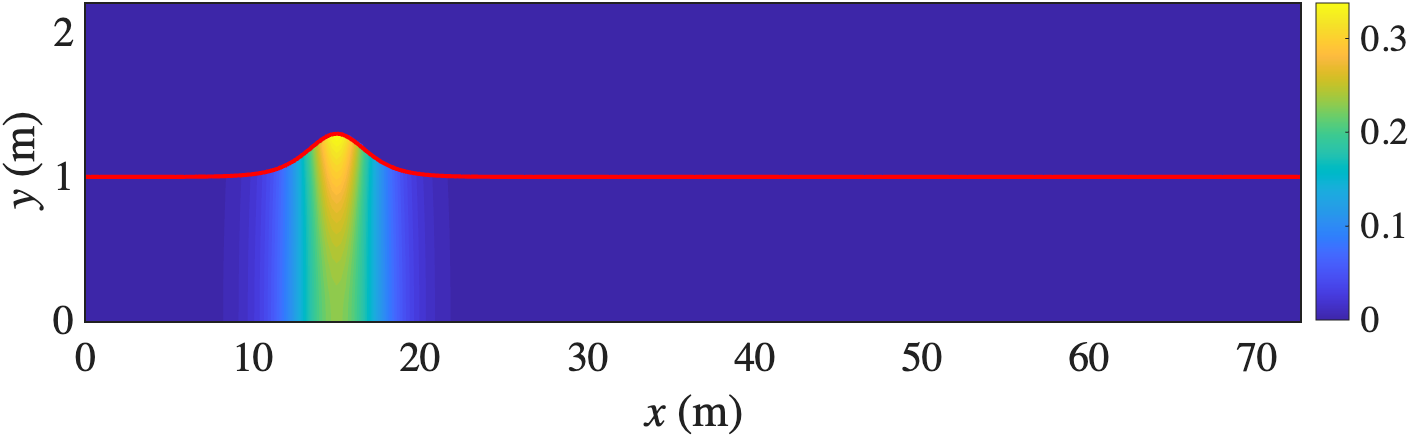}
     \end{subfigure}
     \\
     \begin{subfigure}[b]{0.6\textwidth}
         \centering
         \caption{}
         \includegraphics[width=\textwidth]{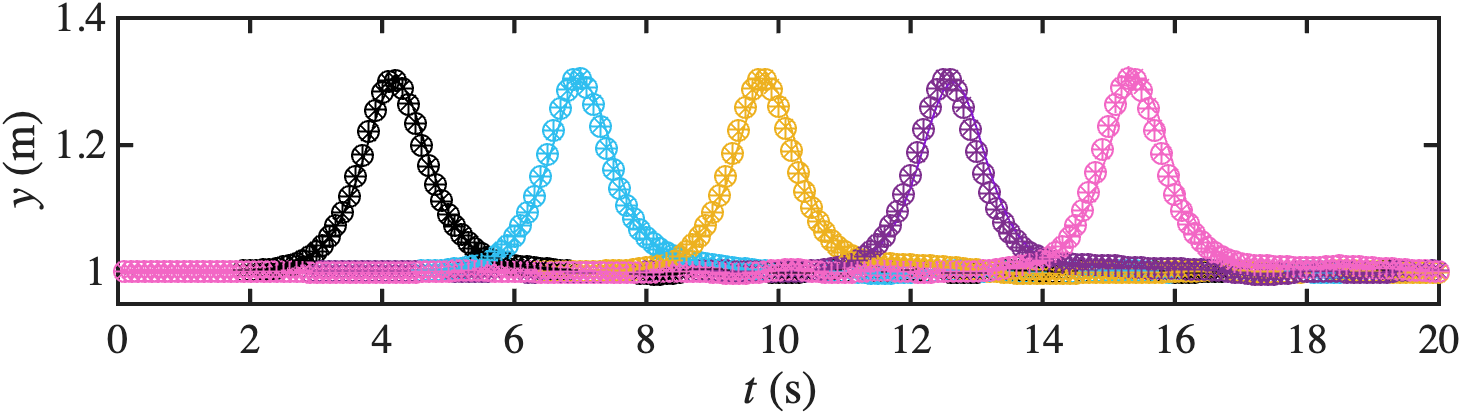}
     \end{subfigure}
        \caption{
        (a) Initial condition for the solitary wave simulation. The velocity magnitude is shown as a colormap, and the interface ($\phi = 0.5$) is indicated by a red solid line.
        (b) Wave elevation profiles of the two-dimensional solitary wave. Symbols represent simulation results using \texttt{isoPhi} (circles), \texttt{gradPhi} (crosses), and \texttt{plicRDF} (diamonds), while solid lines show the theoretical solution from potential flow theory. Colors correspond to probe locations at $x = 30$ (black), $40$ (cyan), $50$ (yellow), $60$ (purple), and $70~\textnormal{m}$ (pink). 
        }
        \label{fig:2D inviscid solitary wave}
\end{figure*}

The velocity and interface fields are initialized using the second-order approximation to solitary wave theory by~\citet{laitone1960second}. The initial wave elevation is given by
\begin{equation}
\eta = \frac{A}{H} \mathrm{S}^2 -\frac{3}{4}\left(\frac{A}{H}\right)^2 \mathrm{S}^2 \left(1-\mathrm{S}^2\right),
\label{eq:2D inviscid solitary wave: initial condition1} 
\end{equation}
and the horizontal and vertical velocity components are
\begin{eqnarray}
&\displaystyle\frac{u}{\sqrt{gH}} = \frac{A}{H}\left\{1-\frac{5}{4}\frac{A}{H}-\frac{3}{2}\frac{A}{H}\left(2y+y^2\right)\right\} \mathrm{S}^2 \nonumber\\
&\displaystyle + \left(\frac{A}{H}\right)^2 \left\{\frac{5}{4} + \frac{9}{4}\left(2y+y^2\right)\right\}\mathrm{S}^4,\\
&\displaystyle\frac{v}{\sqrt{gH}} = \sqrt{3\left(\frac{A}{H}\right)^3}(1+y)\mathrm{S}^2\mathrm{T} \left\{ 1-\frac{7}{8} \right.\nonumber \\
&\displaystyle\left. \frac{A}{H} -\frac{1}{2}\frac{A}{H} (2y+y^2)- \frac{1}{2}\frac{A}{H}\left(1-6y-3y^2\right) \mathrm{S}^2 \right\},
\label{eq:2D inviscid solitary wave: initial condition2}
\end{eqnarray}
where \textcolor{black}{$\mathrm{S} = \text{sech}\left( \widetilde{x} \right)$ and} $\mathrm{T} = \tanh{(\widetilde{x})}$.
The wave coordinate defined as
\begin{equation}
\widetilde{x} = (x-x_0)\sqrt{\frac{3A}{4H}}\left(1-\frac{5A}{8H}\right).
\label{eq:2D inviscid solitary wave: initial condition3} 
\end{equation}
Here, $A$ and $H$ denote the wave amplitude and mean water depth, respectively, and $x_0 = 15~\textnormal{m}$ sets the initial wave peak location. Gravity is denoted by $g = 9.8~\textnormal{m}/\textnormal{s}^2$. A snapshot of the initial configuration is shown in Fig.~\ref{fig:2D inviscid solitary wave}(a). Simulations are carried out up to $t = 20~\textnormal{s}$. 

\subsubsection{Interface regularization effects}
Interface elevations are recorded at five probe locations ($x = 30$, $40$, $50$, $60$, and $70~\textnormal{m}$) and shown in Fig.~\ref{fig:2D inviscid solitary wave}(b). In this inviscid case with zero surface tension, the solitary wave is expected to maintain its shape and kinetic energy as it propagates. Results from all three interface-capturing methods (symbols) closely match the theoretical solution (solid lines), indicating that interface transport is largely unaffected by the choice of method under these idealized conditions.
This conclusion also extends to cases where inertial effects dominate over viscous and capillary forces. For example, we confirmed that under realistic density ratios ($\rho_{\text{water}} / \rho_{\text{air}} \sim O(10^3)$) and small but non-zero surface tension ($\sigma = 0.072~\textnormal{N}/\textnormal{m}$), interface advection accuracy remains robust across all schemes, with errors below $0.2\%$.

However, the effect of the choice of numerical scheme on the low-density (air) phase is more pronounced. To assess this sensitivity, we compare the face fluxes from the \texttt{isoPhi} and \texttt{gradPhi} methods (Eqs.~\eqref{eq:total flux, geometric VOF} and \eqref{eq:total flux, algebraic VOF}).
Due to the coarse grid resolution, \texttt{isoPhi} and \texttt{plicRDF} yield nearly identical results (see Section\ref{sec:result:moving drop}), so we focus on differences only with \texttt{isoPhi}.

\begin{figure*}
     \centering
     \begin{subfigure}[b]{0.8\textwidth}
         \centering
         \caption{}
         \includegraphics[width=\textwidth]{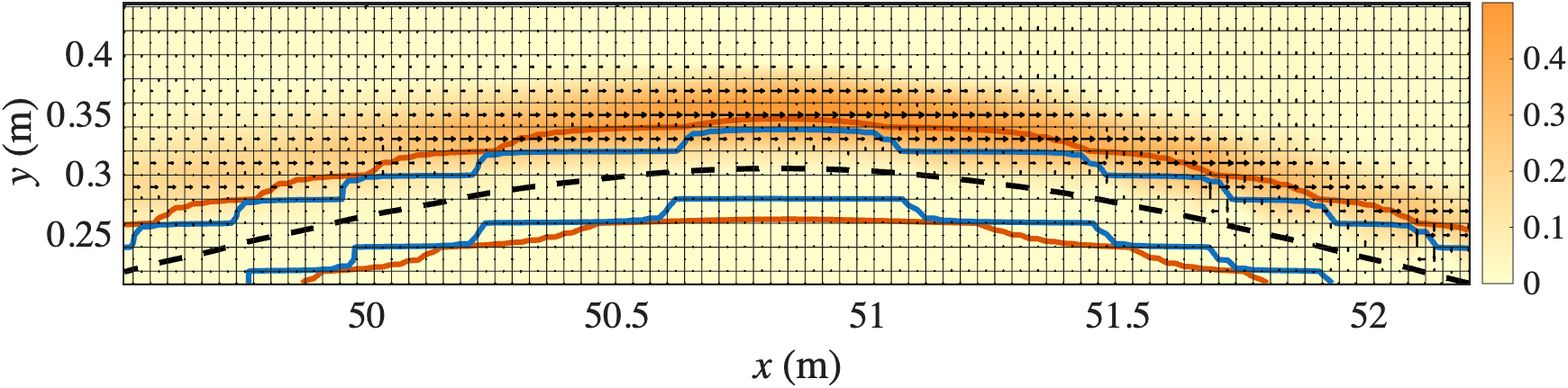}
     \end{subfigure}
     \\
     \begin{subfigure}[b]{0.8\textwidth}
         \centering
         \caption{}
         \includegraphics[width=\textwidth]{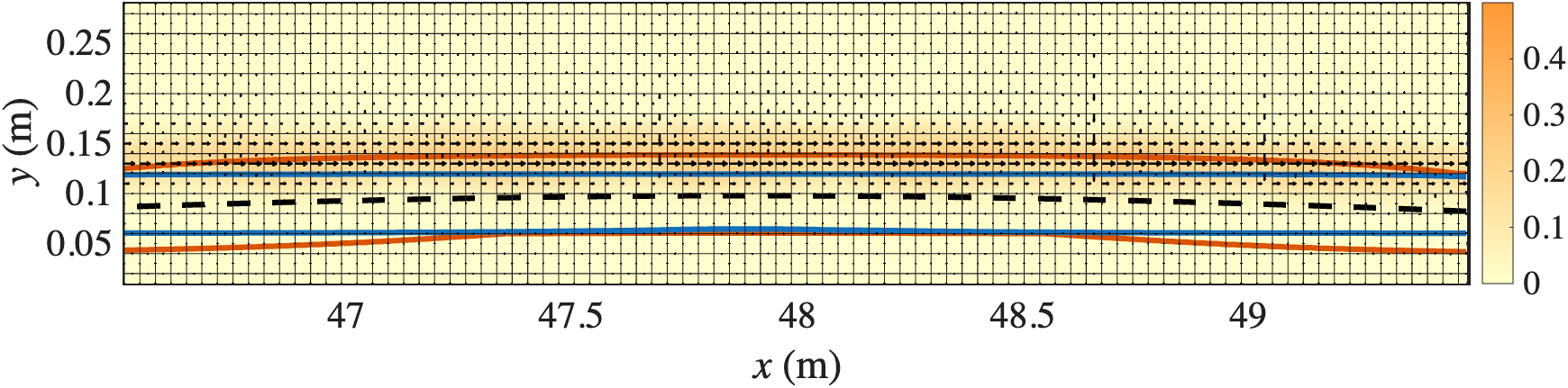}
     \end{subfigure}
     \\
     \begin{subfigure}[b]{0.8\textwidth}
         \centering
         \caption{}
         \includegraphics[width=\textwidth]{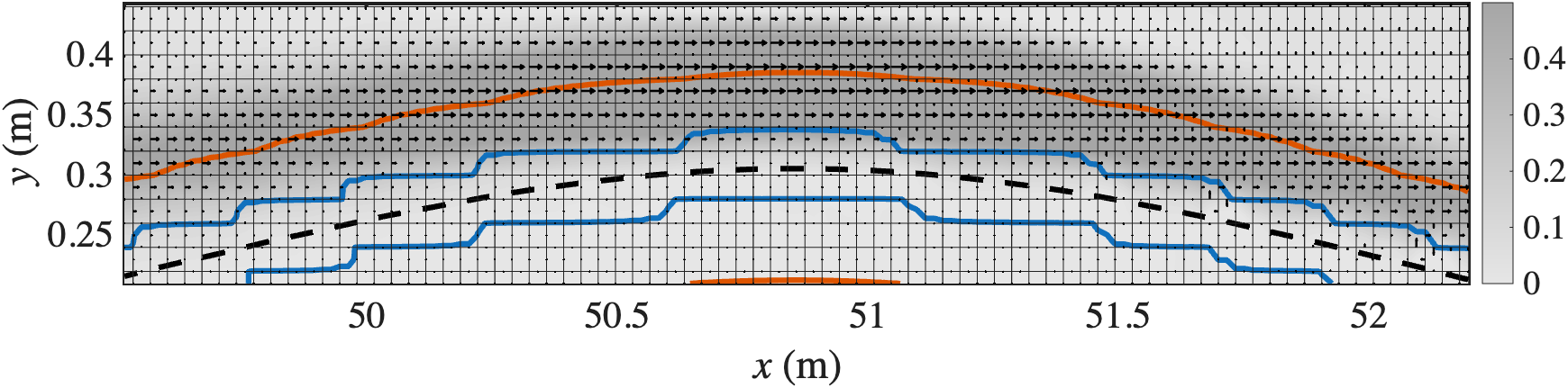}
     \end{subfigure}
     \\
     \begin{subfigure}[b]{0.8\textwidth}
         \centering
         \caption{}
         \includegraphics[width=\textwidth]{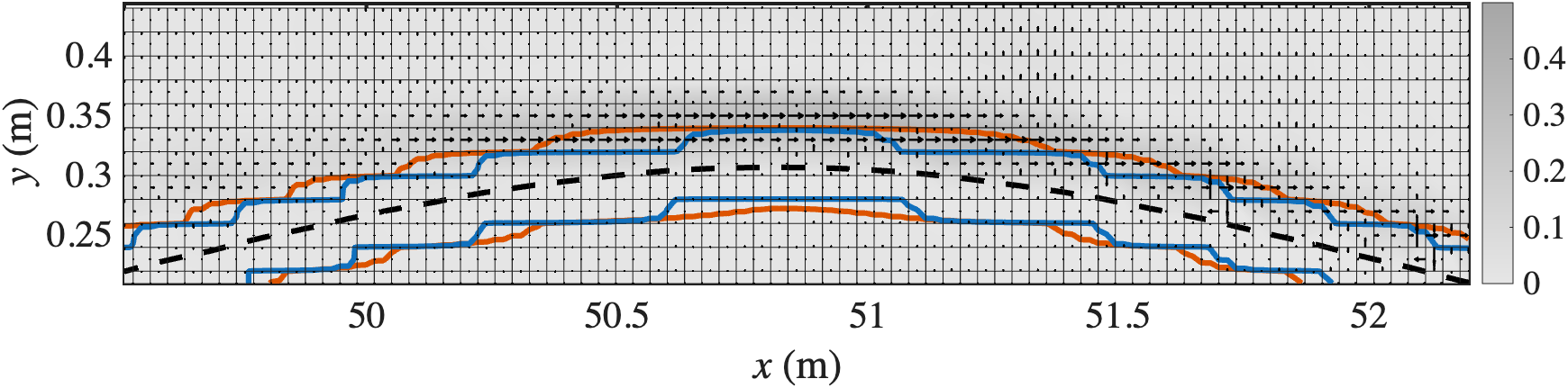}
     \end{subfigure}
    \caption{
    Comparison of face flux differences between \texttt{isoPhi} and \texttt{gradPhi}, calculated as $F_{\textnormal{MULES}} - F_{\textnormal{ISO}}$, near the crest of the solitary wave. Panels show
    (a) $A/H = 0.3$ with $C_\alpha = 1.0$,
    (b) $A/H = 0.1$ with $C_\alpha = 1.0$,
    (c) $A/H = 0.3$ with reduced compression velocity $C_\alpha = 0.1$, and
    (d) $A/H = 0.3$ with increased compression velocity $C_\alpha = 2.0$.
    Flux difference magnitude is shown as a colormap, while face flux vectors are indicated by black arrows at cell faces. Cell face outlines are shown in black. Iso-contours of the phase indicator are displayed for $\phi = 0.5$ (black-dashed), and $\phi = 0.1$, $0.9$ for \texttt{isoPhi} (blue) and \texttt{gradPhi} (red).
    }
    \label{fig:face flux}
\end{figure*}

Fig.~\ref{fig:face flux} shows the flux difference $F_{\text{MULES}} - F_{\text{ISO}}$ near the crest of the solitary wave. For $A = 0.3~\textnormal{m}$ and $H = 1~\textnormal{m}$ (Fig.~\ref{fig:face flux}(a)), a positive flux difference layer is observed above the crest with thickness comparable to the interface. A similar pattern appears for a smaller wave ($A = 0.1~\textnormal{m}$, Fig.~\ref{fig:face flux}(b)). The associated crest velocities are $u_{\text{crest}} = 0.337$ and $0.1~\textnormal{m}/\textnormal{s}$ for $A/H = 0.3$ and $0.1$, respectively. \textcolor{black}{Because a solitary wave is non-periodic, we characterize its length scale using the natural decay length of its $\text{sech}^2$ profile, $\lambda = \left[\sqrt{3A/(4H)}\left(1-5A/(8H)\right)\right]^{-1}$, obtained from the wave coordinate $\widetilde{x} = (x-x_0)/\lambda$ in Eq.~\eqref{eq:2D inviscid solitary wave: initial condition3}, giving $\lambda = 2.59$ and $3.89~\textnormal{m}$ for $A/H = 0.3$ and $0.1$, respectively. Using the water density $\rho_{\text{water}} = 998~\textnormal{kg}/\textnormal{m}^3$, this corresponds to Weber numbers $We = \rho_{\text{water}} u_{\text{crest}}^2 \lambda/\sigma = 4085$ and $540$, Bond numbers $Bo = \rho_{\text{water}} g \lambda^2 /\sigma = 9.15\times10^{5}$ and $2.06\times10^{6}$, and Froude numbers $Fr = u_{\text{crest}}/\sqrt{gH} = 0.108$ and $0.032$, for $A/H = 0.3$ and $0.1$, respectively.} These results suggest that the extra momentum flux introduced in \texttt{gradPhi} is aligned with the wave direction and scales with the advection speed of the interface.
This excess flux originates from the interface compression term, which artificially increases momentum transfer in mixed cells ($0 < \phi < 1$). The magnitude of this effect is governed by the compression velocity in Eq.\eqref{eq: relative velocity definition}.

To confirm this behavior, we conduct additional simulations for $A/H = 0.3$ with varying compression coefficients: $C_\alpha = 0.1$, $1.0$, and $2.0$.
Lower $C_\alpha$ results in a thicker interface and a more pronounced excess flux layer, while higher $C_\alpha$ yields a thinner interface and reduced flux anomaly.
\begin{figure}
     \centering
     \begin{subfigure}[b]{0.9\columnwidth}
         \centering
         \caption{}
         \includegraphics[width=\textwidth]{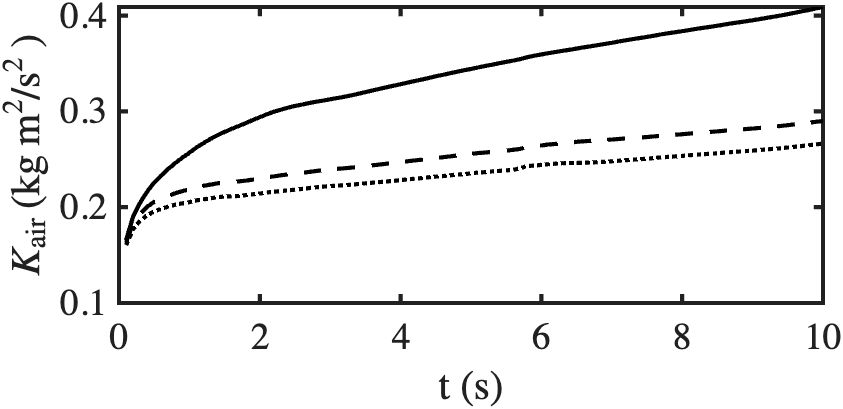}
     \end{subfigure}
     \hfill
     \begin{subfigure}[b]{0.9\columnwidth}
         \centering
         \caption{}
         \includegraphics[width=\textwidth]{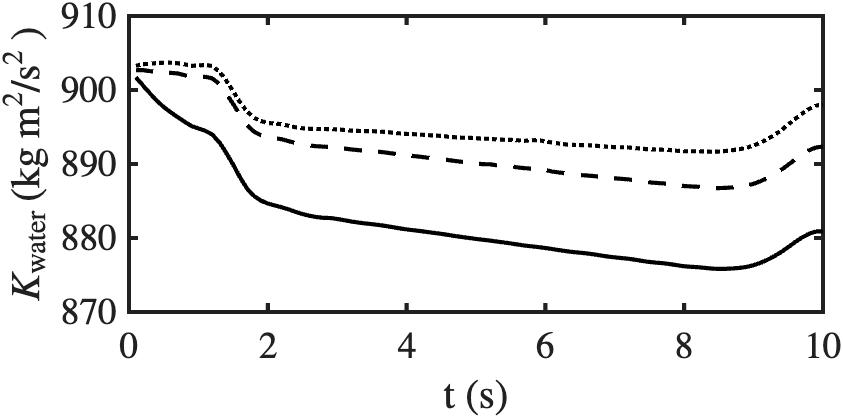}
     \end{subfigure}
        \caption{
        Total kinetic energy in the (a) air phase and (b) water phase during solitary wave simulations. Black lines correspond to different values of the compression velocity coefficient: $C_\alpha = 0.1$ (solid), $C_\alpha = 1.0$ (dashed), and $C_\alpha = 2.0$ (dotted). 
        }
    \label{fig:solitary:kinetic energy}
\end{figure}
This phenomenon is further supported by the kinetic energy distribution across the two phases, as shown in Fig.~\ref{fig:solitary:kinetic energy}. The total kinetic energy of each phase, calculated as $K_{1,2} = \alpha_{1,2} \sum \textcolor{black}{\tfrac{1}{2}\rho u_i^2 \,\mathrm{d}V}$, indicates that lower $C_\alpha$ results in higher kinetic energy in the air phase and more rapid decay in the water phase, reflecting artificial momentum transfer across the interface.

This artificial momentum transfer aligns with observations from Section~\ref{sec:result:moving drop}, where \texttt{gradPhi} shows elevated velocity at $\theta = 0$ and $\pi$ (Fig.~\ref{fig:moving drop error fields}(c)), locations where the interface normal is orthogonal to the advection direction, similar to wave crests. When the grid is under-resolved, such flux artifacts significantly affect fast-moving interfaces.
\todo{Mention about initial condition effects in Figure 13 (b)?}

\subsection{Monochromatic waves with high wave age\label{sec:monochromatic waves with high wave age}}
In this section, we examine a more practical configuration involving simulations of monochromatic waves with high wave age. Using the triple-decomposition framework described in Section~\ref{sec:method:triple decomposition}, we analyze phase-averaged fields and stress distributions for all three interface-capturing methods. Two monochromatic wave scales are considered: a reference scale and a larger scale. This comparison is used to demonstrate how the observations discussed in Section~\ref{sec:result:two-dimensional solitary wave} manifest across different geometric scalings.
%

\begin{table}
\caption{
    Configuration of the (a) base grid and (b) refinement zones. In (a), $y$ indicates the vertical extent of each subdomain, and $N_y$ denotes the number of grid cells in the vertical direction. The total expansion ratio $\Gamma$ represents the ratio between the largest and smallest cell heights. In (b), $\theta$ indicates the refinement level in each Cartesian direction, where $\theta = \log_2(\Delta_{\text{base}}/\Delta_{\text{refined}})$.
    }
    \centering
    \begin{subtable}{1.0\columnwidth}
    \centering
    \caption{}
    \begin{tabular}{c||ccc|ccc|ccc}

        \multirow{2}{*}{} & \multicolumn{3}{c|}{Top} & \multicolumn{3}{c|}{Middle} & \multicolumn{3}{c}{Bottom} \\ \hline 
                         & $y~(\textnormal{m})$ & $N_y$ & $\Gamma$ & $y~(\textnormal{m})$ & $N_y$ & $\Gamma$ & $y~(\textnormal{m})$ & $N_y$ & $\Gamma$ \\ \hline \hline
        $G1$               & $[0.54, 0.84]$ & 50 & 10  & $[0.5, 0.54]$ & 25  & 1  & $[0.0, 0.5]$ & 30 & 40  \\ \hline
        $G2$               & $[0.54, 0.84]$ & 60 & 18  & $[0.5, 0.54]$ & 50  & 1  & $[0.0, 0.5]$ & 40 & 50  \\ \hline
        $G3$               & $[0.54, 0.84]$ & 60 & 35  & $[0.5, 0.54]$ & 100 & 1  & $[0.0, 0.5]$ & 40 & 80   \\ 
    \end{tabular}
    \end{subtable}
    
    \begin{subtable}{1.0\columnwidth}
    \centering
    \caption{}
    \begin{tabular}{c || c | c }
              & $y~(\textnormal{m})$ & $(\theta_x,\theta_y,\theta_z)$  \\
         \hline
         \hline
         $R1$ & $[0.49, 0.58]$ & $(2,0,1)$ \\
         $R2$ & $[0.5, 0.55]$ & $(3,0,2)$ \\
         \hline
    \end{tabular}
    \end{subtable}
    \label{tab:old wave grid configuration}
\end{table}

\subsubsection{Simulation setup}
First, the simulation setup for the reference-scale case is described. The computational domain has dimensions $L_x = 7.0~\textnormal{m}$ in the streamwise direction, $L_y = 0.84~\textnormal{m}$ in the vertical direction, and $L_z = 0.08~\textnormal{m}$ in the spanwise direction. The grid resolution is $N_x = 700$ points in the streamwise direction and $N_z = 10$ points in the spanwise direction. In the vertical direction, the domain is divided into three subdomains: top, middle, and bottom. The top and bottom regions are predominantly occupied by air and water, respectively, while the middle region contains the air–water interface. The middle subdomain is discretized uniformly, whereas the top and bottom subdomains employ stretched grids with increasing cell sizes away from the interface. Details of the grid configuration for each subdomain are provided in Table~\ref{tab:old wave grid configuration}(a).

As an example, in the $G3$ configuration, the middle region uses a uniform grid spacing of $\Delta y = 4.0 \times 10^{-4}~\textnormal{m}$. In the top subdomain, the grid is stretched with a total expansion ratio $\Gamma = \Delta y_{\text{max}} / \Delta y_{\text{min}} = 35$, yielding $\Delta y_{\text{max}} = 1.8 \times 10^{-2}~\textnormal{m}$ near the top boundary and $\Delta y_{\text{min}} = 5.15 \times 10^{-4}~\textnormal{m}$ at the interface with the middle region. The corresponding local cell-to-cell expansion ratio is approximately $1.0621$. This base grid is further refined using local refinement windows, as summarized in Table~\ref{tab:old wave grid configuration}(b). Refinement levels in each direction are denoted by $(\theta_x, \theta_y, \theta_z)$, where $\theta = \log_2(\Delta_{\text{base}}/\Delta_{\text{refined}})$. In total, four combinations of base grid and refinement level are considered: $(G1,R1)$, $(G2,R1)$, $(G3,R1)$, and $(G3,R2)$.

The target wave parameters are a wavelength $\lambda = 0.81~\textnormal{m}$, wave height $H = 0.04~\textnormal{m}$, mean water depth $h = 0.52~\textnormal{m}$, \textcolor{black}{and wave period $T = 0.83~\textnormal{s}$}. The left boundary at $x = 0.0~\textnormal{m}$ serves as the inlet and is divided vertically, with $y \le 0.54~\textnormal{m}$ designated as the wave inlet and $y > 0.54~\textnormal{m}$ as the wind inlet. Similarly, at the right boundary ($x = 7.0~\textnormal{m}$), $y \le 0.54~\textnormal{m}$ defines the wave outlet and $y > 0.54~\textnormal{m}$ defines the wind outlet.

The wave field is prescribed at the wave inlet using first-order Stokes wave theory as a Dirichlet boundary condition, propagating toward the outlet, where an absorbing boundary condition is applied to minimize wave reflections. For the wind field, a constant streamwise velocity of $u = \textcolor{black}{0.05~\textnormal{m/s}}$ is imposed at the wind inlet. To mitigate artificial vortex generation near the interface, a linear vertical interpolation of velocity is applied between the wave crest and the bottom of the wind inlet. A Neumann-type boundary condition is avoided at the wind inlet, as it produces an uncontrolled inflow velocity profile that is highly sensitive to the interface-capturing method.
A slip condition is applied at the top boundary, periodic conditions are imposed in the spanwise direction, and a Neumann-type condition is enforced at the wind outlet.

For the larger-scale configuration, a tenfold geometric scaling is applied to simulate a larger wave case. The scaled wave parameters are a wave period $T = 2.616~\textnormal{s}$, wavelength $\lambda = 8.1~\textnormal{m}$, wave height $H = 0.4~\textnormal{m}$, and mean water depth $h = 5.2~\textnormal{m}$. Consequently, the wave celerity increases from $1.1~\textnormal{m}/\textnormal{s}$ to $3.51~\textnormal{m}/\textnormal{s}$. All domain dimensions and boundary conditions are scaled proportionally. \textcolor{black}{Gravitational acceleration is set to $g = 9.8~\textnormal{m}/\textnormal{s}^2$, directed along $-\hat{y}$, for both the reference- and larger-scale configurations. Using the wave celerity $c$ and wavelength $\lambda$ as characteristic velocity and length scales, together with the water properties $\rho_{\textnormal{water}} = 998~\textnormal{kg}/\textnormal{m}^3$, $\mu_{\textnormal{water}} = 1.0\times10^{-3}~\textnormal{Pa}\cdot\textnormal{s}$, and $\sigma = 0.072~\textnormal{N}/\textnormal{m}$, the corresponding wave Reynolds numbers are $Re = c\lambda\rho_{\textnormal{water}}/\mu_{\textnormal{water}} = 8.89\times10^{5}$ (reference scale) and $2.84\times10^{7}$ (larger scale); the wave Weber numbers are $We = \rho_{\textnormal{water}} c^2 \lambda/\sigma = 1.36\times10^{4}$ and $1.38\times10^{6}$; and the wave Froude numbers are $Fr = c/\sqrt{g\lambda} = 0.390$ and $0.394$, respectively. The near-identical Froude numbers confirm that the larger-scale case is a Froude-scaled counterpart of the reference case.}

\subsubsection{High wave-age waves at small scale}
\begin{figure*}
     \centering
     \begin{subfigure}[b]{0.9\textwidth}
         \centering
         \caption{}
         \includegraphics[width=\textwidth]{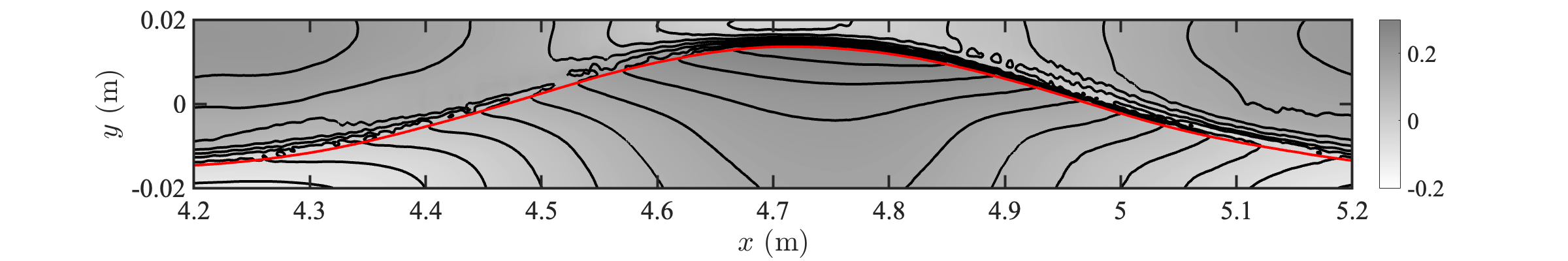}
     \end{subfigure}
     \\
     \begin{subfigure}[b]{0.9\textwidth}
         \centering
         \caption{}
         \includegraphics[width=\textwidth]{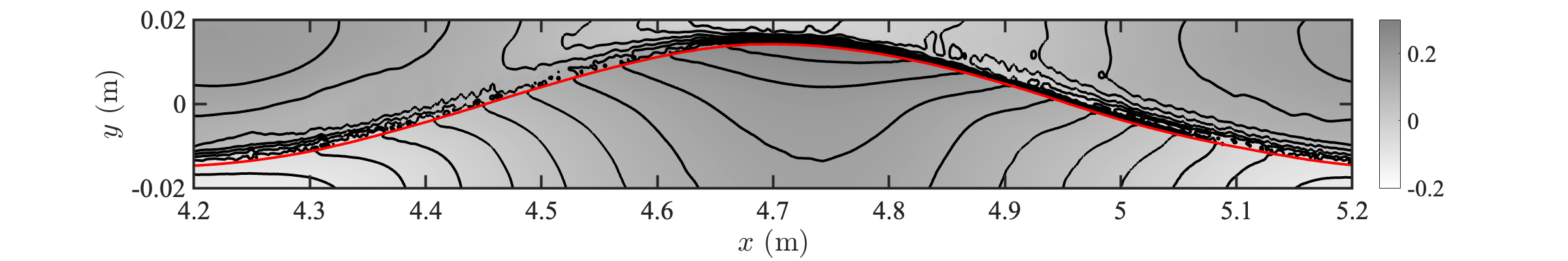}
     \end{subfigure}
     \\
     \begin{subfigure}[b]{0.9\textwidth}
         \centering
         \caption{}
         \includegraphics[width=\textwidth]{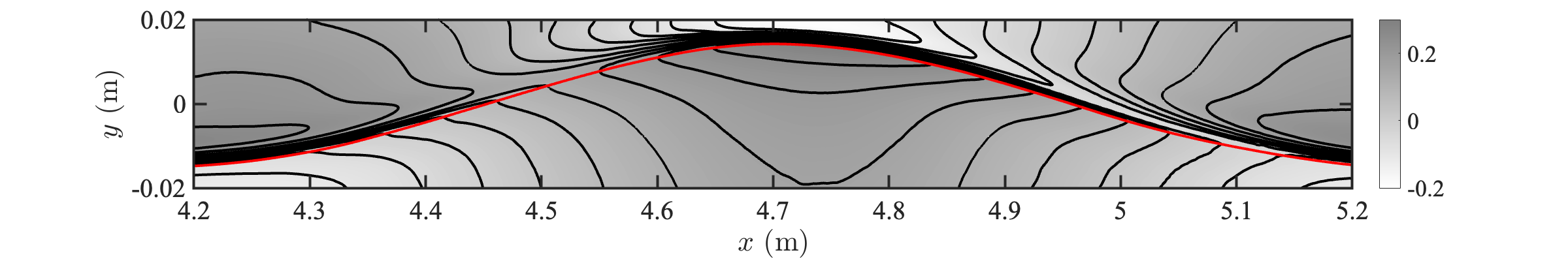}
     \end{subfigure}
     \\
     \begin{subfigure}[b]{0.9\textwidth}
         \centering
         \caption{}
         \includegraphics[width=\textwidth]{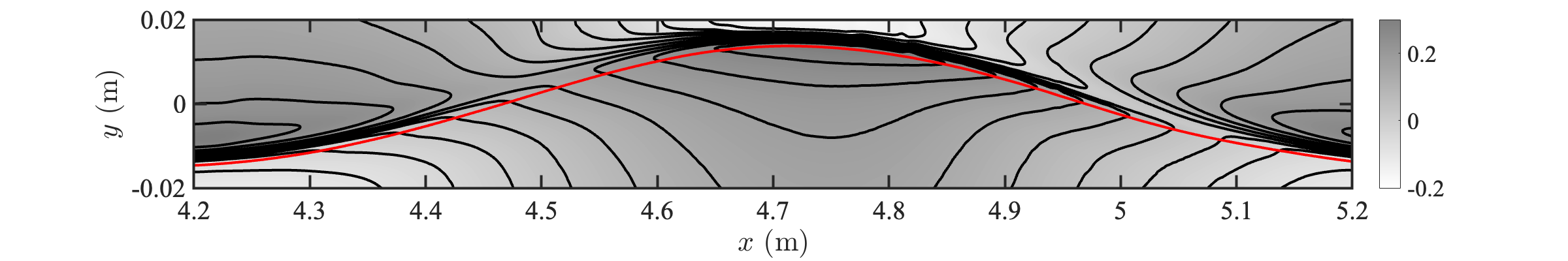}
     \end{subfigure}
        \caption{
        Instantaneous velocity field of monochromatic wave simulations at $t = 20~\textnormal{s}$ in the region $x \in [4.2, 5.2]~\textnormal{m}$.
        Panels show (a) \texttt{isoPhi} with the $(G3,R1)$ grid, (b) \texttt{isoPhi} with the $(G3,R2)$ grid, (c) \texttt{plicRDF}, and (d) \texttt{gradPhi}. The colormap represents velocity magnitude, and the interface ($\phi = 0.5$) is denoted by the red solid line. 
        }
        \label{fig:waveonly:instantaneous velocity fields}
\end{figure*}

We first examine the instantaneous wave profile to investigate the manifestation of spurious currents. Fig.~\ref{fig:waveonly:instantaneous velocity fields} shows the instantaneous velocity fields at $t = 20~\textnormal{s}$ for all interface-capturing methods using the $(G3,R1)$ grid. \textcolor{black}{By this time, the wave has propagated through the domain for approximately $24$ wave periods ($T = 0.83~\textnormal{s}$) for the reference-scale case, and approximately $19$ wave periods ($T = 2.616~\textnormal{s}$) for the larger-scale case discussed in Section~\ref{sec:monochromatic waves with high wave age}, allowing sufficient interaction between the wind and wave fields for the phase-averaged statistics reported below.} For the \texttt{isoPhi} approach, which has been shown to generate the highest spurious current magnitude among the three methods, we observe clear velocity disturbances near the interface within the lighter fluid (air) region. 
In contrast, both \texttt{plicRDF} and \texttt{gradPhi} exhibit significantly smoother instantaneous velocity fields, indicating the absence of grid-scale numerical noise that is prominently observed in \texttt{isoPhi}. However, we emphasize that the smoothness of the velocity field does not inherently guarantee physical accuracy.

To further highlight this effect, we present additional results for \texttt{isoPhi} on a finer $(G3,R2)$ grid. Qualitatively, the finer resolution reveals higher-frequency velocity oscillations and slightly amplified spurious currents, consistent with the curvature error scaling observed in Section~\ref{sec:results:curvature error scaling}. As demonstrated in Sections~\ref{sec:result:moving drop} and \ref{sec:result:two-dimensional solitary wave}, the impact of spurious current becomes more significant when its magnitude is comparable to the characteristic velocity of the system. While large-scale waves with low curvature and negligible surface tension force compared to the inertia force may not be heavily affected, in laboratory-scale or low-Reynolds-number simulations, this influence can be considerable.


\begin{figure*}
     \centering
     \begin{subfigure}[b]{0.9\columnwidth}
         \centering
         \caption{}
         \includegraphics[width=\textwidth]{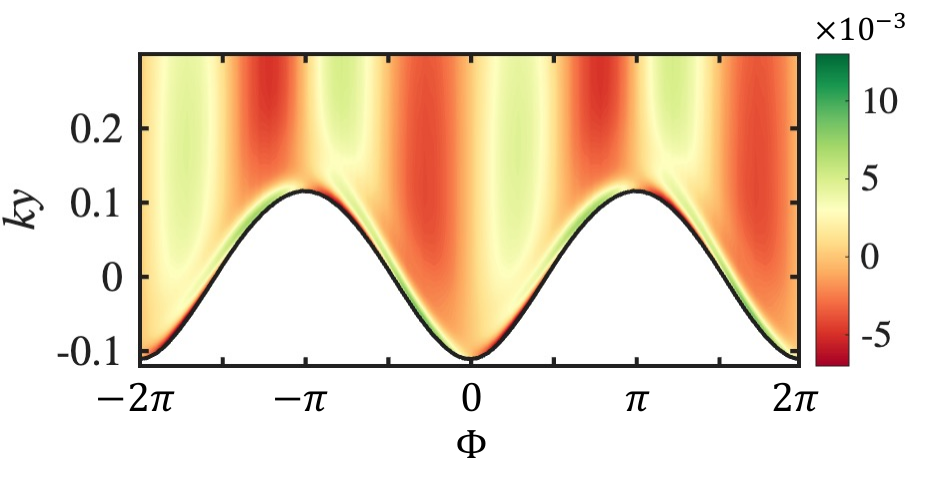}
     \end{subfigure}
     \begin{subfigure}[b]{0.9\columnwidth}
         \centering
         \caption{}
         \includegraphics[width=\textwidth]{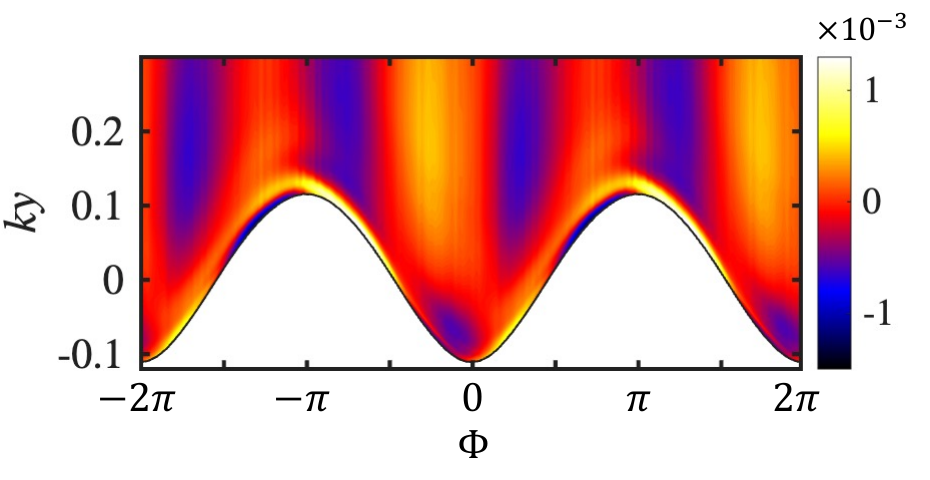}
     \end{subfigure}
     \\
     \begin{subfigure}[b]{0.9\columnwidth}
         \centering
         \caption{}
         \includegraphics[width=\textwidth]{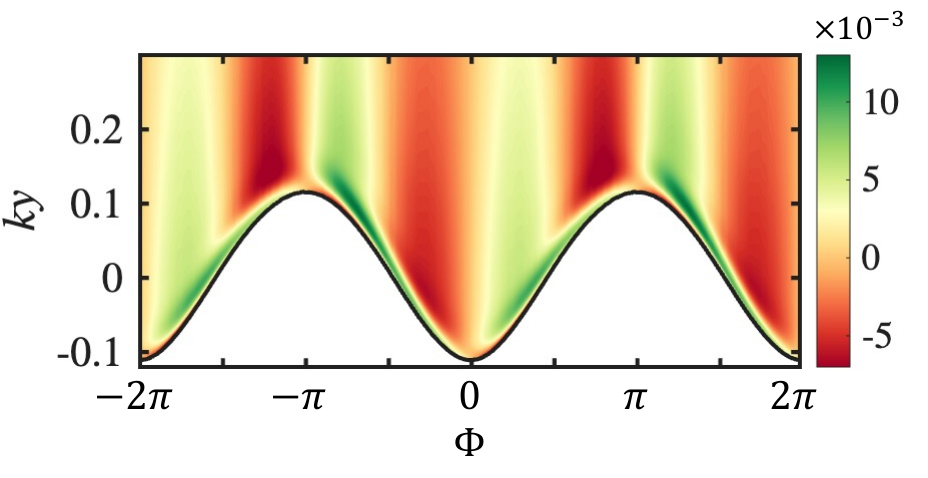}
     \end{subfigure}
     \begin{subfigure}[b]{0.9\columnwidth}
         \centering
         \caption{}
         \includegraphics[width=\textwidth]{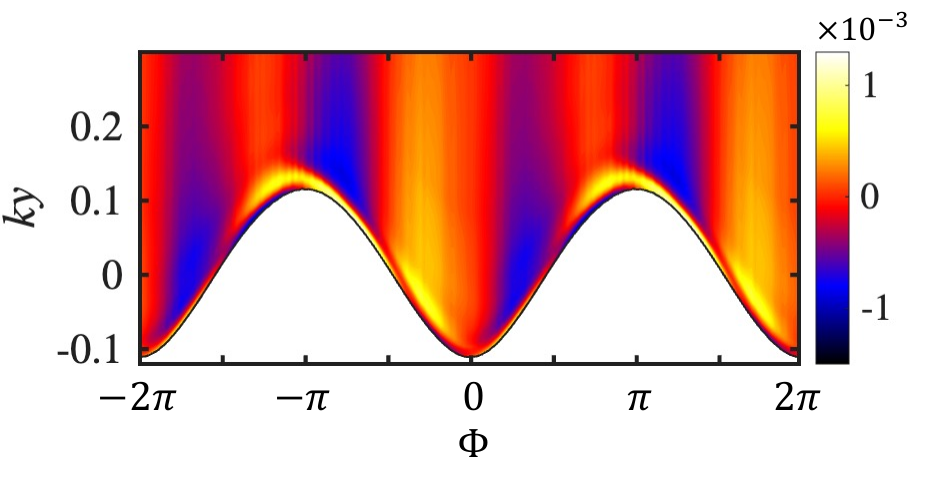}
     \end{subfigure}
     \\
     \begin{subfigure}[b]{0.9\columnwidth}
         \centering
         \caption{}
         \includegraphics[width=\textwidth]{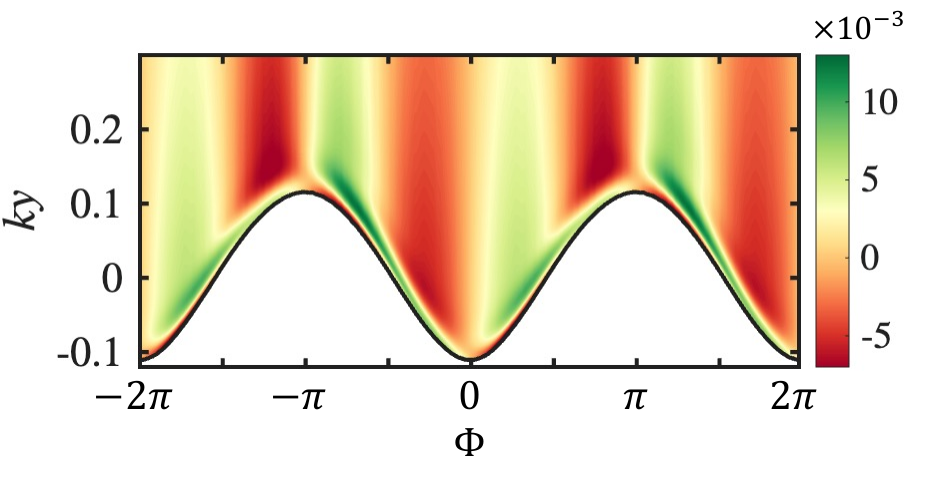}
     \end{subfigure}
     \begin{subfigure}[b]{0.9\columnwidth}
         \centering
         \caption{}
         \includegraphics[width=\textwidth]{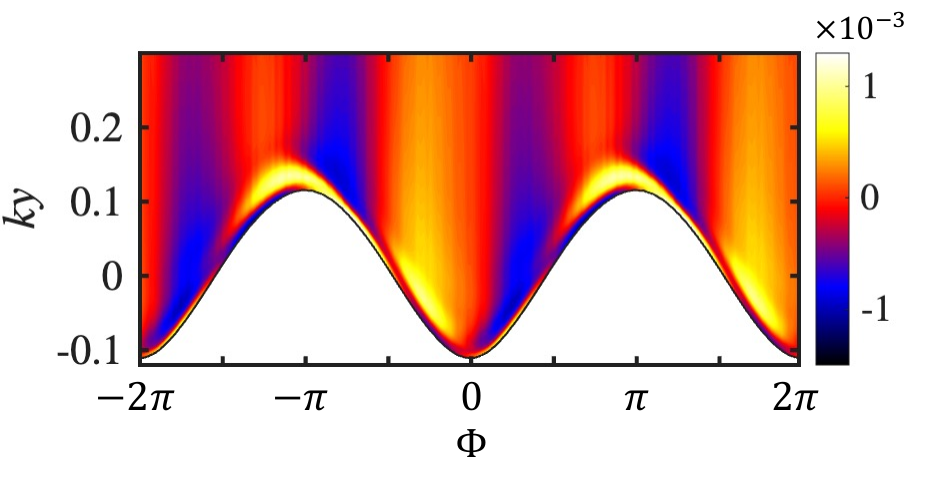}
     \end{subfigure}
        \caption{
        Wave-coherent (first column) and turbulent (second column) Reynolds stress fields for each interface-capturing method.
        Results are presented for (a–b) \texttt{isoPhi}, (c–d) \texttt{plicRDF}, and (e–f) \texttt{gradPhi}. The interface ($\phi = 0.5$) is indicated by black solid lines.}
        \label{fig:mono:stresses}
\end{figure*}
The analysis proceeds with phase-averaged stress fields for the wave-coherent and turbulent components, shown in Fig.~\ref{fig:mono:stresses}.
The wave-coherent stress at this resolution and wave celerity, differences between \texttt{plicRDF} and \texttt{gradPhi} in stress fields are negligible.
However, the \texttt{isoPhi} method exhibits diminished positive stress on both the windward and leeward sides of the wave, consistent with the presence of spurious currents that persist regardless of grid refinement (cf. Section~\ref{sec:results:curvature error scaling}).
For the turbulent stress fields shown in the second column of Fig.~\ref{fig:mono:stresses}, all methods exhibit weak turbulence generation, consistent with the relatively low wave celerity. The \texttt{isoPhi} method fails to capture the positive stress on the leeward side near the troughs and shows only weak asymmetry in the positive stress region beyond the crests. The \texttt{gradPhi} method predicts stronger positive stress than \texttt{plicRDF}, although both exhibit similar spatial distributions.

\subsubsection{Grid convergence of vertical profiles}
\begin{figure}
     \centering
     \begin{subfigure}[b]{0.9\columnwidth}
         \centering
         \caption{}
         \includegraphics[width=\textwidth]{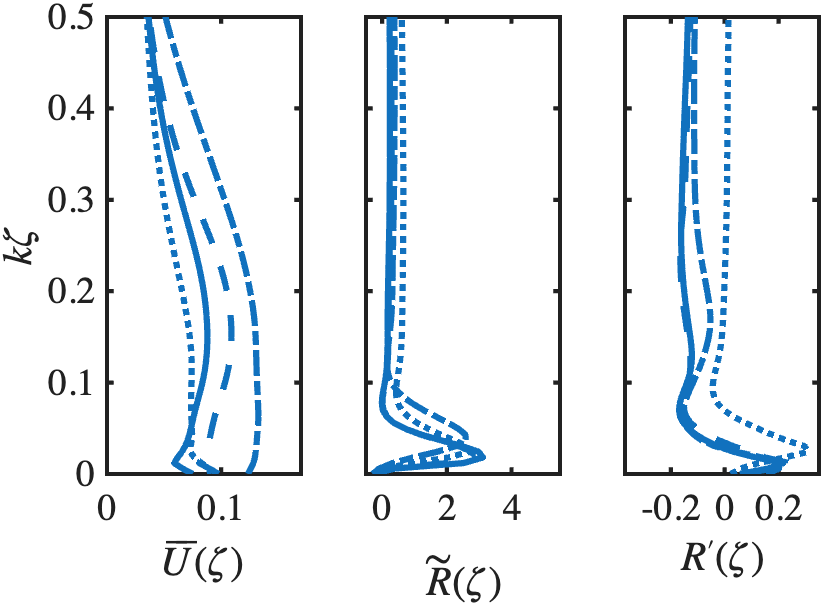}
     \end{subfigure}
     \\
     \begin{subfigure}[b]{0.9\columnwidth}
         \centering
         \caption{}
         \includegraphics[width=\textwidth]{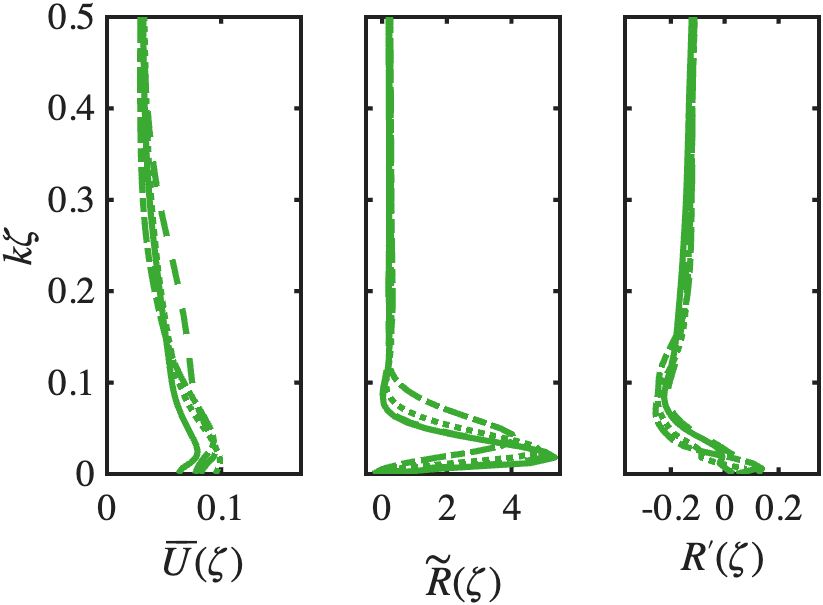}
     \end{subfigure}
     \\
     \begin{subfigure}[b]{0.9\columnwidth}
         \centering
         \caption{}
         \includegraphics[width=\textwidth]{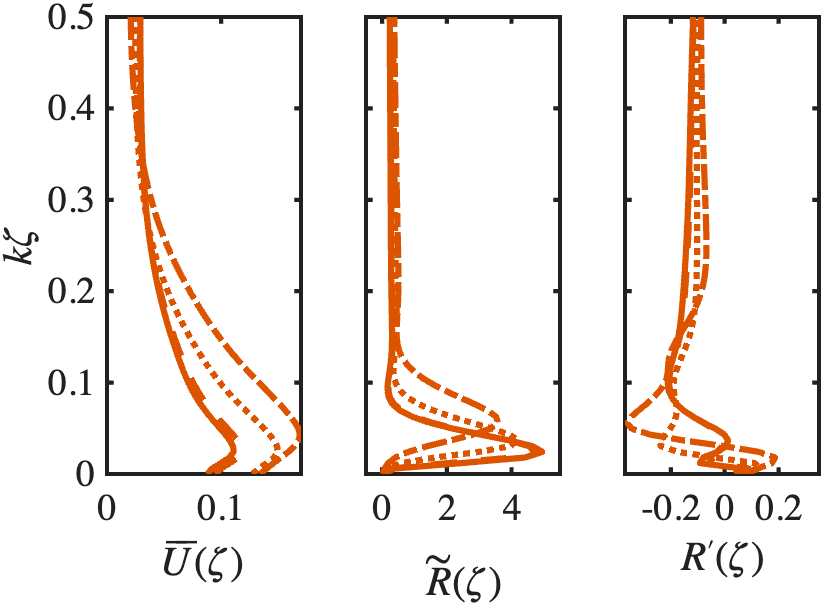}
     \end{subfigure}
        \caption{
        Vertical profiles of the phase-independent mean velocity, wave-coherent stress, and turbulent stress for each column, respectively.
        Subfigures show results from (a) \texttt{isoPhi}, (b) \texttt{plicRDF}, and (c) \texttt{gradPhi}. Line types indicate grid configurations, $(G1,R1)$ (dash-dotted), $(G2,R1)$ (dotted), $(G3,R1)$ (dash), and $(G3,R2)$ (solid). 
        }
        \label{fig:mono:vertical profiles}
\end{figure}
To demonstrate the conclusions from the previous sections, vertical profiles of the phase-independent mean velocity ($\overline{U}$), wave-coherent stress ($\tilde{R}$), and turbulent stress ($R'$) for each interface-capturing method are presented across four grid configurations: $(G1,R1)$, $(G2,R1)$, $(G3,R1)$, and $(G3,R2)$, as shown in Fig.~\ref{fig:mono:vertical profiles}. The finest grid configuration does not include additional refinement in the vertical direction. \textcolor{black}{The mean velocity and stress components in the velocity and stress fields reported in this section are nondimensionalized using a nominal friction velocity $u^* = 1~\textnormal{m/s}$ and its square, respectively. Because this value is not diagnosed from the simulated flow but fixed at unity, this normalization is numerically equivalent to reporting the corresponding dimensional quantities directly. We do not characterize the wind profile in this section using an actual friction velocity, since the wind field here is prescribed as a constant streamwise velocity at the inlet ($u = 0.05~\textnormal{m/s}$, as described above) rather than a logarithmic, friction-velocity-based profile. Furthermore, as shown in Fig.~\ref{fig:mono:vertical profiles}, the phase-independent mean velocity profile does not converge with grid refinement and is highly sensitive to the numerical setup, indicating that it would not provide a robust basis for defining a diagnosed friction velocity even if one were attempted.} Snapshots for phase-averaging were collected within the region $x = 3.8~\textnormal{m}$ to $5.8~\textnormal{m}$.

For the phase-independent mean velocity, \texttt{isoPhi} displays notable kinks, reflecting the influence of spurious currents. These profiles do not exhibit convergence under grid refinement. In contrast, \texttt{gradPhi} shows a clear convergence trend, with a local maximum above the interface becoming more pronounced on finer grids. This behavior is also observed in \texttt{plicRDF}, which exhibits smoother convergence toward a similar profile, albeit with a slightly lower maximum velocity. The larger maximum velocity predicted by \texttt{gradPhi} compared to \texttt{plicRDF} is attributed to the additional flux induced by interface regularization. As the grid is refined, \texttt{gradPhi} exhibits relatively rapid asymptotic convergence, consistent with the findings reported in the previous sections.

For the wave-coherent stress, all methods exhibit similar vertical trends, with local maxima located slightly above the interface that shift closer to it as the resolution increases. The \texttt{isoPhi} method underpredicts the stress magnitude, whereas \texttt{plicRDF} yields the highest values, nearly twice those of \texttt{isoPhi}. Given the negligible wind speed at the inlet, these stress differences correlate with variations in wave height in the phase-averaged fields. The measured wave heights are $3.78\times10^{-2}~\textnormal{m}$ (\texttt{isoPhi}), $3.85\times10^{-2}~\textnormal{m}$ (\texttt{plicRDF}), and $3.75\times10^{-2}~\textnormal{m}$ (\texttt{gradPhi}). The slightly larger wave height predicted by \texttt{plicRDF} (approximately 2\% greater than that predicted by \texttt{gradPhi}) contributes to the corresponding enhancement of the wave-coherent stress. In contrast, the similar wave heights obtained with \texttt{isoPhi} and \texttt{gradPhi} indicate that the stress underprediction in \texttt{isoPhi} originates from contamination by spurious currents.

The turbulent stresses for all three methods exhibit asymptotic convergence, although the converged profiles differ among approaches. Their magnitude is on the order of $O(10^{-4})$ and remains negligible under the present wind–wave conditions. These differences become more pronounced in high wave-age cases with turbulent wind inlet conditions, which are examined in the section below.


\subsubsection{Comparison with large-scale simulations}
\begin{figure}
     \centering
     \begin{subfigure}[b]{0.9\columnwidth}
         \centering
         \caption{}
         \includegraphics[width=\textwidth]{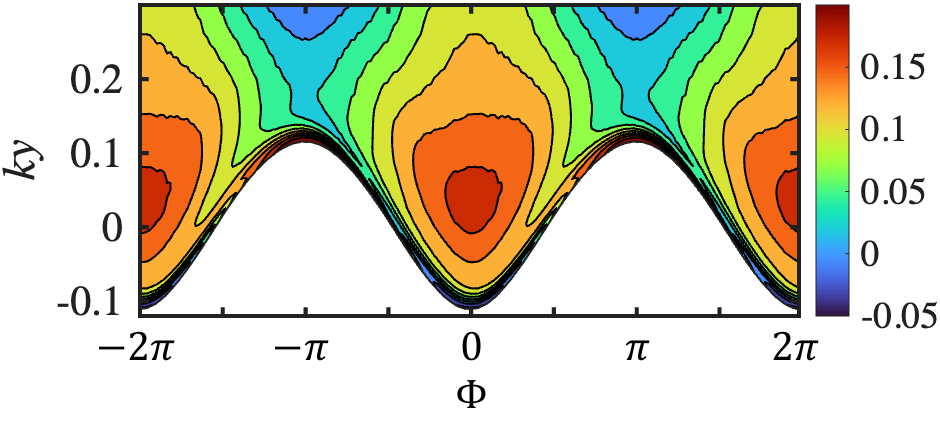}
     \end{subfigure}
     \\
     \begin{subfigure}[b]{0.9\columnwidth}
         \centering
         \caption{}
         \includegraphics[width=\textwidth]{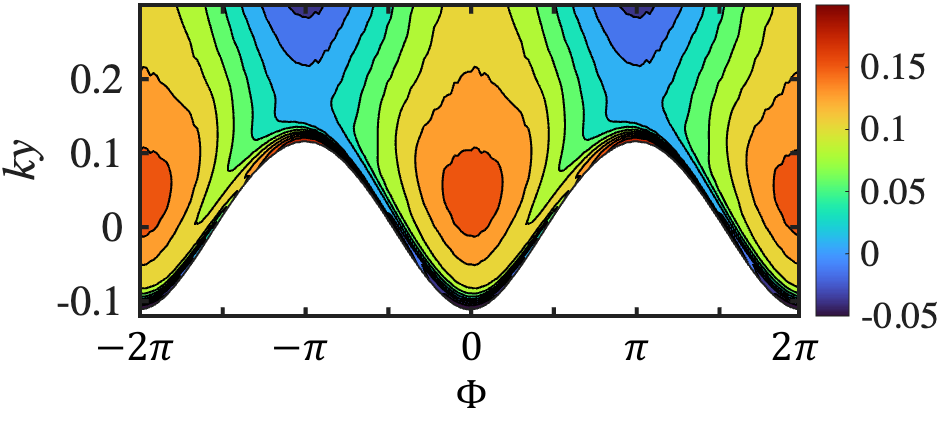}
     \end{subfigure}
     \\
     \begin{subfigure}[b]{0.9\columnwidth}
         \centering
         \caption{}
         \includegraphics[width=\textwidth]{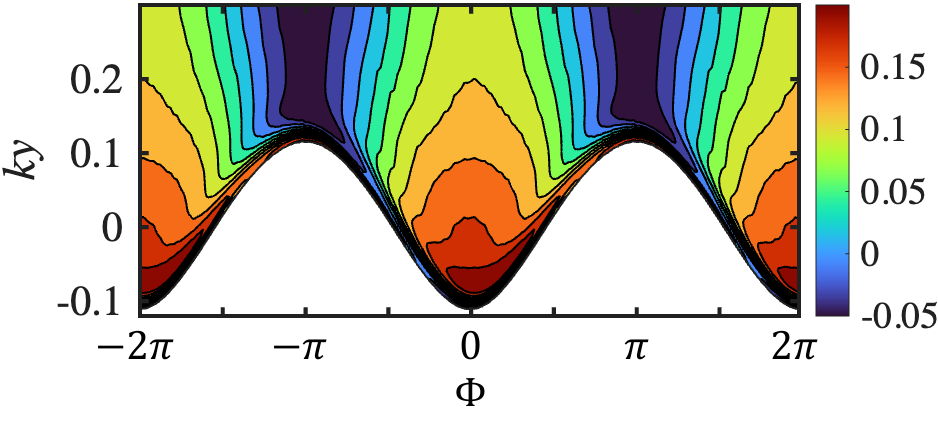}
     \end{subfigure}
     \\
     \begin{subfigure}[b]{0.9\columnwidth}
         \centering
         \caption{}
         \includegraphics[width=\textwidth]{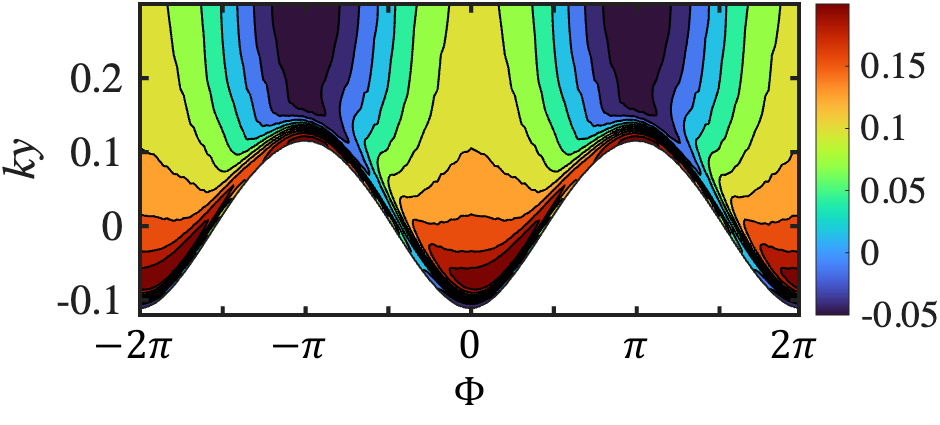}
     \end{subfigure}
        \caption{
        Phase-averaged streamwise velocity field for simulations with different interface-capturing methods.
        Shown are (a) \texttt{isoPhi} with the $(G3,R1)$ grid, (b) \texttt{isoPhi} with the $(G3,R2)$ grid, (c) \texttt{plicRDF}, and (d) \texttt{gradPhi}. The colormap indicates the magnitude of the phase-averaged velocity. 
        }
        \label{fig:mono:phase averaged - coarse grid}
\end{figure}
\begin{figure}
     \centering
     \begin{subfigure}[b]{0.9\columnwidth}
         \centering
         \caption{}
         \includegraphics[width=\textwidth]{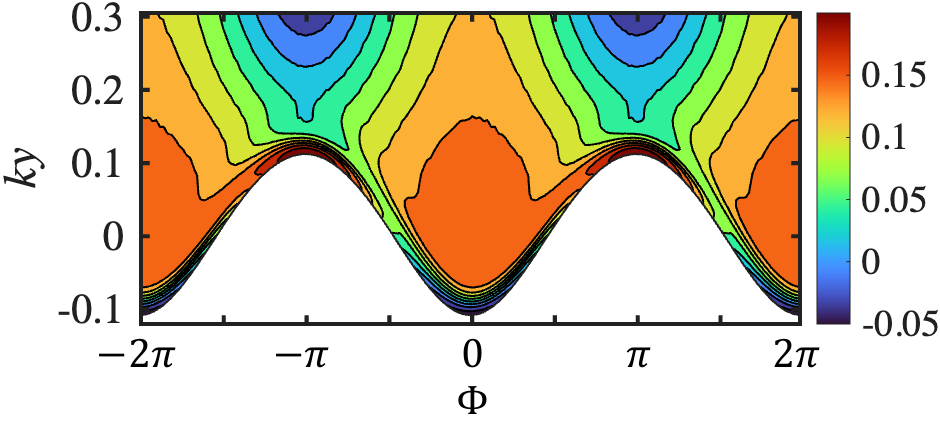}
     \end{subfigure}
     \\
     \begin{subfigure}[b]{0.9\columnwidth}
         \centering
         \caption{}
         \includegraphics[width=\textwidth]{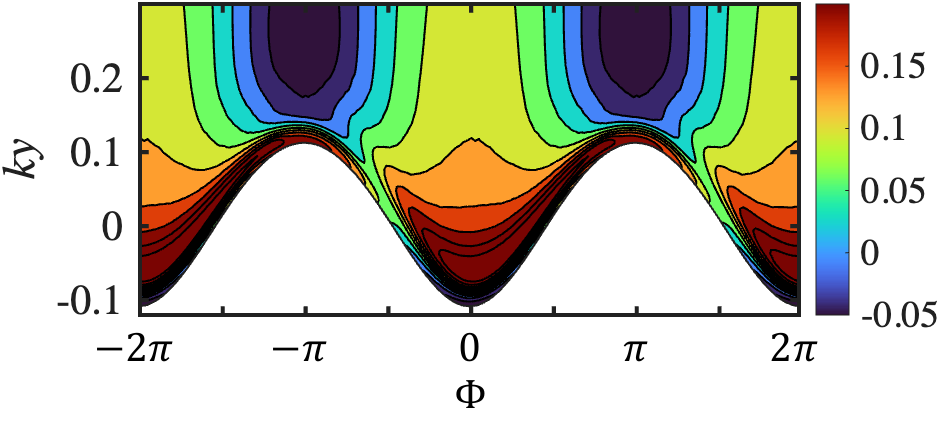}
     \end{subfigure}
     \\
     \begin{subfigure}[b]{0.9\columnwidth}
         \centering
         \caption{}
         \includegraphics[width=\textwidth]{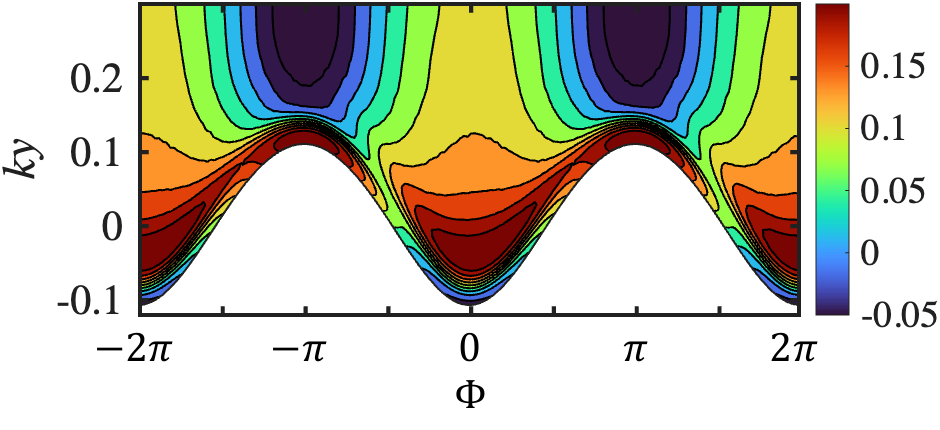}
     \end{subfigure}
        \caption{
        Phase-averaged streamwise velocity fields for geometrically scaled-up monochromatic wave simulations.
        Results are shown for (a) \texttt{isoPhi}, (b) \texttt{plicRDF}, and (c) \texttt{gradPhi}, all using the $(G3,R1)$ grid with a tenfold geometric wave scaling. The colormap represents velocity magnitude.
        }
        \label{fig:mono:phase averaged - large waves}
\end{figure}

Larger-scale waves correspond to a relatively coarser mesh resolution for the wind fields, as well as higher wave celerity compared to the reference case. This configuration therefore provides a suitable framework to examine how the error sources identified in Section~\ref{sec:results:canonical} manifest in the monochromatic wave setup. The phase-averaged streamwise velocity fields are shown in Fig.~\ref{fig:mono:phase averaged - coarse grid} for the reference scale and in Fig.~\ref{fig:mono:phase averaged - large waves} for the larger scale. Both simulations employ the $(G3,R1)$ grid configuration, with the large-wave case having grid spacings that are ten times larger in physical scale.

In Fig.~\ref{fig:mono:phase averaged - coarse grid}(a–b), the \texttt{isoPhi} results show significant deviations from the velocity fields of \texttt{plicRDF} and \texttt{gradPhi} in Fig.~\ref{fig:mono:phase averaged - coarse grid}(c–d), respectively. 
Across all interface-capturing methods, the airflow in proximity to the interface tends to follow the wave-induced motion, characterized by positive streamwise velocity above wave crests and negative velocity above troughs. Further away from the interface, this trend reverses, exhibiting negative velocity above crests and positive velocity above troughs. 
Among the methods, \texttt{isoPhi} demonstrates noticeably smoother transitions toward the maximum level of positive and negative velocity region, suggesting that spurious currents can significantly distort the phase-averaged mean flow when their magnitude becomes comparable to wave-induced velocities.

Phase-averaged streamwise velocity fields for larger scale configuration are shown in Fig.~\ref{fig:mono:phase averaged - large waves}.
In the \texttt{isoPhi} results, the off-interface velocity anomalies observed in the smaller-scale case shift closer to the interface. In contrast, the \texttt{gradPhi} solution exhibits thickened velocity layers near both the crest and trough, with smoother gradients and a broadened region of positive streamwise velocity between successive wave crests. This behavior is consistent with enhanced momentum transport induced by interface regularization.
For \texttt{plicRDF}, thickening of the positive velocity layer is observed near the trough region while maintaining relatively sharp gradients. Compared to the reference scale, where the difference between \texttt{plicRDF} and \texttt{gradPhi} is marginal, a noticeable discrepancy emerges at the larger scale, reflecting the sensitivity of the compression term to the coarser grid resolution.
These findings imply that spurious currents diminish in impact with increased wave scale and interface transport velocity; however, artificial flux layers aligned with interface motion may still arise due to discretization effects.


In conclusion, spurious currents manifest strongly in phase-independent and wave-coherent vertical profiles and can substantially affect velocity prediction in lighter fluids. In coarser grids, artificial compression methods such as \texttt{gradPhi} produce excessive phase-independent velocities aligned with interface motion.
While signs of convergence are observed for each component, additional refinement would be necessary to confirm asymptotic behavior. 

\subsection{Validation against experimental data\label{sec:buckley and veron validation simluation}}
We validate our simulation results against experimental data from \citet{buckley2016} under combined wind-wave conditions.

\subsubsection{Simulation setup}
\todo{note that this portion is overlapping with SNH draft}
The simulation setup, including the wave configuration, follows that described in Section~\ref{sec:monochromatic waves with high wave age}, except for the wind inlet condition. For the wind inlet, we employ the divergence-free synthetic eddy method (DFSEM) proposed by \citet{poletto2013new}, which introduces realistic turbulence by superimposing synthetic eddies on a mean velocity profile. The mean profile follows the standard logarithmic law for wall-bounded flows, expressed as $u(y) = (u^*/\kappa) \log{\left((y + y_0)/y_0\right)}$, where $u^* = 0.041~\textnormal{m}/\textnormal{s}$ is the friction velocity, $\kappa = 0.41$ is the von K\'{a}rm\'{a}n constant, and $y_0 = 1.5 \times 10^{-5}~\textnormal{m}$ is the roughness length of the interface. The Reynolds stress components used for turbulence generation at the inlet are $(R_{xx}$, $R_{xy}$, $R_{xz}$, $R_{yy}$, $R_{yz}$, $R_{zz})$ $=$ $(0.01$, $-0.001681$, $ 0.0$, $0.0025$, $0.0$, $0.0064)~\textnormal{m}^2/\textnormal{s}^2$. The velocity and stresses are normalized by $u^*$ and its square, respectively. \textcolor{black}{The friction velocity $u^* = 0.041~\textnormal{m/s}$ is taken directly from the experimental wind-wave setup of \citet{buckley2016}. Combined with the wave celerity reported in the same reference, $c = 1.3~\textnormal{m/s}$, this corresponds to a wave age $c/u^* \approx 31.7$, consistent with the high-wave-age regime targeted in this comparison. We also note that if the wind field were instead driven over a periodic streamwise domain, the target friction velocity could equivalently be imposed through a prescribed mean pressure gradient; because the present simulation instead uses an inlet/outlet domain, the wind field is prescribed directly at the inlet through the DFSEM condition described above.}

\begin{figure}
    \centering
    \includegraphics[width=1.0\columnwidth]{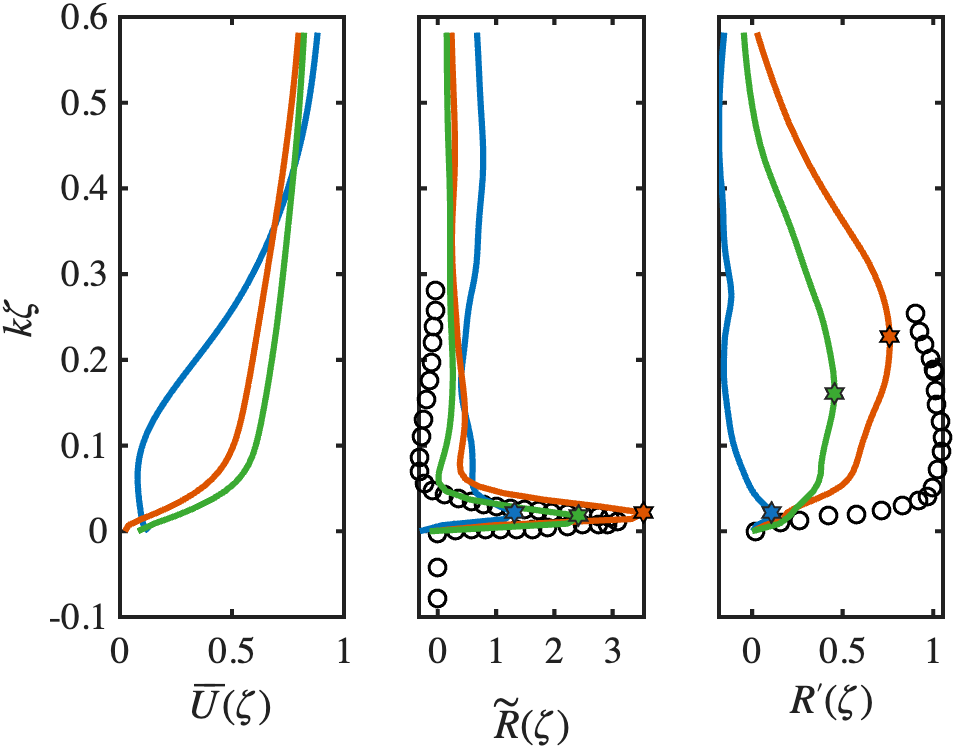}
    \caption{
    Vertical profiles from the old wave simulation, decomposed into (left) phase-independent mean velocity, (middle) wave-coherent stress, and (right) turbulent stress. Lines represent results from \texttt{isoPhi} (blue), \texttt{plicRDF} (green), and \texttt{gradPhi} (red). Black circles indicate experimental reference data from~\citet{buckley2016}. Maximum values of the stress components are indicated by symbols.
    }
    \label{fig:old wave: triple decomposition}
\end{figure}

\subsubsection{Comparison of stress fields}
The comparison of the interface-capturing methods for the vertical profiles is shown in Fig.~\ref{fig:old wave: triple decomposition}. First, no reference data exist for the phase-independent component of the mean velocity. Therefore, it is only compared among the different approaches. Notably, the \texttt{isoPhi} results show a different profile near the interface ($k\zeta < 0.1$), exhibiting an inflection point, whereas the other two approaches display similar profiles.

Next, the experimental data indicate that the wave-coherent stress reaches a maximum value approximately three times the total stress ($\widetilde{R} \approx 3.1$). Although the predicted peak location ($k\zeta \approx 0.018$) is slightly higher than the reference value ($k\zeta \approx 0.011$), all approaches yield maximum values sufficiently close to the reference location, within less than the smallest grid spacing. However, their magnitudes differ: \texttt{isoPhi} predicts $\widetilde{R} \approx 1.31$, \texttt{plicRDF} predicts $\widetilde{R} \approx 2.42$, and \texttt{gradPhi} predicts $\widetilde{R} \approx 3.54$.
The slightly larger maximum value predicted by \texttt{gradPhi} may be attributed to artificial flux near the wave crest, as discussed in Section~\ref{sec:monochromatic waves with high wave age}. We also note that none of the approaches accurately capture the slightly negative stress values beyond the maximum point, farther away from the surface~\citep{sullivan2000simulation}.

For the turbulent stress, all approaches consistently underestimate the magnitude at all vertical locations. The \texttt{isoPhi} approach predicts negative stress immediately away from the interface. Although the \texttt{plicRDF} and \texttt{gradPhi} approaches predict a region of positive stress, both fail to capture the strong gradient near the interface, and their peak magnitudes and locations deviate significantly from the reference data.
     
\begin{figure*}
    \centering
    \begin{subfigure}[b]{0.9\columnwidth}
         \centering
         \caption{}
         \includegraphics[width=\textwidth]{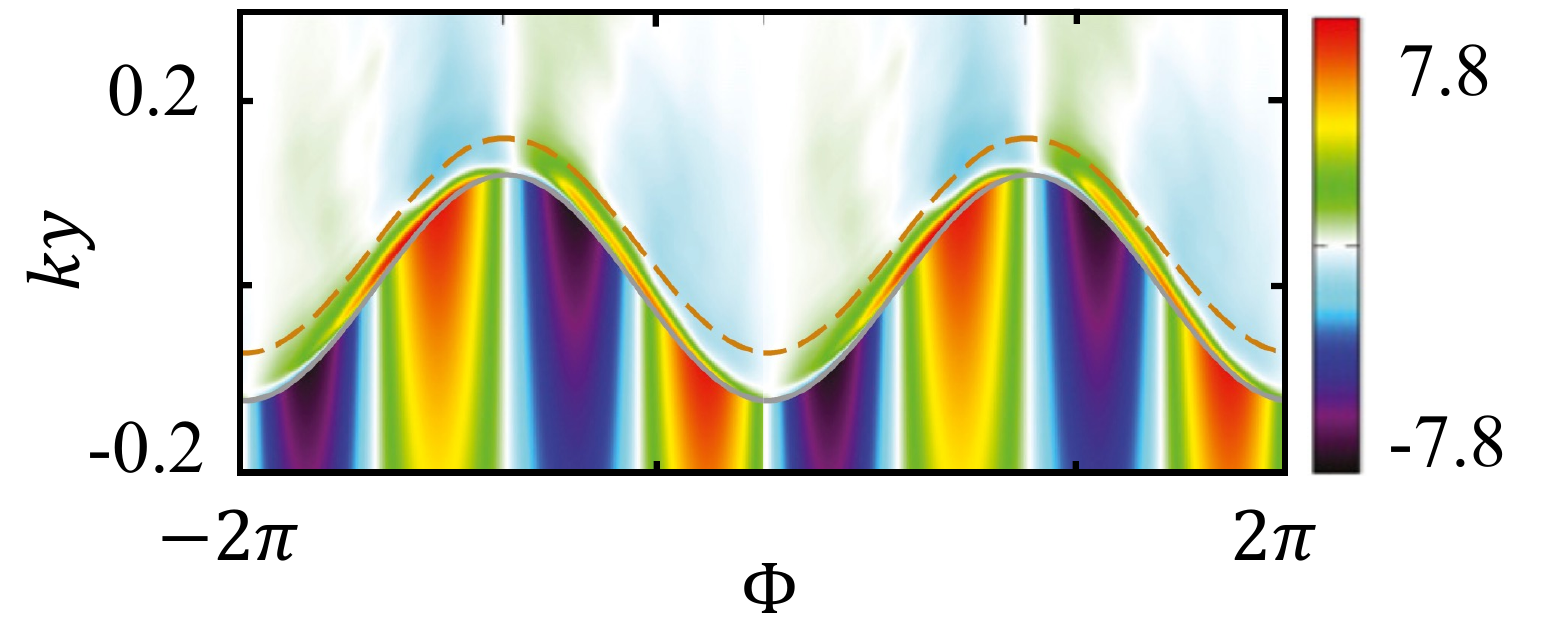}
     \end{subfigure}
     \begin{subfigure}[b]{0.9\columnwidth}
         \centering
         \caption{}
         \includegraphics[width=\textwidth]{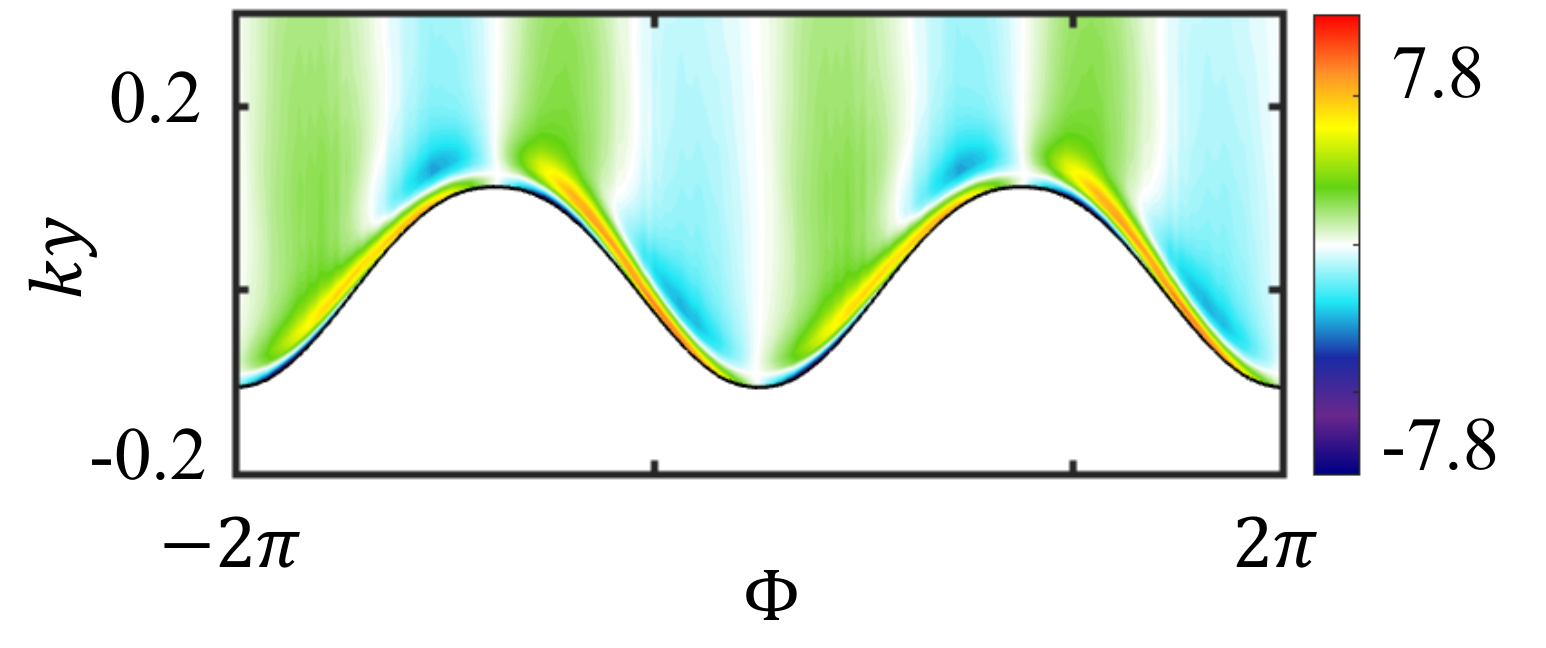}
     \end{subfigure}
     \begin{subfigure}[b]{0.9\columnwidth}
         \centering
         \caption{}
         \includegraphics[width=\textwidth]{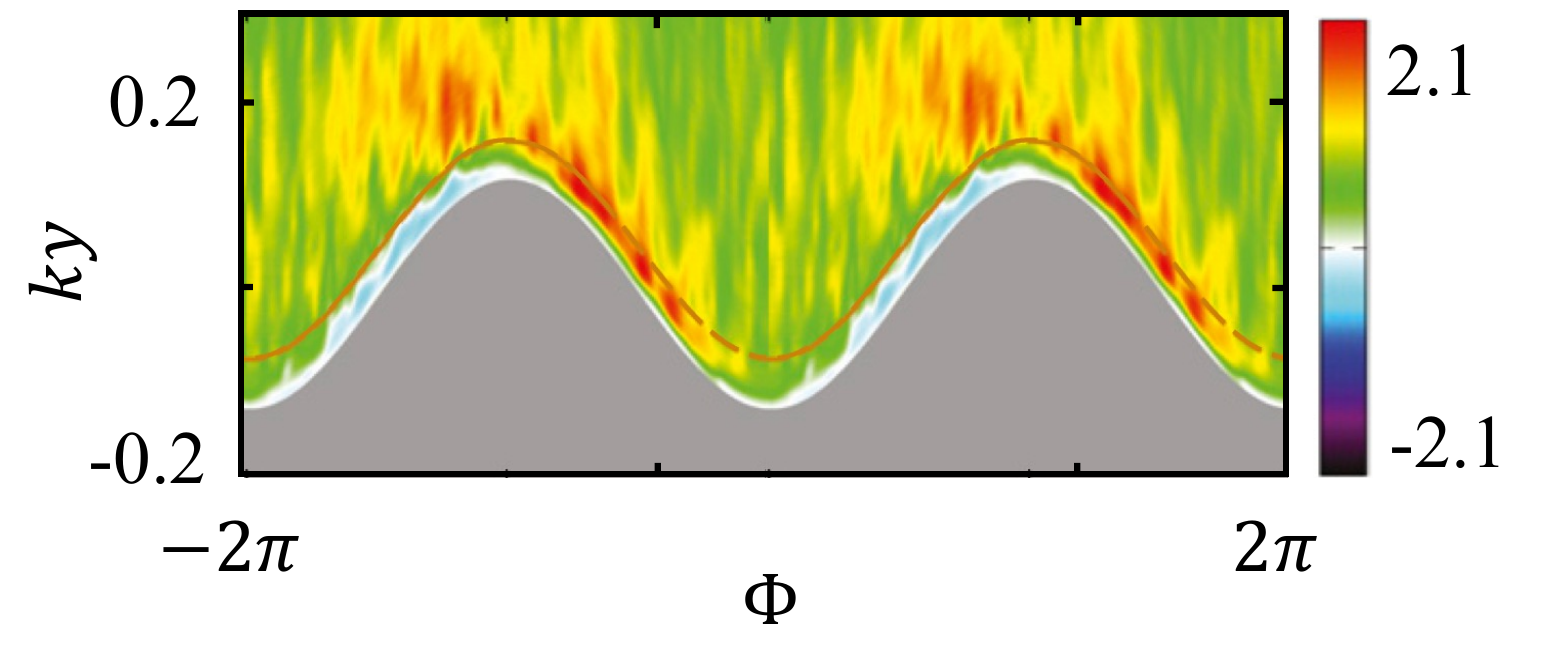}
     \end{subfigure}
     \begin{subfigure}[b]{0.9\columnwidth}
         \centering
         \caption{}
         \includegraphics[width=\textwidth]{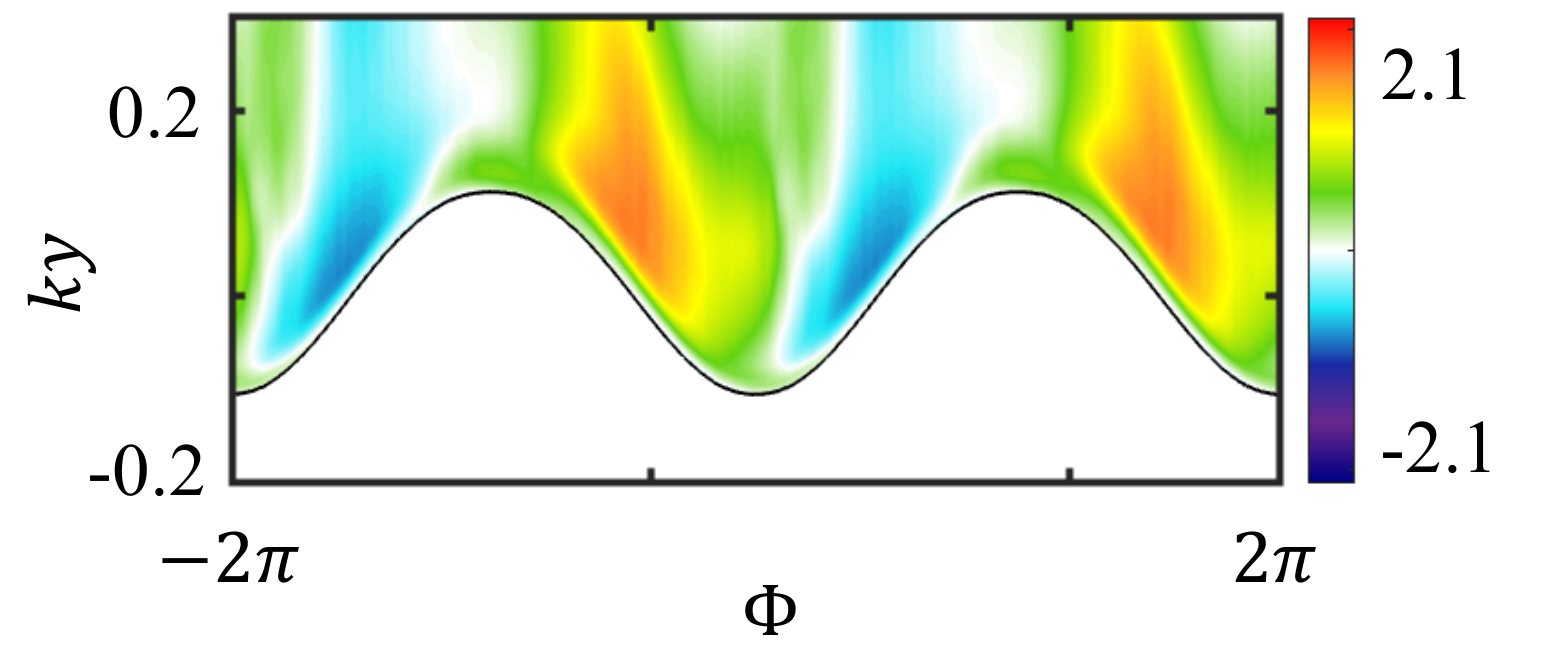}
     \end{subfigure}
    \caption{
    (a,b) Normalized phase-averaged wave-coherent stress and (c,d) turbulent stress from the old wave simulation. Colormaps depict the stress distributions. Simulation results (b,d) are compared with the experimental reference data from \citet{buckley2016}. Gray and orange dashed lines indicate the inner and critical layers, respectively. \textcolor{black}{Only the wind-side (air) fields are shown. The water-side velocity and stress fields closely follow the prescribed inlet Stokes wave profile, consistent with the experimental observations of \citet{buckley2016}, and are therefore omitted to focus on the wind-side response.}
    \textcopyright~American Meteorological Society. Used with permission.
    }
    \label{fig:old wave: buckley comparison}
\end{figure*}

To investigate the origin of this deviation, we compare the spatial distribution of stresses between the simulation and the experimental data. Only the results obtained using the \texttt{plicRDF} method are presented and shown in Fig.~\ref{fig:old wave: buckley comparison}. The comparison of the wave-coherent stress field illustrates that \texttt{plicRDF} successfully identifies the narrow region of positive stress near the interface \citep{belcher1998turbulent}. The maximum positive stress on the windward side is slightly weaker in the simulations, which contributes to the slightly smaller peak observed in the vertical profiles. Nonetheless, the overall stress distribution and its approximate magnitude are satisfactorily captured.

For the turbulent stress components, the simulation qualitatively captures a negative–positive pattern near the interface. \textcolor{black}{However, the simulation misses two critical features evident in the stress distribution in Fig.~\ref{fig:old wave: buckley comparison}. The positive stress near the interface on the leeward side is too weak, and the stress distribution away from the interface is nearly symmetric rather than asymmetric. Together, these two omissions lead to the severe underprediction of the vertical profile in Fig.~\ref{fig:old wave: triple decomposition}.} Similar behavior is observed in the \texttt{gradPhi} results (not shown). This discrepancy may arise from several factors, such as the grid aspect ratio or the accuracy of the wind inlet profile. Recall that this fast-moving wave is expected to generate a positive stress region on the leeward side because the relative wind speed is slower than the wave speed. However, this stress may dissipate too quickly beyond the crest. 
Another possible source of this error is insufficient resolution near the interface, as the smallest grid spacing, $\Delta y_{\text{min}} \approx 10.9l_{\nu}$, is much larger than the viscous length scale $l_\nu = \textcolor{black}{3.66 \times 10^{-5}\,\text{m}}$.
These issues are beyond the scope of the present study.


In conclusion, comparison with the experimental data of \citet{buckley2016} yields several important findings. First, the results reaffirm that spurious currents must be effectively mitigated when the wind and wave velocities are of comparable magnitude, as they significantly distort both the mean and fluctuating components of the flow. In addition, the interface regularization term contributes to the slight overprediction of wave-coherent stress by enhancing the mass flux above the crest under non-negligible wind conditions.
Second, the overall spatial structures of both the wave-coherent and turbulent stress fields are reproduced with reasonable agreement to the experimental measurements, indicating that the dominant flow features are captured. Third, although the current grid resolution resolves the thin stress layers associated with the wave-coherent stress, it fails to reproduce the asymmetry of the turbulent stress. In particular, the stress level on the leeward side remains insufficient to sustain turbulence, and it decays rapidly beyond the crest. The implications of this underprediction require further examination, since resolving the detailed mechanisms of turbulent generation at the interface would substantially increase the computational cost and may limit the practical feasibility of such simulations.

\section{Conclusions\label{sec:conclusions}}

In this study, we conducted a comprehensive investigation into \textcolor{black}{how \textcolor{black}{curvature-estimation} and flux-discretization errors affect wind–wave simulations through the resulting spurious currents and interface regularization,} under high wave-age conditions.
Our findings highlight how two critical numerical factors substantially influence wind–wave simulations. The first factor concerns the generation of spurious currents arising from \textcolor{black}{curvature-estimation} errors. All examined methods exhibit a lack of curvature-error convergence under grid refinement, resulting in persistent spurious velocities in the vicinity of the interface. When the magnitude of these unphysical currents approaches that of the mean flow, they contaminate all components of the triple decomposition, most notably the phase-independent mean velocity. We show that improving curvature estimation by replacing the \texttt{isoPhi} method with the \texttt{plicRDF} method significantly enhances simulation accuracy, reduces wave dissipation, and restores asymptotic convergence with grid refinement.
This issue is particularly relevant to laboratory-scale validation studies, where waves are characterized by relatively low wave celerity and high curvature. Under such conditions, curvature effects become non-negligible. Curvature-induced errors may appear secondary in inertia-dominated wave regimes, where interface transport remains largely unaffected. However, the present results show that they become critical when wind-induced turbulence is considered at comparable length scales.
Although the turbulent component remains underpredicted even at the finest resolution, the overall agreement with the experimental data improves substantially when the RDF-based approach is employed.

The second critical factor is the artificial mass flux introduced by the interface regularization. The \texttt{gradPhi} approach, which is representative of interface-sharpening schemes more broadly, generates an additional flux layer in under-resolved simulations. The thickness of this artificial flux layer is comparable to the numerical interface thickness, and its magnitude scales with the local interface-transport velocity. As a result, even in high-speed flows, this artificial flux becomes non-negligible when the grid resolution is insufficient.
The effects of these factors accumulate across both time and space because of the continuous interaction between the wind and the surface waves throughout the computational domain. Clear signatures of the resulting errors appear in phase-averaged fields and in their corresponding vertical profiles.

\todo{Add comments on what would be diff. for young wave}

\begin{acknowledgments}
This study was supported by the National Science Foundation under Grant No. CMMI-21311961. The authors acknowledge the Stanford Research Computing Center for providing computational resources and technical support that contributed to these results.
\end{acknowledgments}

\appendix

\todo{Read my notes and clarify the difference between RDF and isoAlpha.}

\section{Interface reconstruction\label{sec:method:reconstruction of the interface}}

To utilize the isoAdvector algorithm introduced in Section~\ref{sec:method:geometric VOF}, the interface normal vector $n_{\mathcal{F}(p),i}$ and interface velocity $u_{\mathcal{F}(p),i}$ must be obtained through interface reconstruction. 
Details of the reconstruction methods are described in~\citet{scheufler2019accurate}. For completeness, we summarize two reconstruction procedures here. One is based directly on the phase indicator field and the other utilizes the RDF.

\subsection{$\phi$-Based reconstruction\label{sec:method:isoPhi resconstruction}}
The procedure for identifying the $\phi_p$-isoface is outlined as follows:

\begin{enumerate} 

\item Identify interface cells that satisfy $\epsilon < \phi_p < 1 - \epsilon$, where $\epsilon$ is a specified tolerance used to define interface cells. In this study, we set $\epsilon = 10^{-8}$.

\item Interpolate the phase indicator $\phi$ to the vertices of the cell. The volume fraction of phase1 at vertex $f_v$ is calculated as
\begin{equation}
    f_v = \frac{\sum_{k\in C_v} w_k \phi_k}{\sum_{k \in C_v} w_k},
    \label{eq:vof vertex interpolation, grad-alpha}
\end{equation}
where $w_k$ denotes the inverse distance weights, $k$ is a vertex index, and $C_v$ represents the set of vertices connected to vertex $v$.

\item Iterate on all cell edges to determine the intersection points corresponding to a target volume fraction value $\phi_p^*$. The value of $\phi_p^*$ may be selected from one of the interpolated vertex values $f_v$.

\item For each cell face, the identified intersection points are connected across faces to form an intersection line, e.g., $\overline{II'}$ in Fig.~\ref{fig:isoface schematic}.

\item These intersection lines are then connected to construct the periphery of the polygonal face, for example, the blue plane shown in Fig.~\ref{fig:isoface schematic}. Note that it is not guaranteed to form a planar face.

\item Compute the submerged volume $V(\phi_p^*)$ of the interface cell. Steps 3 to 5 are repeated iteratively, updating $\phi_p^*$ at each iteration, until the condition $V(\phi_p^*)/V_{\text{cv}} = \phi_p$ is satisfied. It is important to note that $\phi_p$ and $\phi_p^*$ are not necessarily equal.

\item Calculate the centroid $x_{\mathcal{F}(p),i}$ and the unit normal vector $n_{\mathcal{F}(p),i}$ of the $\phi_p$-isoface. The isoface is a planar $N_v$-gonal surface composed of $N_v$ vertices, which together form $N_v$ sub-triangles sharing a common average point. The face normal vector is computed as the sum of the normals of the sub-triangles
\begin{equation}
    n_{\mathcal{F}(p),i} = \sum_{k = 1}^{N_v} n_{k,i} = \sum_{k = 1}^{N_v} \frac{1}{2}(x_{k+1,i} - x_{k,i}) \times (\overline{x}_{i} - x_{k,i}),
    \label{eq:normal of polygonal face}
\end{equation}
and the centroid of the isoface is computed as the area-weighted average of the sub-triangle centers
\begin{equation}
    x_{\mathcal{F}(p),i} = 
    \sum_{k = 1}^{N_v} \frac{|n_{k,i}|}{|n_{\mathcal{F}(p),i}|} \left(\frac{x_{k+1,i} + x_{k,i} + \overline{x}_i}{3}\right),
    \label{eq:center of polygonal face}
\end{equation}
where $\overline{x}_i = \frac{1}{N_v} \sum_{k=1}^{N_v} x_{k,i}$ is the geometric center of all vertices, and $n_{k,i}$ is the normal vector of each sub-triangle.

The unit normal vector of the $\phi_p$-isoface is then given by
\begin{equation}
    \hat{n}_{\mathcal{F}(p),i} = \frac{n_{\mathcal{F}(p),i}}{|n_{\mathcal{F}(p),i}|}.
    \label{eq:isoface unit normal}
\end{equation}

\item Decompose the cell into tetrahedra, each sharing the cell center as a common vertex. Once the tetrahedron containing the isoface center $x_{\mathcal{F}(p),i}$ is identified, the interface velocity $u_{\mathcal{F}(p),i}$ is obtained by linear interpolation from the velocity values at the vertices of that tetrahedron.

\end{enumerate}

As mentioned earlier, although geometric VOF-type methods provide a sharp representation of the interface, they often yield inaccurate curvature estimates due to their reliance on the large gradient of the phase indicator across the interface. 

\subsection{RDF-based reconstruction\label{sec:method:plicRDF reconstruction algorithm}}
To mitigate the inaccurate \textcolor{black}{curvature-estimation} issue, a signed-distance-like function has been proposed to reconstruct a smoother auxiliary variable, enabling more accurate curvature estimation and reducing spurious velocity artifacts~\citep{cummins2005estimating}. In this context, \citet{scheufler2019accurate} proposed a residual-based iterative approach to compute the face center $x_{\mathcal{F}(p),i}$ and unit normal vector $n_{\mathcal{F}(p),i}$ more accurately. 

The procedure is as follows:

\begin{enumerate} 

\item Initialize the interface normal $n_{\mathcal{F}(p),i}$ using either the value from the previous time step or the gradient of the phase indicator field.

\item If the interface is sufficiently resolved (e.g., see Eq.~(18) in \citet{scheufler2019accurate}), determine the face center $x_{\mathcal{F}(p),i}$ by following the steps outlined in Section~\ref{sec:method:isoPhi resconstruction}, with the following two modifications:

First, replace the interpolation in Eq.~\eqref{eq:vof vertex interpolation, grad-alpha} with the projected distances to each vertex of cell $p$,
\begin{equation}
    d_k = x_{k,i} n_{\mathcal{F}(p),i},
    \label{eq:plicRDF:distance to vertex}
\end{equation}
where $x_{k,i}$ is the position of vertex $k$ and $n_{\mathcal{F}(p),i}$ is the current estimate of the interface normal.

Second, instead of computing the normal from Eq.~\eqref{eq:normal of polygonal face}, directly use the defined interface normal $n_{\mathcal{F}(p),i}$ for reconstruction.

\item Construct the RDF, $\Psi_p$, at the center of cell $p$ as
\begin{equation}
    \Psi_p = \frac{\sum_q w_{pq} \widetilde{\Psi}_{pq}}{\sum_q w_{pq}},
    \label{eq: RDF construction}
\end{equation}
where the summation is taken over neighboring interface cells $q$. The interface distance contribution from each neighboring cell $q$ is defined as,
\begin{equation}
    \widetilde{\Psi}_{pq} = n_{\mathcal{F}(q),i} \left( x_{p,i} - x_{\mathcal{F}(q),i} \right),
    \label{eq: tilde psi definition}
\end{equation}
where $x_{p,i}$ denotes the centroid of cell $p$, and $x_{\mathcal{F}(q),i}$ and $n_{\mathcal{F}(q),i}$ are the centroid and unit normal of the $\phi$-isoface in neighboring cell $q$, respectively.

The weights $w_{pq}$ are calculated based on the alignment between the interface direction and the distance vector:
\begin{equation}
    w_{pq} = \frac{\left| n_{\mathcal{F}(q),i} \left( x_{p,i} - x_{\mathcal{F}(q),i} \right) \right|^2}{\left| x_{p,i} - x_{\mathcal{F}(q),i} \right|^2}.
    \label{eq:weights to calculate RDF}
\end{equation}

\item Update the interface normal using the gradient of the RDF,
\begin{equation}
    n_{\mathcal{F}(p),i} = \frac{\partial \Psi / \partial x_i}{\left| \partial \Psi / \partial x_i \right|}.
\end{equation}

\item Repeat steps 2 to 4 iteratively until the normal interface change between successive iterations falls below the prescribed tolerance.

\end{enumerate}

Due to the smooth nature of the RDF field, this method provides more reliable and accurate interface normals, particularly in regions where the interface curvature plays a critical role in flow dynamics.

\section{Wave-following coordinate\label{sec:wave-following coordinate systeme}}

The coordinates $(\xi, \zeta)$ define a wave-following reference frame, which adapts to the shape of the surface waves, as illustrated in Fig.~\ref{fig:wave following coordinate}. Various formulations exist for constructing such coordinates, including those proposed by \citet{shen2001} and \citet{buckley2016}. Following the latter, we define the wave-following coordinate system as
\begin{eqnarray}
    \xi(x,z) &=& x - \textnormal{i} \sum_{n} a_n \exp\left[\textnormal{i}(k_n \xi + \phi_n)\right] \exp(-k_n \zeta), \qquad\\
    \zeta(x,z) &=& z - \sum_{n} a_n \exp\left[\textnormal{i}(k_n \xi + \phi_n)\right] \exp(-k_n \zeta), 
    \label{eq:wave-following coordinate} 
\end{eqnarray} 
where $a_n$, $k_n$, and $\phi_n$ represent the amplitude, wavenumber, and phase of the $n$-th Fourier mode, respectively. These coefficients are obtained from the Fourier decomposition of the wave elevation $\eta(\mathbf{x}, t)$, expressed as
\begin{equation} 
\eta(\xi) = \sum_{n} a_n \exp\left[\textnormal{i}(k_n \xi + \phi_n)\right]. 
\end{equation}

The velocity field $\mathbf{u}$ is then interpolated from the original Cartesian coordinate system to the wave-following coordinate system.

For discretized surface points located at $x = x_0$, the corresponding wave-following coordinate is $\xi = \xi_{x_0}$, since $x_0 = \xi_{x_0}$ at $\zeta = 0$ (cf. Eq.\eqref{eq:wave-following coordinate}). The phase domain $\Phi \in [0, 2\pi]$ is uniformly divided into $N_{\text{bins}}$ intervals. Using Eq.\eqref{eq:phase identification}, the instantaneous phase of the wave at each interpolated velocity point $\mathbf{u}(\xi_{x_0}, \zeta)$ is evaluated, and the corresponding velocity data are sorted into the appropriate phase bins. The phase-averaged velocity field $\langle \mathbf{u} \rangle(\xi, \zeta)$ is then obtained by averaging within each bin.



\bibliography{prf}

@string{jcp="J. Comput. Phys."}

@string{jfm="J. Fluid Mech."}

@string{ce="Coast. Eng."}

@string{jpo="J. Phys. Oceanogr."}

@article{sussman2000coupled,
  title={A coupled level set and volume-of-fluid method for computing 3D and axisymmetric incompressible two-phase flows},
  author={Sussman, Mark and Puckett, Elbridge Gerry},
  journal={Journal of computational physics},
  volume={162},
  number={2},
  pages={301--337},
  year={2000},
  publisher={Elsevier}
}

@article{hirt1981volume,
  title={Volume of fluid (VOF) method for the dynamics of free boundaries},
  author={Hirt, Cyril W and Nichols, Billy D},
  journal={Journal of computational physics},
  volume={39},
  number={1},
  pages={201--225},
  year={1981},
  publisher={Elsevier}
}

@article{liu2003phase,
  title={A phase field model for the mixture of two incompressible fluids and its approximation by a Fourier-spectral method},
  author={Liu, Chun and Shen, Jie},
  journal={Physica D: Nonlinear Phenomena},
  volume={179},
  number={3-4},
  pages={211--228},
  year={2003},
  publisher={Elsevier}
}

@article{osher1988fronts,
  title={Fronts propagating with curvature-dependent speed: Algorithms based on Hamilton-Jacobi formulations},
  author={Osher, Stanley and Sethian, James A},
  journal={Journal of computational physics},
  volume={79},
  number={1},
  pages={12--49},
  year={1988},
  publisher={Elsevier}
}

@article{denner2014fully,
  title={Fully-coupled balanced-force VOF framework for arbitrary meshes with least-squares curvature evaluation from volume fractions},
  author={Denner, Fabian and van Wachem, Berend GM},
  journal={Numerical Heat Transfer, Part B: Fundamentals},
  volume={65},
  number={3},
  pages={218--255},
  year={2014},
  publisher={Taylor \& Francis}
}

@article{dodet2019contribution,
  title={The contribution of wind-generated waves to coastal sea-level changes},
  author={Dodet, Guillaume and Melet, Ang{\'e}lique and Ardhuin, Fabrice and Bertin, Xavier and Idier, D{\'e}borah and Almar, Rafael},
  journal={Surveys in Geophysics},
  volume={40},
  number={6},
  pages={1563--1601},
  year={2019},
  publisher={Springer}
}

@article{chini2010impact,
  title={The impact of sea level rise and climate change on inshore wave climate: A case study for East Anglia (UK)},
  author={Chini, Nicolas and Stansby, Peter and Leake, James and Wolf, Judith and Roberts-Jones, Jonah and Lowe, Jason},
  journal={Coastal Engineering},
  volume={57},
  number={11-12},
  pages={973--984},
  year={2010},
  publisher={Elsevier}
}

@article{laitone1960second,
  title={The second approximation to cnoidal and solitary waves},
  author={Laitone, Edmund V},
  journal={Journal of fluid mechanics},
  volume={9},
  number={3},
  pages={430--444},
  year={1960},
  publisher={Cambridge University Press}
}

@article{abadie2015combined,
  title={On the combined effects of surface tension force calculation and interface advection on spurious currents within volume of fluid and level set frameworks},
  author={Abadie, Thomas and Aubin, Joelle and Legendre, Dominique},
  journal={Journal of Computational Physics},
  volume={297},
  pages={611--636},
  year={2015},
  publisher={Elsevier}
}

@article{popinet2018numerical,
  title={Numerical models of surface tension},
  author={Popinet, St{\'e}phane},
  journal={Annual Review of Fluid Mechanics},
  volume={50},
  pages={49--75},
  year={2018},
  publisher={Annual Reviews}
}

@article{gamet2020validation,
  title={Validation of volume-of-fluid OpenFOAM{\textregistered} isoAdvector solvers using single bubble benchmarks},
  author={Gamet, Lionel and Scala, Marco and Roenby, Johan and Scheufler, Henning and Pierson, Jean-Lou},
  journal={Computers \& Fluids},
  volume={213},
  pages={104722},
  year={2020},
  publisher={Elsevier}
}

@book{gardner1989statistical,
  title={Statistical spectral analysis—A nonprobabilistic theory},
  author={Gardner, William A and Robinson, Enders A},
  year={1989}
}

@article{Jofre2015,
author = {Jofre, Llu{\'{i}}s and Borrell, Ricard and Lehmkuhl, Oriol and Oliva, Assensi},
doi = {10.1016/j.jcp.2014.11.009},
issn = {10902716},
journal = {Journal of Computational Physics},
month = {feb},
pages = {269--288},
publisher = {Elsevier Inc.},
title = {{Parallel load balancing strategy for Volume-of-Fluid methods on 3-D unstructured meshes}},
url = {http://dx.doi.org/10.1016/j.jcp.2014.11.009},
volume = {282},
year = {2015}
}

@article{zalesak1979fully,
  title={Fully multidimensional flux-corrected transport algorithms for fluids},
  author={Zalesak, Steven T},
  journal={Journal of computational physics},
  volume={31},
  number={3},
  pages={335--362},
  year={1979},
  publisher={Elsevier}
}

@article{cummins2005estimating,
  title={Estimating curvature from volume fractions},
  author={Cummins, Sharen J and Francois, Marianne M and Kothe, Douglas B},
  journal={Computers \& structures},
  volume={83},
  number={6-7},
  pages={425--434},
  year={2005},
  publisher={Elsevier}
}

@article{scheufler2019accurate,
  title={Accurate and efficient surface reconstruction from volume fraction data on general meshes},
  author={Scheufler, Henning and Roenby, Johan},
  journal={Journal of computational physics},
  volume={383},
  pages={1--23},
  year={2019},
  publisher={Elsevier}
}

@article{francois2006balanced,
  title={A balanced-force algorithm for continuous and sharp interfacial surface tension models within a volume tracking framework},
  author={Francois, Marianne M and Cummins, Sharen J and Dendy, Edward D and Kothe, Douglas B and Sicilian, James M and Williams, Matthew W},
  journal={Journal of Computational Physics},
  volume={213},
  number={1},
  pages={141--173},
  year={2006},
  publisher={Elsevier}
}

@Article{brackbill,
  author = 	 {J. U. Brackbill and D. B. Kothe and C. Zemach},
  title = 	 {A continuum method for modeling surface tension},
  journal = 	jcp,
  volume={100},
  pages={335--354},
  year={1991},
}

@article{cao2021numerical,
  title={A numerical and theoretical study of wind over fast-propagating water waves},
  author={Cao, Tao and Shen, Lian},
  journal={Journal of Fluid Mechanics},
  volume={919},
  pages={A38},
  year={2021},
  publisher={Cambridge University Press}
}

@article{lin2008direct,
  title={Direct numerical simulation of wind-wave generation processes},
  author={Lin, Mei-Ying and Moeng, Chin-Hoh and Tsai, Wu-Ting and Sullivan, Peter P and Belcher, Stephen E},
  journal={Journal of Fluid Mechanics},
  volume={616},
  pages={1--30},
  year={2008},
  publisher={Cambridge University Press}
}

@article{powers2017weather,
  title={The weather research and forecasting model: Overview, system efforts, and future directions},
  author={Powers, Jordan G and Klemp, Joseph B and Skamarock, William C and Davis, Christopher A and Dudhia, Jimy and Gill, David O and Coen, Janice L and Gochis, David J and Ahmadov, Ravan and Peckham, Steven E and others},
  journal={Bulletin of the American Meteorological Society},
  volume={98},
  number={8},
  pages={1717--1737},
  year={2017},
  publisher={American Meteorological Society}
}

@article{makin2005note,
  title={A note on the drag of the sea surface at hurricane winds},
  author={Makin, Vladimir K},
  journal={Boundary-layer meteorology},
  volume={115},
  pages={169--176},
  year={2005},
  publisher={Springer}
}

@article{holthuijsen2012wind,
  title={Wind and waves in extreme hurricanes},
  author={Holthuijsen, Leo H and Powell, Mark D and Pietrzak, Julie D},
  journal={Journal of Geophysical Research: Oceans},
  volume={117},
  number={C9},
  year={2012},
  publisher={Wiley Online Library}
}

@article{andreas1992sea,
  title={Sea spray and the turbulent air-sea heat fluxes},
  author={Andreas, Edgar L},
  journal={Journal of Geophysical Research: Oceans},
  volume={97},
  number={C7},
  pages={11429--11441},
  year={1992},
  publisher={Wiley Online Library}
}

@article{hara2002wind,
  title={Wind forcing in the equilibrium range of wind-wave spectra},
  author={Hara, Tetsu and Belcher, Stephen E},
  journal={Journal of Fluid Mechanics},
  volume={470},
  pages={223--245},
  year={2002},
  publisher={Cambridge University Press}
}

@article{hara2015wave,
  title={Wave boundary layer turbulence over surface waves in a strongly forced condition},
  author={Hara, Tetsu and Sullivan, Peter P},
  journal={Journal of Physical Oceanography},
  volume={45},
  number={3},
  pages={868--883},
  year={2015}
}

@article{roenby2016computational,
  title={A computational method for sharp interface advection},
  author={Roenby, Johan and Bredmose, Henrik and Jasak, Hrvoje},
  journal={Royal Society open science},
  volume={3},
  number={11},
  pages={160405},
  year={2016},
  publisher={The Royal Society}
}

@phdthesis{marquez2013extended,
    author ={M{\'a}rquez Dami{\'a}n, Santiago and Nigro, Norberto Marcelo and Bohorquez Rodr{\'\i}guez de Medina, Patricio and Cantero, Mariano Ignacio and D'Elia, Jorge and Larreteguy, Axel and Storti, Mario Alberto},
    title ={An extended mixture model for the simultaneous treatment of short and long scale interfaces},
    school = {Universidad Nacional del Litoral},
    year = {2013}
}

@article{yang2010direct,
  title={Direct-simulation-based study of turbulent flow over various waving boundaries},
  author={Yang, DI and Shen, Lian},
  journal={Journal of Fluid Mechanics},
  volume={650},
  pages={131--180},
  year={2010},
  publisher={Cambridge University Press}
}

@article{sullivan2000simulation,
  title={Simulation of turbulent flow over idealized water waves},
  author={Sullivan, Peter P and McWilliams, James C and Moeng, Chin-Hoh},
  journal={Journal of Fluid Mechanics},
  volume={404},
  pages={47--85},
  year={2000},
  publisher={Cambridge University Press}
}

@article{song2004numerical,
  title={A numerical study of breaking waves},
  author={Song, Chiyoon and Sirviente, Ana I},
  journal={Physics of fluids},
  volume={16},
  number={7},
  pages={2649--2667},
  year={2004},
  publisher={American Institute of Physics}
}

@article{edson2013exchange,
  title={On the exchange of momentum over the open ocean},
  author={Edson, James B and Jampana, Venkata and Weller, Robert A and Bigorre, Sebastien P and Plueddemann, Albert J and Fairall, Christopher W and Miller, Scott D and Mahrt, Larry and Vickers, Dean and Hersbach, Hans},
  journal={Journal of Physical Oceanography},
  volume={43},
  number={8},
  pages={1589--1610},
  year={2013},
  publisher={American Meteorological Society}
}

@article{Park2018,
  title={Numerical modeling of non-breaking, impulsive breaking, and broken wave interaction with elevated coastal structures: laboratory validation and inter-model comparisons},
  author={Park, H. and Do, T. and Tomiczek, T. and Cox, D. T. and van de Lindt, J. W.},
  journal={Ocean Engineering},
  volume={158},
  pages={78-98},
  year={2018}
}

@article{Huang2009,
  title={Numerical modeling of dynamic wave force acting on {E}cambia bay bridge deck during hurricane {I}van},
  author={Huang, W. and Xiao, H. and Bryan, G.H. and Rotunno, R.},
  journal={Journal of Waterway, Port, Coastal, and Ocean Engineering},
  volume={135},
  issue={4},
  pages={164-175},
  year={2009}
}

@article{hwang2024robust,
  title={A robust phase-field method for two-phase flows on unstructured grids},
  author={Hwang, Hanul and Jain, Suhas S},
  journal={Journal of Computational Physics},
  volume={507},
  pages={112972},
  year={2024},
  publisher={Elsevier}
}

@article{breivik2015surface,
  title={Surface wave effects in the NEMO ocean model: Forced and coupled experiments},
  author={Breivik, {\O}yvind and Mogensen, Kristian and Bidlot, Jean-Raymond and Balmaseda, Magdalena Alonso and Janssen, Peter AEM},
  journal={Journal of Geophysical Research: Oceans},
  volume={120},
  number={4},
  pages={2973--2992},
  year={2015},
  publisher={Wiley Online Library}
}

@article{poletto2013new,
  title={A new divergence free synthetic eddy method for the reproduction of inlet flow conditions for LES},
  author={Poletto, R and Craft, T and Revell, A},
  journal={Flow, turbulence and combustion},
  volume={91},
  pages={519--539},
  year={2013},
  publisher={Springer}
}

@article{Ren2014,
  title={Coupled wind-wave time domain analysis of floating offshore wind turbine based on computational fluid dynamics method},
  author={Ren, N. and Li, Y. and Ou, J.},
  journal={Journal of Renewable and Sustainable Energy},
  volume={6},
  issue={2},
  year={2014}
}

@article{Liu2017,
  title={Establishing a fully coupled {CFD} analysis tool for floating offshore wind turbines},
  author={Liu, Y. and Xiao, Q. and Incecik, A. and Peyrard, C. and Wan, D.},
  journal={Renewable Energy},
  volume={112},
  issue={280-301},
  year={2017}
}

@article{sullivan2014large,
  title={Large-eddy simulation of marine atmospheric boundary layers above a spectrum of moving waves},
  author={Sullivan, Peter P and McWilliams, James C and Patton, Edward G},
  journal={Journal of the Atmospheric Sciences},
  volume={71},
  number={11},
  pages={4001--4027},
  year={2014},
  publisher={American Meteorological Society}
}

@article{iafrati2013modulational,
  title={Modulational instability, wave breaking, and formation of large-scale dipoles in the atmosphere},
  author={Iafrati, A and Babanin, A and Onorato, M},
  journal={Physical review letters},
  volume={110},
  number={18},
  pages={184504},
  year={2013},
  publisher={APS}
}

@article{sullivan2018turbulent,
  title={Turbulent flow over steep steady and unsteady waves under strong wind forcing},
  author={Sullivan, Peter P and Banner, Michael L and Morison, Russel P and Peirson, William L},
  journal={Journal of Physical Oceanography},
  volume={48},
  number={1},
  pages={3--27},
  year={2018},
  publisher={American Meteorological Society}
}

@article{suzuki2013impact,
  title={Impact of breaking wave form drag on near-surface turbulence and drag coefficient over young seas at high winds},
  author={Suzuki, Nobuhiro and Hara, Tetsu and Sullivan, Peter P},
  journal={Journal of physical oceanography},
  volume={43},
  number={2},
  pages={324--343},
  year={2013}
}

@article{belcher1998turbulent,
  title={Turbulent flow over hills and waves},
  author={Belcher, SE and Hunt, JCR},
  journal={Annual Review of Fluid Mechanics},
  volume={30},
  number={1},
  pages={507--538},
  year={1998},
  publisher={Annual Reviews 4139 El Camino Way, PO Box 10139, Palo Alto, CA 94303-0139, USA}
}

@article{banner1998tangential,
  title={Tangential stress beneath wind-driven air--water interfaces},
  author={Banner, Michael L and Peirson, William L},
  journal=jcp,
  volume={364},
  pages={115--145},
  year={1998},
  publisher={Cambridge University Press}
}

@article{banner1976separation,
  title={On the separation of air flow over water waves},
  author={Banner, Michael L and Melville, W Kendall},
  journal=jcp,
  volume={77},
  number={4},
  pages={825--842},
  year={1976},
  publisher={Cambridge University Press}
}

@article{yang2018direct,
  title={Direct numerical simulation of wind turbulence over breaking waves},
  author={Yang, Zixuan and Deng, Bing-Qing and Shen, Lian},
  journal=jcp,
  volume={850},
  pages={120--155},
  year={2018},
  publisher={Cambridge University Press}
}

@article{wu2022revisiting,
  title={Revisiting wind wave growth with fully coupled direct numerical simulations},
  author={Wu, Jiarong and Popinet, St{\'e}phane and Deike, Luc},
  journal={Journal of Fluid Mechanics},
  volume={951},
  pages={A18},
  year={2022},
  publisher={Cambridge University Press}
}

@article{higuera2013,
  title="Realistic wave generation and active wave absorption for Navier--Stokes models: Application to OpenFOAM{\textregistered}",
  author="Higuera, P. and Lara, J.~L. and Losada, I.~J.",
  journal=ce,
  volume="71",
  pages="102--118",
  year="2013",
  publisher="Elsevier"
}

@article{cimarelli2023,
  title={On wind--wave interaction phenomena at low Reynolds numbers},
  author={Cimarelli, A. and Romoli, F. and Stalio, E.},
  journal={Journal of Fluid Mechanics},
  volume={956},
  pages={A13},
  year={2023},
  publisher={Cambridge University Press},
  doi={10.1017/jfm.2023.4}
}

@article{jacobsen2012waves2foam,
  title={A wave generation toolbox for the open-source CFD library: OpenFoam{\textregistered}},
  author={Jacobsen, N.~G. and Fuhrman, D.~R. and Freds{\o}e, J.},
  journal={International Journal for Numerical Methods in Fluids},
  volume={70},
  number={9},
  pages={1073--1088},
  year={2012},
  publisher={Wiley}
}

@article{buckley2016,
  title={Structure of the airflow above surface waves},
  author={Buckley, M. P. and Veron, F.},
  journal=jpo,
  volume={46},
  number={5},
  pages={1377--1397},
  year={2016}
}

@article{shen2001,
  title={Large-eddy simulation of free-surface turbulence},
  author={Shen, L. and Yue, D. K. P.},
  journal=jfm,
  volume={440},
  pages={75--116},
  year={2001},
  publisher={Cambridge University Press}
}

@article{renardy2002prost,
  title={PROST: a parabolic reconstruction of surface tension for the volume-of-fluid method},
  author={Renardy, Yuriko and Renardy, Michael},
  journal={Journal of computational physics},
  volume={183},
  number={2},
  pages={400--421},
  year={2002},
  publisher={Elsevier}
}

@article{aboukhedr2018simulation,
  title={Simulation of micro-flow dynamics at low capillary numbers using adaptive interface compression},
  author={Aboukhedr, Mahmoud and Georgoulas, Anastasios and Marengo, Marco and Gavaises, Manolis and Vogiatzaki, Konstantina},
  journal={Computers \& Fluids},
  volume={165},
  pages={13--32},
  year={2018},
  publisher={Elsevier}
}

@article{hoang2013benchmark,
  title={Benchmark numerical simulations of segmented two-phase flows in microchannels using the Volume of Fluid method},
  author={Hoang, Duong A and van Steijn, Volkert and Portela, Luis M and Kreutzer, Michiel T and Kleijn, Chris R},
  journal={Computers \& Fluids},
  volume={86},
  pages={28--36},
  year={2013},
  publisher={Elsevier}
}

@article{georgoulas2015numerical,
  title={Numerical investigation of quasi-static bubble growth and detachment from submerged orifices in isothermal liquid pools: The effect of varying fluid properties and gravity levels},
  author={Georgoulas, Anastasios and Koukouvinis, P and Gavaises, Manolis and Marengo, Marco},
  journal={International Journal of Multiphase Flow},
  volume={74},
  pages={59--78},
  year={2015},
  publisher={Elsevier}
}

@article{mulbah2022review,
  title={A review of VOF methods for simulating bubble dynamics},
  author={Mulbah, Christian and Kang, Can and Mao, Ning and Zhang, Wei and Shaikh, Ali Raza and Teng, Shuang},
  journal={Progress in Nuclear Energy},
  volume={154},
  pages={104478},
  year={2022},
  publisher={Elsevier}
}

@article{lafaurie1994modelling,
  title={Modelling merging and fragmentation in multiphase flows with SURFER},
  author={Lafaurie, Bruno and Nardone, Carlo and Scardovelli, Ruben and Zaleski, St{\'e}phane and Zanetti, Gianluigi},
  journal={Journal of Computational Physics},
  volume={113},
  number={1},
  pages={134--147},
  year={1994},
  publisher={Elsevier}
}

@article{theillard2021volume,
  title={A volume-preserving reference map method for the level set representation},
  author={Theillard, Maxime},
  journal={Journal of Computational Physics},
  volume={442},
  pages={110478},
  year={2021},
  publisher={Elsevier}
}

@article{rycroft2020reference,
  title={Reference map technique for incompressible fluid--structure interaction},
  author={Rycroft, Chris H and Wu, Chen-Hung and Yu, Yue and Kamrin, Ken},
  journal={Journal of Fluid Mechanics},
  volume={898},
  pages={A9},
  year={2020},
  publisher={Cambridge University Press}
}

\end{document}